\documentclass[12pt]{article}
\usepackage{epsfig}

\usepackage{mathrsfs}
\usepackage[T1]{fontenc}
\usepackage{mathpazo}
\usepackage{setspace}
\usepackage{amsfonts}
\usepackage{amssymb}
\usepackage{amsmath}
\usepackage{epsfig}
\usepackage{latexsym}
\usepackage{color}
\usepackage{graphicx}
\usepackage{nicefrac}
\usepackage[latin1]{inputenc}
\usepackage{pstricks}
\usepackage{slashed}

\def\hybrid{\topmargin -20pt    \oddsidemargin 0pt
        \headheight 0pt \headsep 0pt
        \textwidth 6.25in       
        \textheight 9.5in       
        \marginparwidth .875in
        \parskip 5pt plus 1pt   \jot = 1.5ex}

\hybrid

\catcode`\@=11

\def\marginnote#1{}
\newcount\hour
\newcount\minute
\newtoks\amorpm
\hour=\time\divide\hour by60 \minute=\time{\multiply\hour by60
\global\advance\minute by-\hour}
\edef\standardtime{{\ifnum\hour<12 \global\amorpm={am}%
        \else\global\amorpm={pm}\advance\hour by-12 \fi
        \ifnum\hour=0 \hour=12 \fi
        \number\hour:\ifnum\minute<10 0\fi\number\minute\the\amorpm}}
\edef\militarytime{\number\hour:\ifnum\minute<10
0\fi\number\minute}

\def\draftlabel#1{{\@bsphack\if@filesw {\let\thepage\relax
   \xdef\@gtempa{\write\@auxout{\string
      \newlabel{#1}{{\@currentlabel}{\thepage}}}}}\@gtempa
   \if@nobreak \ifvmode\nobreak\fi\fi\fi\@esphack}
        \gdef\@eqnlabel{#1}}
\def\@eqnlabel{}
\def\@vacuum{}
\def\draftmarginnote#1{\marginpar{\raggedright\scriptsize\tt#1}}

\def\draft{\oddsidemargin -.5truein
        \def\@oddfoot{\sl preliminary draft \hfil
        \rm\thepage\hfil\sl\today\quad\militarytime}
        \let\@evenfoot\@oddfoot \overfullrule 3pt
        \let\label=\draftlabel
        \let\marginnote=\draftmarginnote
   \def\@eqnnum{(\theequation)\rlap{\kern\marginparsep\tt\@eqnlabel}%
\global\let\@eqnlabel\@vacuum}  }

\def\preprint{\twocolumn\sloppy\flushbottom\parindent 2em
        \leftmargini 2em\leftmarginv .5em\leftmarginvi .5em
        \oddsidemargin -.5in    \evensidemargin -.5in
        \columnsep .4in \footheight 0pt
        \textwidth 10.in        \topmargin  -.4in
        \headheight 12pt \topskip .4in
        \textheight 6.9in \footskip 0pt
        \def\@oddhead{\thepage\hfil\addtocounter{page}{1}\thepage}
        \let\@evenhead\@oddhead \def\@oddfoot{} \def\@evenfoot{} }

\def\numberbysection{\@addtoreset{equation}{section}
        \def\theequation{\thesection.\arabic{equation}}}

\def\underline#1{\relax\ifmmode\@@underline#1\else
        $\@@underline{\hbox{#1}}$\relax\fi}

\def\titlepage{\@restonecolfalse\if@twocolumn\@restonecoltrue\onecolumn
     \else \newpage \fi \thispagestyle{empty}\c@page\z@
        \def\thefootnote{\fnsymbol{footnote}} }

\def\endtitlepage{\if@restonecol\twocolumn \else \newpage \fi
        \def\thefootnote{\arabic{footnote}}
        \setcounter{footnote}{0}}  

\catcode`@=12 \relax

\def\figcap{\section*{Figure Captions\markboth
        {FIGURECAPTIONS}{FIGURECAPTIONS}}\list
        {Figure \arabic{enumi}:\hfill}{\settowidth\labelwidth{Figure
999:}
        \leftmargin\labelwidth
        \advance\leftmargin\labelsep\usecounter{enumi}}}
 \relax
\def\tablecap{\section*{Table Captions\markboth
        {TABLECAPTIONS}{TABLECAPTIONS}}\list
        {Table \arabic{enumi}:\hfill}{\settowidth\labelwidth{Table
999:}
        \leftmargin\labelwidth
        \advance\leftmargin\labelsep\usecounter{enumi}}}
 \relax
\def\reflist{\section*{References\markboth
        {REFLIST}{REFLIST}}\list
        {[\arabic{enumi}]\hfill}{\settowidth\labelwidth{[999]}
        \leftmargin\labelwidth
        \advance\leftmargin\labelsep\usecounter{enumi}}}
 \relax
\makeatletter
\newcounter{pubctr}
\def\publist{\@ifnextchar[{\@publist}{\@@publist}}
\def\@publist[#1]{\list
        {[\arabic{pubctr}]\hfill}{\settowidth\labelwidth{[999]}
        \leftmargin\labelwidth
        \advance\leftmargin\labelsep
        \@nmbrlisttrue\def\@listctr{pubctr}
        \setcounter{pubctr}{#1}\addtocounter{pubctr}{-1}}}
\def\@@publist{\list
        {[\arabic{pubctr}]\hfill}{\settowidth\labelwidth{[999]}
        \leftmargin\labelwidth
        \advance\leftmargin\labelsep
        \@nmbrlisttrue\def\@listctr{pubctr}}}
 \relax
\makeatother
\newskip\humongous \humongous=0pt plus 1000pt minus 1000pt

\newif\ifdtup

\relax

\def\be{\begin{equation}}
\def\ee{\end{equation}}
\def\ba{\begin{eqnarray}}
\def\ea{\end{eqnarray}}

\renewcommand{\theequation}{\thesection.\arabic{equation}}

\newcommand{\eqn}[1]{(\ref{#1})}

\author{
  \begin{minipage}{.97\linewidth}
    \vspace{0cm}
    \begin{center}
      \begin{small}
        \textbf{Ioannis Bakas}\footnote{bakas@ajax.physics.upatras.gr} ${\ }^1$ and
        \textbf{Dieter L\"ust}\footnote{dieter.luest@lmu.de} ${\ }^{2,3,4}$
      \end{small}
    \end{center}
    \vspace{0.5cm}
    \hspace{2cm}\begin{minipage}{.7\linewidth}
     {\it \begin{footnotesize}
    \begin{itemize}
        \item[${}^1$] Department of Physics, University of Patras\\
       GR--26500 Patras, Greece
      \item[${}^2$] Max-Planck-Institut f\"ur Physik\\
       F\"ohringer Ring 6, D--80805 M\"unchen, Germany
       \item[${}^3$] Arnold-Sommerfeld-Center f\"ur Theoretische Physik\\
        Department f\"ur Physik, Ludwig-Maximilians-Universit\"at M\"unchen\\
        Theresienstra\ss e 37, D--80333 M\"unchen, Germany
       \item[${}^4$] Theory Division, Department of Physics, CERN\\
        CH--1211 Geneva 23, Switzerland
       \end{itemize}
     \end{footnotesize}}
    \end{minipage}
    \vspace{0.5cm}
  \end{minipage}
}

\title{\vspace{-0.1cm}
 \boldmath \begin{LARGE}
    \textbf{Axial anomalies of Lifshitz fermions}
  \end{LARGE} \unboldmath
}

\begin{document}

\renewcommand{\thepage}{\arabic{page}}
\setcounter{page}{1}


\begin{titlepage}
  \maketitle
  \thispagestyle{empty}

  \vspace{-13.2cm}
  \begin{flushright}
    LMU-ASC 07/11, ~~
    MPP-2011-35, ~~
    CERN-PH-TH/2011-051
  \end{flushright}

  \vspace{11cm}

  \begin{center}
    \textsc{Abstract}\\
  \end{center}
  We compute the axial anomaly of a Lifshitz fermion theory with anisotropic scaling
  $z=3$ which is minimally coupled to geometry in $3+1$ space--time dimensions. We find
  that the result is identical to the relativistic case using path integral methods. An
  independent verification is provided by showing with spectral methods that the
  $\eta$--invariant of the Dirac and Lifshitz fermion operators in three dimensions are equal.
  Thus, by the integrated form of the anomaly, the index of the Dirac operator still accounts
  for the possible breakdown of chiral symmetry in non--relativistic theories of gravity.
  We apply this framework to the recently constructed gravitational instanton backgrounds
  of Ho\v{r}ava--Lifshitz theory and find that the index is non--zero provided that
  the space--time foliation admits leaves with harmonic spinors. Using Hitchin's
  construction of harmonic spinors on Berger spheres, we obtain explicit results
  for the index of the fermion operator on all such gravitational instanton backgrounds
  with $SU(2) \times U(1)$ isometry. In contrast to the instantons of Einstein gravity,
  chiral symmetry breaking becomes possible in the unimodular phase of Ho\v{r}ava--Lifshitz
  theory arising at $\lambda = 1/3$ provided that the volume of space is bounded from below
  by the ratio of the Ricci to Cotton tensor couplings raised to the third power. Some
  other aspects of the anomalies in non--relativistic quantum field theories are also discussed.

\end{titlepage}

\tableofcontents


\section{Introduction}
\setcounter{equation}{0}

Since the discovery of the axial anomaly and its profound success in explaining the
two--photon decay mode of the $\pi^0$ meson \cite{bell, adler}, there has been
a lot of activity in the study of anomalous symmetries in quantum field theory and
their geometrical and topological origin (see, for instance, \cite{jackiw}
for a recent brief overview of the subject and its history) as well as in model
building\footnote{It is beyond the scope of the present work to give a comprehensive
account of the many important contributions made over the years by several groups of
people, as we will only focus on the axial anomaly in its simplest form. Several expository
contributions on the subject can be found, for instance, in \cite{bardeen}.}. The axial anomaly
is accounted perturbatively by the one--loop fermion contribution (triangle diagram)
of the two--photon coupling to an axial current, as in spinor electrodynamics, but
it also has an alternative description in terms of the non--invariance of the Euclidean
path integral measure under chiral rotations of the fermions coupled to a background
gauge field \cite{fuji} (see also the textbook \cite{suzuki}). This framework was soon
afterwards generalized to fermions coupled to a background metric by computing the
gravitational corrections to the axial anomaly associated to the two--graviton
decay mode of the $\pi^0$ meson \cite{salam, eguchi1, nielsen}, which also enjoys a path
integral derivation \cite{fuji, suzuki} as for the case of gauge fields (see also
\cite{gaume} for the structure of gravitational anomalies in general dimensions). These
results refer to the so called local form of the anomaly.

Another important development is related to the integrated form of axial anomaly
which can be non--zero when the background gauge field is non--Abelian and
topologically non--trivial. In this case, the four--dimensional Dirac operator
exhibits normalizable zero modes with unequal number of positive and negative
chirality solutions (denoted by $n_{\pm}$ respectively) in the background of an
instanton configuration. Thus, there is chiral symmetry breaking induced by
instantons that may in turn lead to baryon and lepton number violation in
Yang--Mills theories \cite{hooft}. This novel possibility is a physical manifestation
of the Atiyah--Singer index theorem \cite{AS} (but see also the textbooks
\cite{gilka, ezra}) that equates the index of the
four--dimensional Dirac operator, namely $n_+ - n_-$, to the topological charge $k$
of the background gauge field on the compactified Euclidean space-time $S^4$.
(In the actual instanton solutions one has $n_+ = k > 0$ and $n_- = 0$, whereas for
the anti--instantons the situation is reversed, i.e., $n_- = - k > 0$ and $n_+ = 0$.)

Likewise, the integrated form of the gravitational correction to the axial anomaly
is a topological invariant given by (one eighth of) the Hirzebruch signature of
space-time, which can be non-zero on topologically
non--trivial gravitational backgrounds and may lead to similar results for chiral
symmetry breaking by the index theorem of Dirac operator. This is certainly
true for the Dirac operator on $K3$, which is the unique compact self--dual
gravitational instanton without boundaries \cite{K3a, K3b}; in this case there are
two covariantly constant negative helicity modes and no positive helicity modes
so that $n_+ - n_- = -2$. However, gravitational instantons with boundaries do
not admit any normalizable zero modes of the Dirac operator, provided that
the Ricci scalar curvature of space--time is non--negative (as it is by Einstein's field
equations when the cosmological constant is non--negative), and by a general result
based on Lichnerowicz's theorem \cite{lichn} the index of Dirac operator vanishes
on such backgrounds \cite{gibbons}. An
alternative derivation of the same result is based on the Atiyah--Patodi--Singer
index theorem \cite{APS, melrose, gilka} taking into proper account the boundary terms
of space--time from the asymptotic regions
\cite{hawking, extra1, romer, eguchi2, pope, extra2, extra3, extra4}.
Thus, although the signature of space--time is non--zero, the index of the
Dirac operator turns out to be zero for all gravitational instanton backgrounds with
boundaries, in which case there is no chiral symmetry breaking induced by
quantum tunneling effects. For a systematic exposition of all these matters
we refer the interested reader to the report \cite{eguchi3}. More explanations
will also be given in the text.

In this paper we examine the chiral anomaly for non--relativistic fermions
of Lifshitz type that are coupled to background fields in four space-time dimensions.
Our investigation was prompted by the current activity on non-relativistic theories
of gravity with anisotropic scaling, serving as alternatives to the ultraviolet
completion of Einstein theory of gravitation with higher order spatial derivative
terms \cite{horava1, horava2}. The resulting theory sacrifices relativistic
invariance for perturbative renormalizability at short distances and despite criticism
related to the presence of an unphysical mode and the difficulty to flow to ordinary
gravity at large distances (see, for instance, \cite{blas, thomas} and references therein,
as well as \cite{hennea}),
it provides an interesting framework that should be given the benefit of the doubt to play
a useful role in quantum gravity (see also \cite{horava3} for a more recent proposal that
defies some of the criticism). Coupling the theory to Lifshitz fermions with
the same anisotropic scaling in space and time, as for gravity, is a step that we
take in this work and which, apparently, has not been considered in the literature so
far. The computations we present in the following provide the gravitational contribution
to the axial anomaly at a Lifshitz point and, as such, they generalize rather naturally
the results on the axial anomaly for Lifshitz fermions coupled to gauge field backgrounds
\cite{wadia}. In either case, the local form of the anomaly turns out to be identical
to the relativistic case. These results are surprising at first sight, but clearly they
should be related to the topological character of the anomaly that is proportional to
the characteristic classes ${\rm Tr}(F \wedge F)$ and ${\rm Tr}(R \wedge R)$ for gauge
and metric fields, respectively. They are also quite general since they treat the gauge and
metric fields as background without reference to their field equations. As such, they
provide the local form of the axial anomaly at a Lifshitz point and they are independent
of the particular form of the bosonic action. This can make them useful for applications in
other fields of current research that incorporate ideas of quantum criticality.

The integrated form of the axial anomaly at a Lifshitz point is also identical to the
relativistic case, and, therefore, it equals the index of the relativistic Dirac
operator in a given background. This will be shown explicitly by proving an index theorem
for the Dirac--Lifshitz operator and by computing the $\eta$--invariant of the operator on
the three--dimensional leaves of space--time foliation. These results can be regarded as an
alternative proof that the anomalies are the same in both theories. The index is also inert to
the back--reaction of fermions on the gauge and metric fields. Thus, it makes good sense to
compute the index on the background of an instanton solution, as in the relativistic case.
Here, we pass directly to $(3+1)$--dimensional Ho\v{r}ava--Lifshitz gravity\footnote{The
other interesting case of a Lifshitz type gauge theory (in the spirit proposed in
\cite{horava4}) coupled to Lifshitz fermions and the integrated form of the axial
anomaly in the background of gauge fields will not be treated at all, for a good reason
that will be discussed later,
even though it looks more elementary. Hopefully, we will turn to this and related issues
elsewhere.} coupled to Lifshitz fermions with anisotropic scaling $z=3$ using its
simplest form (said to satisfy the detailed balance condition \cite{horava1, horava2})
that exhibits instanton solutions in the Euclidean regime, \cite{BBLP}.
We will find explicit examples that are capable to produce a non--vanishing index and,
hence, lead to chiral symmetry breaking by gravitational instanton effects. These
instantons are very different in nature from the gravitational instanton solutions of
ordinary Einstein theory in that they are not self--dual spaces; they resemble more the
instanton solutions of point particle systems
that interpolate continuously between different vacua as the Euclidean time runs from
$-\infty$ to $+\infty$ and they have the topology $\mathbb{R} \times \Sigma_3$
(e.g., $\Sigma_3 \simeq S^3$ in the typical examples that we will consider later). These
configurations are also chiral in that they exist for one choice of orientation
on $\Sigma_3$ and not for the other, but as it turns out by explicit computation they
should be "sufficiently chiral" to allow for non--vanishing index (the precise meaning of
these words will be given in due course). This possibility will
be made available in the phase of Ho\v{r}ava--Lifshitz theory that exhibits
spatial unimodular invariance in sharp contrast to ordinary Einstein gravity that never
allows for chiral symmetry breaking induced by instantons. The
results can be regarded as an application of our instanton solutions \cite{BBLP} to
non--relativistic theories of gravitation.

The material of this paper is organized as follows: In section 2, we present
our results for the axial anomaly of a Lifshitz fermion theory which is minimally
coupled to gauge and metric backgrounds and show that in either case they are
the same with the relativistic theory of Dirac fermions, hereby extending the
results of Dhar et.al. \cite{wadia} to the gravitational form of the anomaly. In section
3, we present a short overview of Ho\v{r}ava--Lifshitz gravity in $3+1$ dimensions
and describe its instantons in terms of eternal solutions of a certain geometric
flow equation called Ricci--Cotton flow, following \cite{BBLP}. We also summarize
all instanton solutions with $SU(2)$ isometry that
will be used as working examples in the applications. In section 4, we compute the
index of the Dirac--Lifshitz operator on these gravitational instanton backgrounds by
taking proper account of the boundary terms. This index counts the difference of
positive and negative chirality zero modes of a Lifshitz fermion and it turns out
to be non-zero on certain instanton backgrounds in the unimodular phase of the
theory, which arises when the parameter $\lambda$ of superspace assumes
a particular value (to be explained later) that is special to the deep ultra--violet
regime of the theory. In this context, we also obtain a critical value for the
couplings, equivalently for the volume of space, beyond which zero modes become possible.
The index is always zero in all other phases of the theory. The index can become
non--zero for geometrical rather than topological reasons (this is a special feature
associated  to the presence of harmonic spinors in three dimensions according to
Hitchin \cite{hitchin}, but see also \cite{bar}),
which explains the lower bound on the volume of space that will be obtained later.
Although the results are derived using the class of $SU(2)$ symmetric solutions, they
serve as prototype for the more general situation.
In section 5, we compare our results with the index of the Dirac operator on
gravitational instanton backgrounds of ordinary Einstein theory focusing, in
particular, to instantons with $SU(2)$ isometry for which direct comparison
can be easily made. It is reconfirmed that the index is zero in that case.
In section 6, we present our conclusions and discuss some of the implications
of anomalies in quantum field theories of Lifshitz type. Directions for future work are
also outlined.

Finally, there are four appendices with some technical material. Appendix A
refers to the algebra of Dirac matrices in flat Euclidean space $\mathbb{R}^4$ and
contains some useful mathematical identities that involve the trace of products
of gamma--matrices. They will all be needed for the evaluation of the axial anomalies.
Appendix B contains some geometrical apparatus that is needed for the coupling
of fermions to geometry and the evaluation of the anomaly based on the notion of
geodesic interval and the associated Synge--DeWitt tensors \cite{witt1, synge} (but see
also \cite{hada}).
Appendix C summarizes the Euclidean path integral method used to obtain the local
form of the axial anomaly in the background of gauge and metric fields in relativistic
quantum field theory, following \cite{fuji, suzuki}. It sets up the notation and framework
for carrying out the same computation in the non--relativistic theory of Lifshitz
fermions. Appendix D summarizes some mathematical details related to harmonic spinors
on $S^3$ equipped with homogeneous metrics, following \cite{hitchin, bar}, which are
needed for computing the index of the fermion operator on our gravitational instanton
backgrounds. Explicit results are included for the case of Berger spheres that support
gravitational instantons with $SU(2) \times U(1)$ group of isometries.

\section{Axial anomalies of Lifshitz fermions}
\setcounter{equation}{0}

In this section, we consider the non--relativistic theory of Lifshitz fermions
and discuss some general properties including the symmetry of axial rotations
in the massless case. We also set up the formalism and compute the quantum anomaly
to the divergence of the corresponding axial current in close analogy with the
relativistic Dirac theory. The generalization to fermions $\psi$ with mass $m$ can
be easily made by adding the classical term $m \bar{\psi} \gamma_5 \psi$ to the
divergence of the axial current, as usual, but this will not be discussed further.

\subsection{Lifshitz fermions and axial symmetry}

The theory of Lifshitz fermions in $3+1$ space--time dimensions is defined by
the following action
\be
S = \int dt d^3x \sqrt{|{\rm det} G |} ~ \bar{\psi} ~i \gamma^{\mu}
\mathbb{D}_{\mu} \psi
\label{startipoint}
\ee
that resembles that of a Dirac fermion $\psi$. The conjugate field is defined, as usual,
by $\bar{\psi} = \psi^{\dagger} \gamma^0$. The action is generally taken in
curved space-time with local coordinates $(t, x^i)$ (assuming the presence of a
privileged time direction) and metric $G_{\mu \nu}$ with signature $-+++$,
whereas $\gamma^{\mu}$ are the standard Dirac matrices satisfying the anti--commutation
relations
\be
[\gamma^{\mu} , ~ \gamma^{\nu}]_+ = 2 G^{\mu \nu} ~.
\ee
The difference, however, comes in the definition of the operator $\mathbb{D}_{\mu}$,
which is of mixed order,
\be
\mathbb{D}_0 = D_t ~, ~~~~~ \mathbb{D}_i = {1 \over 2} \left(D_i (-D^2) +
(-D^2) D_i \right) ,
\label{lifsh}
\ee
where $D_{\mu}$ are the usual covariant derivatives expressing the minimal coupling
of the fermion field $\psi$ to the background geometry and/or background gauge
fields and $D^2 = D_i D^i$. Clearly, this is a non--relativistic
theory\footnote{Note that the theory of Lifshitz fermions we are considering
here is rather different from that of non--relativistic fermions based on
Galilean--covariant field theory \cite{hagen}. The latter can be obtained from
a relativistic massless fermion theory in one dimension higher by performing a
Scherk--Schwarz reduction along a null direction and they exhibit different
behavior.} with anisotropic scaling $z=3$, which is seen by assigning
scaling dimensions $[L] = -1$, $[T] = -3$ and $[\psi] = 3/2$. Then, the analogue
of the Dirac equation and its conjugate equation read as follows,
\be
\gamma^{\mu} \mathbb{D}_{\mu} \psi = 0 ~, ~~~~~~
(\mathbb{D}_{\mu} \bar{\psi}) \gamma^{\mu} = 0 ~.
\ee

The action \eqn{startipoint} is naturally defined on space--times that are
topologically of the form $\mathbb{R} \times \Sigma_3$ using the privileged time
direction and the corresponding foliation by three--dimensional spatial slices
$\Sigma_3$. Then, the space--time metric $G_{\mu \nu}$ takes the general form
\be
ds^2 = -N^2 dt^2 +g_{ij}\left(dx^i +N^i dt\right) \left(dx^j +N^j dt\right)
\ee
using lapse and shift functions $N (t, x)$ and $N_i (t, x)$, respectively, and
$g_{ij} (t, x)$, as in
the ADM (Arnowitt--Deser--Misner) formulation of Einstein gravity (see, for instance,
\cite{adm}). The metric on $\Sigma_3$ is $g_{ij}$ providing the
induced metric on the leaves of the foliation, i.e., $g_{ij} = G_{ij}$, whereas
$G^{ij} = g^{ij} - \hat{N}^i \hat{N}^j$ with $\hat{N}^i = N^i / N$. We also
have $\sqrt{|{\rm det} G |} = N \sqrt{{\rm det} g}$. Here, we confine
ourselves to the so called {\em projectable} case, meaning that the lapse function $N$
associated to the freedom of time reparametrization is restricted to be a function
of $t$ alone, whereas the shift functions $N_i$ associated to diffeomorphisms of
$\Sigma_3$ can depend on all space--time coordinates. This is natural in all Lifshitz
theories that do not exhibit general coordinate invariance but only a restricted
space--time symmetry associated to foliation preserving diffeomorphisms
$\tilde{t} = \tilde{t}$, ${\tilde{x}}^i = {\tilde{x}}^i (t, x)$. Furthermore, in
order to simplify certain aspects of the presentation, and, also, in view of the
applications we will make in subsequent sections, we adopt the choice of lapse and
shift functions
\be
N(t) = 1 ~, ~~~~~~ N_i (t, x) = 0
\ee
unless stated otherwise. Setting $N(t)=1$ amounts to using proper time in the projectable
case and $N_i (t, x) = 0$ amounts to having $G^{ij} = g^{ij}$ so that the spatial space--time
indices can be lowered and raised using the three--dimensional metric on $\Sigma_3$ without
worrying about off--diagonal space--time terms.

Next, we examine the symmetries of the classical action \eqn{startipoint}. First of all there
is a non--relativistic variant of the vector current $J^{\mu} (t, x) = \bar{\psi} \gamma^{\mu}
\psi$ associated to the invariance of the action under $\psi \rightarrow {\rm exp}(i \alpha)
\psi$ and $\bar{\psi} \rightarrow {\rm exp}(-i \alpha) \bar{\psi}$, but its details will not
be relevant to the present work. In any case, the vector symmetry is also present quantum
mechanically, since there is no anomaly to obstruct its conservation law.
Also, since the mass of the fermions is zero, the action \eqn{startipoint} is invariant under
chiral rotations,
\be
\psi \rightarrow e^{i \alpha \gamma_5} \psi ~, ~~~~~
\bar{\psi} \rightarrow \bar{\psi} e^{i \alpha \gamma_5} ~,
\label{chirot}
\ee
using $\gamma_5 = (i/24) \epsilon_{\mu \nu \kappa \lambda}
\gamma^{\mu} \gamma^{\nu} \gamma^{\kappa} \gamma^{\lambda}$ (it takes the standard form
$\gamma_5 = i \gamma^0 \gamma^1 \gamma^2 \gamma^3$ in terms of tangent space--time indices
associated to a local Lorentz frame),
as for the relativistic case. It follows that there is an associated
axial current $J_5^{\mu}$, which is classically conserved by virtue
of the equations of motion,
\be
\nabla_{\mu} J_5^{\mu} (t, x) \equiv {1 \over \sqrt{|{\rm det} G |}}
~ \partial_{\mu} \left(\sqrt{|{\rm det} G |} ~ J_5^{\mu} \right) = 0 ~,
\ee
where its time and space components are chosen to be
\be
J_5^0 (t, x) = \bar{\psi} \gamma^0 \gamma_5 \psi
\ee
and
\ba
J_5^i (t, x) & = & {1 \over 2} \Big[ (D_j \bar{\psi}) \gamma^j \gamma_5 (D^i \psi) +
(D^i \bar{\psi}) \gamma^j \gamma_5 (D_j \psi) +
\bar{\psi} \gamma^i \gamma_5 (-D^2 \psi) + \nonumber\\
& & ~~~ (-D^2 \bar{\psi}) \gamma^i \gamma_5 \psi -
\bar{\psi} \gamma^j \gamma_5 (D^i D_j \psi) -
(D^i D_j \bar{\psi}) \gamma^j \gamma_5 \psi \Big] ~.
\ea
These are the components of a real current that should be compared to the
corresponding components of the axial current $J_5^{\mu} (t, x) = \bar{\psi}
\gamma^{\mu} \gamma_5 \psi$ of the relativistic Dirac theory, which are, of course,
much simpler. The situation is more analogous to the current conservation law
associated to the time dependent Schr\"odinger equation in non--relativistic
quantum mechanics.

Before we proceed further, an important remark is in order regarding the uniqueness
of the Dirac--Lifshitz operator. There is a factor ordering ambiguity in the construction
of its spatial components because the covariant derivatives that we raise to the third power
do not commute -- their commutator is proportional to the field strength of the background
field (it is the Riemann curvature in the case of a gravitational field). Another choice
that could be made equally well is provided by
\be
\mathbb{D}_i^{\prime} = - D_j D_i D^j ~,
\label{lliffsh}
\ee
in which case the spatial components of the associated axial current should be replaced by
\be
{J_5^{\prime}}^i (t, x) = (D_j \bar{\psi}) \gamma^i \gamma_5 (D^j \psi) -
\bar{\psi} \gamma^j \gamma_5 (D_j D^i \psi) -
(D_j D^i \bar{\psi}) \gamma^j \gamma_5 \psi
\ee
to ensure current conservation with respect to the new (prime) Dirac--Lifshitz equation. Note,
however, that this ambiguity is irrelevant for the purposes of the present work, since
$\mathbb{D}_i$ and $\mathbb{D}_i^{\prime}$ differ from each other by a total derivative
term of the background field strength, which can not affect the local form of the anomaly,
since its functional form is protected by topology. Throughout this paper we will adopt the
particular choice \eqn{lifsh}, even though the alternative choice \eqn{lliffsh} looks simpler
as it admits a simpler current. This will not affect much the intermediate steps of the
calculation of the axial anomaly\footnote{Actually, in the paper \cite{wadia} that
only discusses the gauge field contribution to the axial anomaly of Lifshitz fermions, the
issue of factor ordering is not addressed at all (not even the Hermiticity properties of the
Dirac--Lifshitz operator) because this turns out to be irrelevant for the intermediate steps
of the calculation, and, of course, for the final result.}. Finally, we note for completeness
that other factor orderings also provide viable choices for the Dirac--Lifshitz operator
(for instance, one may choose the Weyl--ordered third order operator
$\mathbb{D}_i^{\prime \prime} = (\mathbb{D}_i + \mathbb{D}_i^{\prime})/2$), but they all have
the same degeneracy.

We can perform path integral quantization of the Lifshitz fermions
coupled to background gauge or metric fields and examine the fate
of axial current conservation. We will derive the corresponding axial
anomaly following the method of Fujikawa \cite{fuji, suzuki} by taking the
theory in the Euclidean domain. First, using the standard Noether
procedure, the fermionic action transforms as follows under the
chiral rotation \eqn{chirot} with parameter $\alpha (t, x)$,
\be
S \rightarrow S + \int dt d^3x \sqrt{|{\rm det} G |} ~ \alpha (t, x)
\nabla_{\mu} J_5^{\mu} (t, x) ~.
\label{action}
\ee
In the quantum theory, this transformation is compensated by the
Jacobian of the path integral measure for the fermions
$({\cal D} \bar{\psi}) ({\cal D} \psi)$ and their combination yields
the quantum correction to the axial current conservation law.

In the Euclidean regime, where we will work from now on,
the signature of space--time is assumed to be $++++$ (opposite to the
$----$ convention often adopted in the literature). For this, we perform
the Wick rotation $t \rightarrow it$ and $\gamma_0 \rightarrow i \gamma_0$
so that $(\gamma^{\mu})^{\dagger} = \gamma^{\mu}$ for all space--time indices.
In the Euclidean regime, the Dirac operator $i \gamma^{\mu} D_{\mu}$,
as well as the Dirac--Lifshitz operator $i \gamma^{\mu} \mathbb{D}_{\mu}$ we
are considering here are both Hermitian, whereas $\gamma_5$ becomes $\gamma_5 =
- \gamma_0 \gamma_1 \gamma_2 \gamma_3$ and it anti--commutes, as usual, with
the Dirac matrices, i.e., $[\gamma_5 , ~ \gamma^{\mu}]_+ = 0$, and
$(\gamma_5)^{\dagger} = \gamma_5$. The Euclidean space $\gamma_5$ follows from
its Lorentzian counterpart by letting $\gamma_5 \rightarrow i \gamma_5$; this
should be properly accounted when comparing \eqn{action} with the transformation
of the fermionic measure that is most conveniently described in the Euclidean
regime. The algebra of Dirac matrices in $\mathbb{R}^4$ together with their
trace identities that will be relevant for the computation of the axial anomalies
are summarized in Appendix A.

The path integral over the fermions with Euclidean action $S_{\rm E}$ is defined as
\be
Z = \int ({\cal D} \bar{\psi}) ({\cal D} \psi) e^{-S_{\rm E}}
\ee
setting Planck's constant equal to 1.
For now and later use, we consider the complete set of eigen--functions
$\varphi_n$ of the Hermitian operator $i \gamma^{\mu} \mathbb{D}_{\mu}$,
\be
(i \gamma^{\mu} \mathbb{D}_{\mu}) \varphi_n (t, x) = \lambda_n
\varphi_n (t, x) ~,
\ee
which are normalized as
\be
\int dt d^3x \sqrt{{\rm det} G} ~ \varphi_n^{\dagger} (t, x)
\varphi_m (t, x) = \delta_{n m} ~.
\ee
Then, a general fermion configuration is decomposed into eigen--states
as $\psi (t,x) = \sum_n a_n \varphi_n (t, x)$ and $\bar{\psi} (t,x) =
\sum_n \varphi_n^{\dagger} (t, x) \bar{b}_n$, using independent elements
$a_n$ and $\bar{b}_n$ of the Grassmann algebra, whereas the fermionic path
integral measure is formally written as $({\cal D} \bar{\psi}) ({\cal D} \psi)
= \prod_m \bar{b}_m \prod_n a_n$. Under the chiral rotation \eqn{chirot},
the coefficients $a_n$ and $\bar{b}_n$ can be easily seen to transform as
$a_n \rightarrow \sum_m C_{n m} a_m$ and $\bar{b}_n \rightarrow
\sum_m C_{n m} \bar{b}_m$, where
\be
C_{n m} = \int dt d^3x \sqrt{{\rm det} G} ~ \varphi_n^{\dagger} (t, x)
e^{i \alpha (t, x) \gamma_5} \varphi_m (t, x) ~.
\ee
As a result, the fermionic path integral measure picks up a Jacobian factor
and transforms as
\be
({\cal D} \bar{\psi}) ({\cal D} \psi) \rightarrow ({\cal D} \bar{\psi})
({\cal D} \psi) ~ {\rm exp} \left(-2i \int dt d^3x \sqrt{{\rm det} G} ~
\alpha (t, x) \sum_n \varphi_n^{\dagger} (t, x) \gamma_5 \varphi_n (t, x)
\right) .
\label{measure}
\ee
Consequently, the primitive local form of the axial anomaly follows by
combining the result of transformations \eqn{action} and \eqn{measure} in the
Euclidean regime and it reads
\be
\nabla_{\mu} J_5^{\mu} (t, x) = 2 \sum_n \varphi_n^{\dagger} (t, x)
\gamma_5 \varphi_n (t, x) ~.
\label{primitive}
\ee

Note that all steps outlined above are identical to the path integral
formulation of the axial anomaly for Dirac fermions followed by simply
replacing $D_{\mu}$ by $\mathbb{D}_{\mu}$. However, the actual evaluation
is more tricky and it requires computing the formal sum (trace of $\gamma_5$)
shown in \eqn{primitive} over the complete set of eigen--functions of the
interacting theory. In all cases, this sum is ill--defined, as it formally
diverges, and appropriate regularization is required to extract the local
form of the anomaly. The standard procedure is to regularize the large
eigen--values, i.e., $|\lambda_n| \leq M^z$ (accounting also for their scaling dimension),
and then obtain finite result for the right--hand side of equation \eqn{primitive} by
computing
\be
A(t, x) = \lim_{M \rightarrow \infty} \Big[\sum_n \varphi_n^{\dagger} (t, x)
\gamma_5 ~ e^{- \lambda_n^2 / M^{2z}} \varphi_n (t, x) \Big] ~.
\label{limita}
\ee
Note the appearance of the anisotropy scaling parameter $z$ in the cut--off
$M$, which is $z=1$ for Dirac fermions and $z=3$ for the Lifshitz fermions
we are considering here.

The final expression for the anomaly depends upon the background gauge and/or
metric field via the corresponding field strength and it should be gauge
invariant. It should also be topological density so that $A(t, x)$ can be
locally written as total derivative term to account for the anomaly in the
divergence of the axial current \eqn{primitive} and, as such, its form
is very restricted (up to an overall factor). The actual terms that contribute
to the evaluation of the anomaly depend crucially on the operator and the
associated  eigen--values and functions $\lambda_n$ and $\varphi_n$ that enter
into \eqn{limita}; in a diagrammatic approach to the same problem different
loop diagrams contribute to the answer, depending on the available
couplings to the background fields. Nevertheless, the outcome is proportional to
${\rm Tr} (F \wedge F)$ for the gauge and ${\rm Tr} (R \wedge R)$ for the
metric field contribution to the axial anomaly, as required on topological
grounds that severely constrain the form of anomalous terms. It can also be seen
without much effort that the other topological density in four dimensions,
${\rm Tr} (R \wedge {}^{\star} R)$, can not possibly contribute to the answer for
it has an excess of anti--symmetric Levi--Civita tensors.

In the following, we present the explicit computation of the anomaly terms
\eqn{limita} for the case of Lifshitz fermions coupled to background
gauge and metric fields and show that they are the same, including the
overall numerical factors, as for the Dirac fermions that are summarized in
Appendices B and C. This particular result is not obvious from the beginning and has
important consequences to the integrated form of the anomaly. Here, we choose
to work with Fujikawa's path integral method for evaluating the anomaly and leave
to the interested reader the diagrammatic interpretation of the individual
terms that are contributing to the final answer.

\subsection{Gauge field contribution to the anomaly}

We will first compute the axial anomaly of Lifshitz fermions coupled to an
Abelian or non--Abelian gauge field, thus reconfirming the results reported
in \cite{wadia}. Our presentation includes all intermediate steps of the
calculation and parallels the derivation of the axial anomaly for Dirac
fermions (see, in particular, Appendix C.1 for comparison). According to
equation \eqn{limita}, we have to compute the regularized sum
\be
A(t, x) = \lim_{M \rightarrow \infty} \Big[\sum_n \varphi_n^{\dagger} (t, x)
\gamma_5 ~ e^{- (i\gamma^{\mu} \mathbb{D}_{\mu})^2 / M^6} \varphi_n (t, x)
\Big] ~,
\ee
using the coupled Dirac--Lifshitz operator $i \gamma^{\mu} \mathbb{D}_{\mu}$ in
flat space--time which is minimally coupled to the external gauge field with
$D_{\mu} = \partial_{\mu} - iA_{\mu}$. The explicit gauge field
dependence will be extracted using the plane wave basis of solutions of the free
Dirac--Lifshitz operator. Thus, we use the alternative expression
\be
A(t, x) = \lim_{M \rightarrow \infty} {\rm Tr} \int {d^4 k \over (2\pi)^4} ~
\gamma_5 ~ e^{-i k_{\mu} x^{\mu}} e^{- (i\gamma^{\mu} \mathbb{D}_{\mu})^2 / M^6}
e^{i k_{\mu} x^{\mu}}
\label{minorikoi}
\ee
and perform the computation acting with the sixth order operator
$(i\gamma^{\mu} \mathbb{D}_{\mu})^2$ on the plane waves. The trace is taken
over anything available on the right--hand side.

By definition of the Dirac--Lifshitz operator \eqn{lifsh}, we have the
following relation,
\be
(i\gamma^{\mu} \mathbb{D}_{\mu})^2 = - \mathbb{D}_{\mu} \mathbb{D}^{\mu} -
{1 \over 4} [\gamma^{\mu} , ~ \gamma^{\nu}] [\mathbb{D}_{\mu} , ~
\mathbb{D}_{\nu}] ~.
\label{mimiko}
\ee
The commutator term is not the field strength of the gauge field, as usual, but an
operator valued quantity with "electric" and "magnetic" components
\be
[\mathbb{D}_{0} , ~ \mathbb{D}_{i}] = {i \over 2} \left(F_{0i} D^2 +
D^2 F_{0i} + F_{0l} D^l D_i + D_iD^l F_{0l} + D_i F_{0l} D^l + D^l F_{0l} D_i
\right)
\label{minio1}
\ee
and
\ba
[\mathbb{D}_{j} , ~ \mathbb{D}_{k}] & = & -{i \over 4} \left(F_{jk} (D^2)^2 +
(D^2)^2 F_{jk} + F_{lk} D^l D^2 D_j + D_j D^2 D^l F_{lk} +
\right. \nonumber\\
& & \left. ~~~ F_{jl} D^l D_k D^2 + D^2 D_k D^l F_{jl} + D^2 F_{lk} D^l D_j +
D_j D^l F_{lk} D^2 + \right. \nonumber\\
& & \left. ~~~ D^2 F_{jl} D^l D_k + D_k D^l F_{jl} D^2 + D_j F_{lk} D^l D^2 +
D^2 D^l F_{lk} D_j + \right. \nonumber\\
& & \left. ~~~ D^l F_{jl} D_k D^2 + D^2 D^l F_{jl} D_k + D_k F_{jl} D^l D^2 +
D^2 D_k F_{jl} D^l + \right. \nonumber\\
& & \left. ~~~ D^l F_{lk} D^2 D_j + D_j D^2 F_{lk} D^l + 2 D^2 F_{jk} D^2 \right) ,
\label{minio2}
\ea
respectively. Here, the derivatives act as operators on anything that appears
on the right. Thus, many more terms will participate in the evaluation of the
axial anomaly compared to the relativistic case.

The action of the interacting Dirac--Lifshitz operator on the plane waves
amounts to replacing $D_{\mu}$ by $D_{\mu} + ik_{\mu}$ everywhere, so that we may
formally write
\be
{\rm exp} \left(-{(i \gamma^{\mu} \mathbb{D}_{\mu})^2 \over M^6} \right)
e^{ik_{\mu} x^{\mu}} = e^{ik_{\mu} x^{\mu}} {\rm exp}
\left(-{\left(i \gamma^{\mu} (\mathbb{D}_{\mu} + i \mathbb{K}_{\mu}) \right)^2
\over M^6} \right)
\ee
when the plane wave factor ${\rm exp} (ik_{\mu} x^{\mu})$ passes to the far left
and eventually cancels the other factor ${\rm exp} (-ik_{\mu} x^{\mu})$ appearing in
expression \eqn{minorikoi}. Here, we set for notational convenience
\be
\mathbb{D}_0 + i \mathbb{K}_0 = D_0 + i k_0 ~, ~~~~~
\mathbb{D}_i + i \mathbb{K}_i = - {1 \over 2} \left((D_i + i k_i) (D + i k)^2
+ (D + i k)^2 (D_i + i k_i) \right) .
\label{notaconve}
\ee
Obviously this $k$--shift produces many terms in the exponential, but the most
relevant ones are easily selected by rescaling $k_0$ to $M^3 k_0$
and $k_i$ to $Mk_i$ (due to $z=3$ anisotropy) and expanding the result around
${\rm exp}[-k_0^2 - (k_i k^i)^3]$ in power series of $1/M$. Setting $k^2 = k_i k^i$,
we have, in particular,
\ba
A(t, x) & = & \lim_{M \rightarrow \infty} M^6 ~ {\rm Tr} \int {d^4 k \over (2\pi)^4}
~ e^{-k_0^2 - k^6} ~ \gamma_5 ~ {\rm exp} \Big\{-{i \over 2 M^4}
[\gamma^0 , ~ \gamma^i] \left(F_{0i} k^2 + 2 F_{0l} k_i k^l \right)  \nonumber\\
& &  ~~~~~~~~~ + {i \over 4 M^2} [\gamma^j , ~ \gamma^k] \left(F_{jk} k^4 +
4 F_{jl} k_k k^l k^2 \right) + \cdots \Big\} ~ ,
\ea
where $\cdots$ denote all other subleading operator terms that arise from \eqn{mimiko}
by expanding the shifted covariant derivatives; it can be easily seen that these
do not contribute to the final result. Taking into account the trace identities of
products of gamma--matrices and, in particular, ${\rm Tr}(\gamma_5 \gamma^0 \gamma^i
\gamma^j \gamma^k) = - 4 \epsilon^{0ijk}$ for Euclidean space gamma--matrices, it turns
out that only the quadratic term in the power series expansion of the exponential
gives a non--vanishing contribution, whereas all other terms either have zero trace or
they vanish as $M \rightarrow \infty$. Thus, we arrive at the expression
\be
A(t, x) = - 2 \epsilon^{0ijk} \int {d^4 k \over (2\pi)^4} ~ e^{-k_0^2 - k^6}
{\rm Tr} [(F_{0i} k^2 + 2 F_{0l} k_i k^l) (F_{jk} k^4 +
4 F_{jm} k_k k^m k^2)]
\ee
that provides a finite contribution to the anomaly when the regulator is finally removed.

Next, we perform the Gaussian integration over $k_0$, picking
up a factor of $\sqrt{\pi}$, and introduce a unit three--momentum vector with components
$\hat{k}_i$ (i.e., $k_i = k \hat{k}_i$) so that
\ba
A(t, x) & = & - {1 \over 8 \pi^{7/2}} ~ \epsilon^{0ijk} \int d^3 k ~ e^{-k^6}
k^6 ~ {\rm Tr} \left(F_{0i} F_{jk} + 4 F_{0i} F_{jm} \hat{k}_k \hat{k}^m
\right. \nonumber\\
& & \left. + 2 F_{0l} F_{jk} \hat{k}_i \hat{k}^l + 8 F_{0l} F_{jm} \hat{k}_i \hat{k}^l
\hat{k}_k \hat{k}^m \right) .
\label{kothon}
\ea
To complete the calculation we introduce spherical coordinates
in the space of three--dimensional momenta of length $k$ and note that
\be
\int d^3 k ~ e^{-k^6} k^6 = 4\pi \int_0^{\infty} dk ~ k^8 e^{-k^6} =
{4\pi \over 3} \int_0^{\infty} dx ~ x^2 e^{-x^2} = {1 \over 3} \pi^{3/2}
\label{rua1}
\ee
and
\be
\int d^3 k ~ e^{-k^6} k^6 ~ \hat{k}_i \hat{k}_j = {4 \pi \over 3}
~ \delta_{ij} \int_0^{\infty} dk ~ k^8 e^{-k^6} =
{1 \over 9} \pi^{3/2} \delta_{ij} ~.
\label{rua2}
\ee
The last identity differs from the first in the angular integration and it
can be readily verified using the orthogonal components of the unit vector $\hat{k}_1 =
{\rm sin} \theta {\rm sin} \phi$, $\hat{k}_2 = {\rm sin} \theta {\rm cos} \phi$
and $\hat{k}_3 = {\rm cos} \theta$ in momentum space.
Clearly, the last term in expression \eqn{kothon} vanishes, since it is
contracted with the totally anti--symmetric symbol $\epsilon^{0ijk}$ (one should
simply rename the indices $i$ and $k$), whereas the remaining terms in the trace are all
proportional to $F_{0i} F_{jk}$ after performing the integration over $k$.

Collecting all terms together, we obtain the final result for the axial anomaly of
Dirac--Lifshitz fermions in the background of gauge fields (they can be Abelian or
non--Abelian), which equals $2A(t, x)$,
\be
\partial_{\mu} j_5^{\mu} = - {1 \over 4 \pi^2} \epsilon^{0ijk} {\rm Tr}(F_{0i} F_{jk})
= - {1 \over 8 \pi^2} {\rm Tr} (F \wedge F) ~.
\ee
It is identical to the relativistic case, as advertised before, even though it was
obtained by combining more terms with a different weight.

\subsection{Metric field contribution to the anomaly}

Next, we obtain the gravitational contribution to the axial anomaly of
Lifshitz fermions, which is new and constitutes one of our main results. We refer
the reader to Appendices B and C.2 for the notation and comparison with the
relativistic case. Throughout this subsection, Latin
letters $a, b, c, d$ from the beginning of the alphabet are used to denote tangent
space--time indices, whereas capital Latin letters $I, J, K, L$ from the middle of the
alphabet are reserved for the tangent space indices (not to be confused with the space
indices that are denoted by small case Latin letters $i, j, k, l$). We also assume that
the space--time is of the form $\mathbb{R} \times \Sigma_3$ and it comes equipped with a
metric $G_{\mu \nu}$ satisfying the projectability condition, as explained before.

Starting from the general expression \eqn{limita}, the primitive form of the anomaly
is conveniently expressed as
\ba
A(t, x) & = & \lim_{M \rightarrow \infty} \sum_n \varphi_n^{\dagger}
(t, x) \gamma_5 ~ e^{- (i\gamma^{\mu} \mathbb{D}_{\mu})^2 / M^6} \varphi_n (t, x)
\nonumber\\
& = & \lim_{M \rightarrow \infty} \lim_{x \rightarrow x^{\prime}}
{\rm Tr} \int {d^4 k \over (2\pi)^4} ~
\gamma_5 ~ e^{- (i\gamma^{\mu} \mathbb{D}_{\mu})^2 / M^6}
e^{i k_{\mu} \nabla^{\mu} \sigma (x, x^{\prime})}
\ea
using the Dirac--Lifshitz operator $i \gamma^{\mu} \mathbb{D}_{\mu}$ with
$D_{\mu} = \partial_{\mu} + (1/8) [\gamma_a , ~ \gamma_b] {\omega_{\mu}}^{ab}$
that is minimally coupled to geometry via the spin connection.
Here, the Dirac gamma--matrices are expressed as $\gamma_{\mu} = {e^a}_{\mu}
\gamma_a$ using tangent space--time indices and $[\gamma_a , ~ \gamma_b]_+ =
2 \delta_{ab}$. The trace is taken on anything that is available on the right--hand
side of the equation. We also use the analogue of plane wave basis in curved space that
naturally involves the notion of geodesic interval $\sigma (x, x^{\prime})$, as it was
originally defined and used in the literature \cite{witt1, synge, hada}, in order to
extract the background field dependence of the anomaly.

As before, using the definition \eqn{lifsh} of the Dirac--Lifshitz operator, we have
the following relation,
\be
(i\gamma^{\mu} \mathbb{D}_{\mu})^2 = - \mathbb{D}_{\mu} \mathbb{D}^{\mu} -
{1 \over 4} [\gamma^{\mu} , ~ \gamma^{\nu}] [\mathbb{D}_{\mu} , ~
\mathbb{D}_{\nu}] ~.
\ee
The commutator terms are operator valued involving the components of the Riemann
curvature and they turn out to be
\ba
[\mathbb{D}_0 , ~\mathbb{D}_i] & = & - {1 \over 16} [\gamma_a , ~ \gamma_b] \left(
{R^{ab}}_{0i} D^2 + D^2 {R^{ab}}_{0i} + {R^{ab}}_{0l} D^l D_i + \right. \nonumber\\
& & \left. ~~~ D_i D^l {R^{ab}}_{0l} + D_i {R^{ab}}_{0l} D^l + D^l {R^{ab}}_{0l} D_i
\right)
\label{makios1}
\ea
and
\ba
[\mathbb{D}_j , ~ \mathbb{D}_k] & = & {1 \over 32} [\gamma_c , ~ \gamma_d]
\left({R^{cd}}_{jk} (D^2)^2 +
(D^2)^2 {R^{cd}}_{jk} + {R^{cd}}_{lk} D^l D^2 D_j + \right. \nonumber\\
& & \left. D_j D^2 D^l {R^{cd}}_{lk} + {R^{cd}}_{jl} D^l D_k D^2 +
D^2 D_k D^l {R^{cd}}_{jl} + D^2 {R^{cd}}_{lk} D^l D_j + \right. \nonumber\\
& & \left. D_j D^l {R^{cd}}_{lk} D^2 + D^2 {R^{cd}}_{jl} D^l D_k +
D_k D^l {R^{cd}}_{jl} D^2 + D_j {R^{cd}}_{lk} D^l D^2 + \right. \nonumber\\
& & \left. D^2 D^l {R^{cd}}_{lk} D_j + D^l {R^{cd}}_{jl} D_k D^2 +
D^2 D^l {R^{cd}}_{jl} D_k + D_k {R^{cd}}_{jl} D^l D^2 + \right. \nonumber\\
& & \left. D^2 D_k {R^{cd}}_{jl} D^l + D^l {R^{cd}}_{lk} D^2 D_j + D_j D^2
{R^{cd}}_{lk} D^l + 2 D^2 {R^{cd}}_{jk} D^2 \right) ,
\label{makios2}
\ea
thus providing the metric field analogue of the electric and magnetic components
of the field strength \eqn{minio1} and \eqn{minio2}, respectively. Here,
$D^2 = D_i D^i$ and the derivatives act as operators on anything that appears
on their right. Putting these together, we obtain a relation for
$(i\gamma^{\mu} \mathbb{D}_{\mu})^2$ that generalizes \eqn{finalrel} to the
case of Dirac--Lifshitz operator, but the result is quite lengthy; there is no
simple analogue of Lichnerowicz's formula for higher order fermion operators.
Only those terms that can contribute to the axial anomaly will be selected later.

Acting with the Dirac--Lifshitz operator on ${\rm exp} (ik_{\mu} \nabla^{\mu}
\sigma (x, x^{\prime}))$ amounts to replacing $D_{\mu}$ by $D_{\mu} + i
\Delta_{\mu}$ everywhere with
\be
\Delta_{\mu} (x, x^{\prime}) = k_{\nu}
\nabla_{\mu} \nabla^{\nu} \sigma (x, x^{\prime}) ~.
\ee
Indeed, one has
\be
D_{\mu} ~ e^{ik_{\mu} \nabla^{\mu} \sigma (x, x^{\prime})} =
e^{ik_{\mu} \nabla^{\mu} \sigma (x, x^{\prime})} (D_{\mu} + i \Delta_{\mu}) ~.
\ee
This relation and its higher derivative generalizations will be used later in explicit
calculations. For example,
a closely related identity that involves the action of two covariant derivatives on
curved space waves is
\ba
& & [D_{\mu} , ~ D_{\nu}] ~  e^{ik_{\mu} \nabla^{\mu} \sigma (x, x^{\prime})} =
e^{ik_{\mu} \nabla^{\mu} \sigma (x, x^{\prime})} [D_{\mu} + i \Delta_{\mu} , ~
D_{\nu} + i \Delta_{\nu}] = \nonumber\\
& & ~~~~~~~~~~~ e^{ik_{\mu} \nabla^{\mu} \sigma (x, x^{\prime})} \left({1 \over 8}
[\gamma_a , ~ \gamma_b] ~ {R^{ab}}_{\mu \nu} + i \nabla_{\mu} \Delta_{\nu} - i
\nabla_{\nu} \Delta_{\mu} \right) .
\label{pouthoun}
\ea
All terms should be accounted properly before taking the coincidence limit. For now it only
suffices to note that since $\nabla^{\mu} \sigma (x, x^{\prime})$ vanishes
in the limit $x \rightarrow x^{\prime}$, we have
\be
\lim_{x \rightarrow x^{\prime}} {\rm exp} \left(-{(i \gamma^{\mu}
\mathbb{D}_{\mu})^2 \over M^6} \right) e^{ik_{\mu} \nabla^{\mu}
\sigma (x, x^{\prime})} = \lim_{x \rightarrow x^{\prime}} {\rm exp}
\left(-{\left(i \gamma^{\mu} (\mathbb{D}_{\mu} + i \mathbb{K}_{\mu})
\right)^2 \over M^6} \right) ,
\ee
after passing ${\rm exp}(ik_{\mu} \nabla^{\mu} \sigma (x, x^{\prime}))$ to the far left
and setting for notational convenience
\be
\mathbb{D}_0 + i \mathbb{K}_0 = D_0 + i \Delta_0 ~, ~~~~~
\mathbb{D}_i + i \mathbb{K}_i = - {1 \over 2} \left((D_i + i \Delta_i) (D + i \Delta)^2
+ (D + i \Delta)^2 (D_i + i \Delta_i) \right) .
\label{notaconve7}
\ee
Then, the primitive form of the anomaly takes the rather simple looking form
\be
A(t, x) = \lim_{M \rightarrow \infty} \lim_{x \rightarrow x^{\prime}} {\rm Tr} \int
{d^4 k \over (2\pi)^4} ~ \gamma_5 ~ {\rm exp}
\left(-{\left(i \gamma^{\mu} (\mathbb{D}_{\mu} + i \mathbb{K}_{\mu})
\right)^2 \over M^6} \right)
\label{prothipog}
\ee
that needs to be evaluated carefully before removing the point split and the
cutoff $M$ from the integral.

Note that the operators \eqn{notaconve7} resemble \eqn{notaconve} introduced for the
computation of the axial anomaly in the background of a gauge field, justifying
the use of the same notation. There is an
importance difference, however, apart from the fact that the coupling is now taken
with respect to the background geometry. The quantity $\Delta_{\mu} (x, x^{\prime})$
equals $k_{\mu}$ in the limit $x \rightarrow x^{\prime}$ provided that there are no
more derivatives acting on it. The operators \eqn{notaconve7} naturally involve terms
with up to two derivatives of $\Delta_{\mu} (x, x^{\prime})$ that need to be extracted
before taking the limit $x \rightarrow x^{\prime}$. There can also be more derivatives acting
on $\Delta_{\mu} (x, x^{\prime})$ when the power series expansion of the exponential is
employed for the computation of the anomaly. Thus, the Synge--DeWitt tensors
\cite{witt1, synge}, which are the multiple covariant derivatives of the geodesic interval
$\sigma (x, x^{\prime})$ in the coincidence limit $x \rightarrow x^{\prime}$, are expected
to play essential role in the calculation (see Appendix B.2 for the mathematical details).
As it turns out, Synge--DeWitt tensors with up to four derivatives of $\sigma$ can and
will contribute to the calculation of the gravitational anomaly of $z=3$ Lifshitz fermions,
whereas for ordinary Dirac fermions the corresponding tensors involve up to two derivatives of
$\sigma$, as in the quantity $\Delta_{\mu} (x, x^{\prime})$. Thus, in the latter case, it is
legitimate to replace $\Delta_{\mu}$ by $k_{\mu}$ from the very beginning and proceed with
the calculation, as described in Appendix C.2, without worrying much about the intricacies of
the point split method on curved spaces, which, otherwise, can lead to an error.  In a nut--shell,
one may say that the use of the geodesic interval for evaluating the anomaly is rather "cosmetic"
for relativistic fermions, serving only the rigorous derivation of the final result,
whereas for non--relativistic fermions it has very essential role. We will say more about
this in due course.

The actual computation of the axial anomaly proceeds in several steps that are sketched in the
following. First of all , we rescale $\Delta_0$ to
$M^3 \Delta_0$ and $\Delta_i$ to $M \Delta_i$. This is not necessarily equivalent to rescaling
$k_0$ to $M^3 k_0$ and $k_i$ to $M^3 k_i$, which is naturally implied by the anisotropic
scaling of the Lifshitz theory. Note in this respect that if we had rescaled the time and space
components of both $\Delta_{\mu}$ and $k_{\mu}$ as just described, the rescaled quantities would
have been related to each other as
\ba
\Delta_0 (x, x^{\prime}) & = & k_0 \nabla_0 \nabla^0 \sigma (x, x^{\prime}) +
{1 \over M^2} k_i \nabla_0 \nabla^i \sigma (x, x^{\prime}) ~, \\
\Delta_i (x, x^{\prime}) & = & M^2 k_0 \nabla_i \nabla^0 \sigma (x, x^{\prime}) +
k_j \nabla_i \nabla^j \sigma (x, x^{\prime}) ~.
\ea
Although $\nabla_0 \nabla^i \sigma$ and $\nabla_i \nabla^0 \sigma$ vanish in the limit
$x \rightarrow x^{\prime}$ (irrespective of the choice of shift functions $N_i$ in the
ADM decomposition of the four--dimensional metric), their multiple covariant derivatives do
not vanish in general. The scaling of $\Delta_{\mu}$ is the same as $k_{\mu}$ provided
that there are no derivatives acting on it prior to the coincidence limit. On the other hand,
the scaling of the multiple derivatives of $\Delta_{\mu}$ do not follow the scaling of
$k_{\mu}$ because there are additional terms with anomalous scaling which are given by the
appropriate components of the Synge--DeWitt tensors. Such terms can and will become relevant
in the calculations. Using
the notation for the Synge--DeWitt tensors given in Appendix B.2, i.e., $[\nabla_{\mu}
\nabla_{\nu} \cdots \nabla_{k} \sigma]$, we have, in particular, the following anomalous
scaling relations,
\ba
\lim_{x \rightarrow x^{\prime}} (\nabla_{\mu} \nabla_{\nu} \Delta_0 (x, x^{\prime})) & = &
k_0 [\nabla_{\mu} \nabla_{\nu} \nabla_0 \nabla^0 \sigma] +
{1 \over M^2} k_i [\nabla_{\mu} \nabla_{\nu} \nabla_0 \nabla^i \sigma] ~, \\
\lim_{x \rightarrow x^{\prime}} (\nabla_{\mu} \nabla_{\nu} \Delta_i (x, x^{\prime})) & = &
M^2 k_0 [\nabla_{\mu} \nabla_{\nu} \nabla_i \nabla^0 \sigma] +
k_j [\nabla_{\mu} \nabla_{\nu} \nabla_i \nabla^j \sigma] ~.
\label{scala9}
\ea
Here, we only give the result for terms that involve two derivatives of
$\Delta_{\mu}(x, x^{\prime})$. Terms with only one derivative vanish identically in the
coincidence limit since the third order Synge--DeWitt tensors are identically zero, i.e.,
$[\nabla_{\mu} \nabla_{\mu} \nabla_{k} \sigma] = 0$. The fourth order Synge--DeWitt tensors
$[\nabla_{\mu} \nabla_{\mu} \nabla_{k} \nabla_{\lambda} \sigma]$ are proportional to the
Riemann curvature tensor and their form will be employed later to treat these and other
relevant terms. When there are more than two derivatives acting on $\Delta_{\mu}$ the
resulting tensors are non--zero, in general, but such terms will not be encountered in
the present work.

For now, we adopt the rescaling of $\Delta_0$ and $\Delta_i$ as a good book keeping
device to start organizing the power series expansion of the Dirac--Lifshitz operator and the
exponential of its square and worry later about the possible relevance of the anomalous scaling
terms when passing to the rescaled momenta $k_0$ and $k_i$. In terms of the rescaled quantities,
the square of the $\Delta$--shifted Dirac--Lifshitz operator arising in \eqn{prothipog},
\be
-\left(i \gamma^{\mu} (\mathbb{D}_{\mu} + i \mathbb{K}_{\mu})
\right)^2 = (\mathbb{D}_{\mu} + i \mathbb{K}_{\mu})(\mathbb{D}^{\mu} + i \mathbb{K}^{\mu})
+ {1 \over 4} [\gamma^{\mu} , ~ \gamma^{\nu}] [\mathbb{D}_{\mu} + i \mathbb{K}_{\mu} , ~
\mathbb{D}_{\nu} + i \mathbb{K}_{\nu}] ~,
\ee
involves several terms that are conveniently organized in powers of $1/M$ as outlined below
step by step.

First, we expand the Lifshitz analogue of the $\Delta$--shifted Bochner Laplacian, setting
$\Delta^2 = \Delta_i \Delta^i$,
\ba
& & {1 \over M^6} (\mathbb{D}_{\mu} + i \mathbb{K}_{\mu})(\mathbb{D}^{\mu} +
i \mathbb{K}^{\mu}) = - \Delta_0 \Delta^0 - \Delta^6 + {i \over M^3}
[2 \Delta^0 D_0 + \nabla_0 \Delta^0] \nonumber\\
& & ~~~~~~~~ + {3i \over M} [2 \Delta^4 \Delta^n D_n + \Delta^4
\nabla_n \Delta^n + 4 \Delta^2 \Delta^n \Delta ^m \nabla_n \Delta_m]
\nonumber\\
& & ~~~~~~~~~ + {1 \over M^2} [3 \Delta^4 D^2 + 12 \Delta^2 \Delta^n \Delta^m D_n D_m +
12 A^n D_n + B] \nonumber\\
& & ~~~~~~~~~ + {\rm irrelevant ~ terms} ~,
\label{kasoulai}
\ea
where the coefficient functions shown in the third line turn out to be
\be
A^n = \Delta^2 \Delta_m (\nabla^n \Delta^m + \nabla^m \Delta^n) + \Delta^2
\Delta^n \nabla_m \Delta^m + 2 \Delta^n \Delta^m \Delta^r \nabla_m \Delta_r ~,
\ee
\ba
B & = & \Delta^2 \Delta^n (5 \nabla^2 \Delta_n + 5 \nabla_n \nabla_m \Delta^m +
2 \nabla_m \nabla_n \Delta^m) + 8 \Delta^n \Delta^m \Delta^r \nabla_n \nabla_m
\Delta_r + \nonumber\\
& & \Delta^2 [2(\nabla_n \Delta^m)(\nabla_m \Delta^n) +3(\nabla_n \Delta^n)
(\nabla_m \Delta^m) + 4(\nabla_n \Delta_m)(\nabla^n \Delta^m)] + \nonumber\\
& & \Delta^n \Delta^m [12(\nabla_n \Delta_m) (\nabla_r \Delta^r) + 7(\nabla_n \Delta_r)
(\nabla_m \Delta^r) + 7 (\nabla_r \Delta_n)(\nabla^r \Delta_m) + \nonumber\\
& & 10 (\nabla_n \Delta_r)
(\nabla^r \Delta_m)] ~.
\ea
The first line of equation \eqn{kasoulai} contains all terms associated to
$(D_0 + i \Delta_0)(D^0 + i \Delta^0)$ apart from $D_0 D^0$ which is of order $1/M^6$
and it is irrelevant. The subleading terms of order $1/M^3$ or higher that arise from
$(\mathbb{D}_i + i \mathbb{K}_i)(\mathbb{D}^i + i \mathbb{K}^i)$ are also irrelevant and
they are omitted from the expansion. It can be seen that they do not contribute to the
anomaly based on the criterion that will be given below.

Next, we expand the operator $[\gamma^{\mu} , ~ \gamma^{\nu}] [\mathbb{D}_{\mu} + i
\mathbb{K}_{\mu} , ~ \mathbb{D}_{\nu} + i \mathbb{K}_{\nu}]$ in powers of $1/M$ using
the expressions \eqn{makios1} and \eqn{makios2} with $D_{\mu}$ shifted to
$D_{\mu} + i \Delta_{\mu}$. We obtain, in particular, the following result for the
electric commutator after rescaling $\Delta_0$ and $\Delta_i$,
\ba
& & {1 \over 2M^6} [\gamma^0 , ~ \gamma^i] [\mathbb{D}_0 + i
\mathbb{K}_0 , ~ \mathbb{D}_i + i \mathbb{K}_i] = {1 \over 16M^4}
[\gamma^0 , ~ \gamma^i] [\gamma_a , ~ \gamma_b] \left({R^{ab}}_{0i} \Delta^2 +
2 {R^{ab}}_{0l} \Delta_i \Delta^l \right) \nonumber\\
& & ~~~~~~~~~ - {i \over 16 M^5} [\gamma^0 , ~ \gamma^i] [\gamma_a , ~ \gamma_b]
\left(2{R^{ab}}_{0i} \Delta_l D^l + 2{R^{ab}}_{0l} \Delta^l D_i
+ 2{R^{ab}}_{0l} \Delta_i D^l + \right. \nonumber\\
& & \left. ~~~~~~~~~~~~~~~~~~~~~~~~~~~
{R^{ab}}_{0i} (\nabla_n \Delta^n) + {R^{ab}}_{0l} (\nabla_i \Delta^l) +
{R^{ab}}_{0l} (\nabla^l \Delta_i) \right) \nonumber\\
& & ~~~~~~~~~ - {1 \over 16 M^6} [\gamma^0 , ~ \gamma^i] [\gamma_a , ~ \gamma_b]
\left({R^{ab}}_{0i} D^2 + {R^{ab}}_{0l} D^l D_i +
{R^{ab}}_{0l} D_i D^l \right) \nonumber\\
& & ~~~~~~~~~ - {1 \over 2} [\gamma^0 , ~ \gamma^i] \left({i \over M} C_1 + {1 \over M^2} C_2
- {i \over 2M^3} C_3 - {1 \over M^4} C_4 + {i \over 2M^5} C_5 \right) + \cdots ~,
\label{katra1}
\ea
where $\cdots$ include the terms that involve one or two derivatives of the Riemann
curvature tensor, which, in fact, are irrelevant for the axial anomaly. All other terms are
relevant for the computation and for this reason we give the explicit form of the coefficient
functions $C_i$,
\be
C_1 = \Delta^2 \nabla_i \Delta_0 + 2 \Delta_i \Delta^n \nabla_n \Delta_0 ~,
\ee
\ba
C_2 & = & 2(\Delta_i \nabla^n \Delta_0 + \Delta^n \nabla_i \Delta_0) D_n + 2
(\nabla_n \Delta_0) \Delta^n D_i + \Delta^n (\nabla_i \nabla_n \Delta_0 + \nabla_n
\nabla_i \Delta_0) + \nonumber\\
& & \Delta_i \nabla^2 \Delta_0 +
(\nabla_i \Delta_0) (\nabla_n \Delta^n) + (\nabla_n \Delta_0) (\nabla_i \Delta^n +
\nabla^n \Delta_i) ~,
\ea
\ba
C_3 & = & 2(\nabla_i \Delta_0) D^2 + 2 (\nabla^n \Delta_0) (D_i D_n + D_n D_i) +
2 (\nabla^n \nabla_i \Delta_0 + \nabla_i \nabla^n \Delta_0)
D_n + \nonumber\\
& & 2 (\nabla^2 \Delta_0) D_i + \nabla^2 \nabla_i \Delta_0 + \nabla_i \nabla^2 \Delta_0
+ 2 \Delta^2 \nabla_0 \Delta_i + 4 \Delta_i \Delta^n \nabla_0 \Delta_n ~,
\ea
\ba
C_4 & = & 2 (\Delta^n \nabla_0 \Delta_i + \Delta_i \nabla_0 \Delta^n) D_n + 2 (\nabla_0 \Delta^n)
\Delta_n D_i + \Delta_i \nabla_n \nabla_0 \Delta^n + (\nabla_0 \Delta_i)
(\nabla_n \Delta^n) + \nonumber\\
& & \Delta^n (\nabla_i \nabla_0 \Delta_n + \nabla_n \nabla_0 \Delta_i) +
(\nabla_0 \Delta^n) (\nabla_i \Delta_n + \nabla_n \Delta_i) ~,
\ea
\ba
C_5 & = & 2 (\nabla_0 \Delta_i) D^2 + 2 (\nabla_0 \Delta^n) (D_i D_n +
D_n D_i) + 2 (\nabla_i \nabla_0 \Delta^n + \nabla^n \nabla_0 \Delta_i) D_n +
\nonumber\\
& & 2 (\nabla_n \nabla_0 \Delta^n) D_i + \nabla^2 \nabla_0 \Delta_i +
\nabla_i \nabla_n \nabla_0 \Delta^n ~.
\ea

It should be emphasized that the results shown above follow from equation \eqn{makios1} by
simply replacing $D_{\mu} \rightarrow D_{\mu} + i \Delta_{\mu}$ as well as
\be
{1 \over 8} [\gamma_a , ~ \gamma_b] ~ {R^{ab}}_{\mu \nu} \rightarrow {1 \over 8}
[\gamma_a , ~ \gamma_b] ~ {R^{ab}}_{\mu \nu} + i (\nabla_{\mu} \Delta_{\nu} -
\nabla_{\nu} \Delta_{\mu})
\label{moustoule}
\ee
based on \eqn{pouthoun}. If there were no more derivatives acting on it, the
coincidence limit of $\nabla_{\mu} \Delta_{\nu}$ would have been zero, rendering the
excess terms in \eqn{moustoule} obsolete. Note, however, that the multiple derivatives
of $\Delta_{\mu}$ no not vanish in the coincidence limit, in general, and they should
be accounted properly before letting $x \rightarrow x^{\prime}$ in all terms that
originate from the operators $D_i {R^{ab}}_{0l}$, $D^2 {R^{ab}}_{0i}$, etc. The same
remarks apply to the terms that arise from equation \eqn{makios2} and they are discussed
next.

The magnetic commutator term admits the following expansion in powers of
$1/M$, using the rescaled variables $\Delta_0$ and $\Delta_i$,
\ba
& & {1 \over 4M^6} [\gamma^j , ~ \gamma^k] [\mathbb{D}_j + i
\mathbb{K}_j , ~ \mathbb{D}_k + i \mathbb{K}_k] =
{1 \over 32M^2} [\gamma^j , ~ \gamma^k] [\gamma_c , ~ \gamma_d]
\left({R^{cd}}_{jk} \Delta^4 + 4 {R^{cd}}_{jl} \Delta_k \Delta^l \Delta^2
\right) \nonumber\\
& & ~~~~~~ + {i \over 4M} [\gamma^j , ~ \gamma^k] \left(\Delta^4 (\nabla_j \Delta_k -
\nabla_k \Delta_j) + 4 \Delta_k \Delta^l \Delta^2 (\nabla_j \Delta_l - \nabla_l \Delta_j)
\right) \nonumber\\
& & ~~~~~~ + {1 \over 2M^2} [\gamma^j , ~ \gamma^k] \Big\{(\nabla_j \Delta_k -
\nabla_k \Delta_j) [2\Delta^n \Delta^2 D_n + \nabla_n (\Delta^n \Delta^2)] +
(\nabla_j \Delta_l - \nabla_l \Delta_j) \times \nonumber\\
& & ~~~~~~~~~~~~~~~~ [4\Delta^n \Delta_k \Delta^l D_n +
2 \Delta_k \Delta^2 D^l + 2 \Delta^l \Delta^2 D_k + 2 \nabla_n (\Delta^n \Delta_k \Delta^l)
+ \nabla_k (\Delta^l \Delta^2) + \nonumber\\
& & ~~~~~~~~~~~~~~~~ \nabla^l (\Delta_k \Delta^2)] + \Delta^n
\Delta^2 \nabla_n (\nabla_j \Delta_k - \nabla_k \Delta_j) + 2 \Delta^n \Delta_k \Delta^l
\nabla_n (\nabla_j \Delta_l - \nabla_l \Delta_j) + \nonumber\\
& & ~~~~~~~~~~~~~~~~ \Delta_k \Delta^2 \nabla^l (\nabla_j \Delta_l - \nabla_l \Delta_j) +
\Delta^l \Delta^2 \nabla_k (\nabla_j \Delta_l - \nabla_l \Delta_j) \Big\} \nonumber\\
& & ~~~~~~ + {\rm irrelevant ~ terms},
\label{katra2}
\ea
where the irrelevant terms are of order $1/M^3$ or higher, as all terms of the same order
that arise in the power series expansion of $(\mathbb{D}_i + i \mathbb{K}_i)
(\mathbb{D}^i + i \mathbb{K}^i)/M^6$.

Next, we use the Campbell--Baker--Hausdorff formula for any two operators
$X$ and $Y$, which is conveniently written here as
\ba
e^Y & = & e^{-X} {\rm exp} \left(X + Y + {1 \over 2} [X , ~ Y] + {1 \over 12}
[X , ~ [X , ~ Y]] - {1 \over 12} [Y , ~ [X , ~ Y]] \right.\nonumber\\
& & \left. ~~~~~~~~~~~~~~ - {1 \over 24} [Y , ~ [X , ~ [X , ~ Y]]] + {\rm higher ~
commutator ~ terms} \right)
\label{CBHf}
\ea
and choose, in particular,
\be
X = \Delta_0 \Delta^0 + \Delta^6 , ~~~~~~~ Y = - {1 \over M^6} (\mathbb{D}_{\mu} +
i \mathbb{K}_{\mu})(\mathbb{D}^{\mu} + i \mathbb{K}^{\mu}) ~.
\ee
For practical reasons and in view of the applications that will be discussed in
subsequent sections, we consider four--dimensional metrics with vanishing shift
functions, $N_i (t, x)= 0$, in the proper time gauge, $N(t) =1$, so that we can
simply set $\Delta_0 \Delta^0 = \Delta_0^2$ here and in the following.
In any case, \eqn{CBHf} allows to pull out a factor ${\rm exp}(-\Delta_0^2 - \Delta^6)$
on the left, which is derivative free, so that we can safely assert irrespective of the
remaining operator terms that
\be
\lim_{x \rightarrow x^{\prime}} e^{-\Delta_0^2 - \Delta^6} = e^{-k_0^2 - k^6}
\ee
after rescaling $k_0$ to $M^3 k_0$ and $k_i$ to $Mk_i$ and setting $k_i k^i = k^2$.
Adopting this rescaling everywhere, we find that $A(t, x)$ takes the form
\ba
A (t, x) & = & \lim_{M \rightarrow \infty} M^6 ~ {\rm Tr} \int {d^4 k \over (2\pi)^4}
~ e^{-k_0^2 - k^6} ~ \gamma_5 ~ \lim_{x \rightarrow x^{\prime}} {\rm exp} \Big\{X + Y +
{1 \over 2} [X , ~ Y] \nonumber\\
& & + {1 \over 12} [X , ~ [X , ~ Y]] - {1 \over 12} [Y , ~ [X , ~ Y]]
- {1 \over 24} [Y , ~ [X , ~ [X , ~ Y]]] + \cdots \Big\}
\label{tatouli}
\ea
which is the basis for all subsequent manipulations. Note that the successive commutators
of $X$ and $Y$ appearing inside the curly bracket will generate many more terms beyond
those appearing in $X$ to all orders of $1/M$. We simply have to select the relevant ones,
but they are too many to display them here. These complications do not arise in the
relativistic case, since the coincidence limit can be taken safely at an early stage
of the calculation. Even if we had used the Campbell--Baker--Hausdorff formula there,
the additional terms associated to successive commutators would have been irrelevant
at the end.

The rest of the computation proceeds by series expansion of the exponential operator in
powers of $1/M$, as in a Dyson expansion. The relevant terms for the anomaly are selected
by the following criteria:
\begin{itemize}
\item they should be of order $1/M^6$, and
\item they should involve quadratic curvature terms or
sufficient number of powers of the derivative operator that can yield quadratic curvature
terms at the end of the calculation, and
\item they should involve sufficient even number of gamma
matrices (at least four) so that their trace together with $\gamma_5$ can be non--zero,
and, finally,
\item they should contain even powers of $k_0$ so that their integral with
${\rm exp} (-k_0^2)$ can also be non--zero (recall that $k_0$ runs from $-\infty$ to $+\infty$).
\end{itemize}
Note at this end that all other terms of order $1/M^6$ that do not involve four derivatives
will not be capable to produce a topological density in four space--time dimensions; it can also
be verified by direct computation that all such terms cancel against each other. Terms of the
higher order obviously give zero in the limit $M \rightarrow \infty$, whereas terms of lower
order ought to cancel for, otherwise, the anomaly would be infinity; it can also be verified
by direct calculations such terms indeed cancel order by order.

The first three criteria are obvious and they are similar to the ones used for the
evaluation of the gravitational contribution to the axial anomaly of relativistic fermions,
apart from the fact that the relevant terms there are of order $1/M^4$ rather than $1/M^6$
by the difference in time scaling. Then, because of this difference, and the higher order
structure of the Dirac--Lifshitz operator, the terms contributing to the anomaly turn out to
be very different from those
arising in the relativistic case. For example, as will be seen later, terms that involve products
of up to eight gamma matrices with $\gamma_5$ contribute to the final answer, whereas in the
relativistic case only the trace of $\gamma_5$ with four gamma matrices comes into play. Most
importantly, higher order derivatives of the geodesic interval are generated
(either in $X$, as summarized by adding together the individual contributions \eqn{kasoulai},
\eqn{katra1} and \eqn{katra2}, or in the successive commutators of $X$ with $Y$ or in the
subsequent expansion of the exponential operator shown in \eqn{tatouli}) and they all
contribute to the axial anomaly of Lifshitz fermions in various ways.

The fourth criterion looks rather superfluous at first sight, but, in fact, it
helps to select some additional terms that could have been easily missed otherwise. Recall
at this point the mismatch in the scaling of $\Delta_0$ and $\Delta_i$ as compared to
$k_0$ and $k_i$ when multiple derivatives act on $\Delta_{\mu} (x, x^{\prime})$ (see, in
this respect, equation \eqn{scala9}). Certain terms that are seemingly of order $1/M^8$,
like the cross product of $\Delta^0 D_0 /M^3$ appearing in the first line of \eqn{kasoulai}
and ${R^{ab}}_{0l} \nabla_i \Delta^l /M^5$ appearing in the third line of \eqn{katra1}, yield
the following contribution in the coincidence limit
\be
{1 \over M^8} \lim_{x \rightarrow x^{\prime}} {R^{ab}}_{0l} \Delta_0 \nabla_0 \nabla_i
\Delta^l = {1 \over M^6} {R^{ab}}_{0l} k_0^2 [\nabla_0 \nabla_i \nabla^l \nabla^0 \sigma]
+ {\cal O}\left({1 \over M^8}\right) ,
\ee
which is non--zero. Such terms are effectively of order $1/M^6$, they contain four derivatives,
they have sufficient number of gamma matrices ($[\gamma^0 , ~ \gamma^i][\gamma_a , ~ \gamma_b]$
in the particular example we are discussing here, following from \eqn{katra1}), and, most
importantly, they contain an even number of powers of $k_0$ ($k_0^2$ in the present example)
so that the integral over $k_0$ is non--zero. One has to extract carefully all such terms
from the power series expansion of the exponential, although they are more rare compared
to the other terms for which the mismatch in the scaling is irrelevant. Finally, the fourth
criterion is also used to eliminate all cross terms of $\Delta^0 D_0 /M^3$ with the
${\cal O} (1/M^3)$ terms arising in the expansions \eqn{kasoulai} or \eqn{katra2} and alike,
because they contain odd powers of $k_0$ in the coincidence limit. This is the reason why such
higher order terms were omitted from the expansions \eqn{kasoulai} and \eqn{katra2} as irrelevant.

The various terms that meet all four criteria are still too many to display them one by one.
Here, we will only sketch the steps we have taken for their determination by grouping them
together in different classes. The intermediate details are straightforward but
very cumbersome and they are left to the interested reader to complete.

In the course of the calculation we also need the following integrals over the space of
three--momenta (whereas the integration over $k_0$ is very simple),
\be
I_n \equiv \int d^3 k ~ e^{-k^6} k^{6n} = {(2n-1)!! \over 3 \cdot 2^{n-1}} \pi^{3/2} ,
\label{skoubi1}
\ee
as well as the curved space analogue of equation \eqn{rua2}, which now reads
\be
\int d^3 k ~ e^{-k^6} k^{6n} ~ \hat{k}_I \hat{k}_J =
{I_n \over 3} ~ \delta_{IJ} ~.
\label{skoubi2}
\ee
The proof is easily done by introducing tangent space indices $I$, $J$ (to be distinguished
from the spatial indices $i$, $j$ on the slices $\Sigma_3$ of
$\mathbb{R} \times \Sigma_3$), so that the unit vectors $\hat{k}_i$ that are defined by
$k_i = k \hat{k}_i$ are decomposed with respect to the dreibeins associated to the metric
$G_{ij} = \delta_{IJ} {e^I}_{i} {e^J}_{j}$ as follows, $\hat{k}_i = {e^I}_i
\hat{k}_{I}$ (see also Appendix C.2 for the corresponding integral in the space of
four--momenta). The rest proceeds as in flat space, choosing, in particular, the components
of $\hat{k}_{I}$ as $\hat{k}_1 = {\rm sin} \theta {\rm sin} \phi$,
$\hat{k}_2 = {\rm sin} \theta {\rm cos} \phi$ and $\hat{k}_3 = {\rm cos} \theta$ in terms of
spherical coordinates in momentum space.

Likewise, we evaluate the following integrals
that will all be needed in the course of the calculation (in fact, up to $n=3$),
\be
\int d^3 k ~ e^{-k^6} k^{6n} ~ \hat{k}_I \hat{k}_J \hat{k}_K \hat{k}_L =
{I_n \over 15} ~ \left(\delta_{IJ} \delta_{KL} + \delta_{IK} \delta_{JL} +
\delta_{IL} \delta_{JK} \right)
\label{skoubi3}
\ee
and
\ba
\int d^3 k ~ e^{-k^6} k^{6n} ~ \hat{k}_I \hat{k}_J \hat{k}_K \hat{k}_L
\hat{k}_M \hat{k}_N & = &
{I_n \over 105} ~ \Big[\delta_{IJ} \left(\delta_{KL} \delta_{MN} + \delta_{KM} \delta_{LN}
+ \delta_{KN} \delta_{LM} \right) \nonumber\\
& & ~~~ + \delta_{IK} \left(\delta_{JL} \delta_{MN} + \delta_{JM} \delta_{LN}
+ \delta_{JN} \delta_{LM} \right) \nonumber\\
& & ~~~ + \delta_{IL} \left(\delta_{JK} \delta_{MN} + \delta_{JM} \delta_{KN}
+ \delta_{JN} \delta_{KM} \right)  \nonumber\\
& & ~~~ + \delta_{IM} \left(\delta_{JK} \delta_{LN} + \delta_{JL} \delta_{KN}
+ \delta_{JN} \delta_{KL} \right)  \nonumber\\
& & ~~~ + \delta_{IN} \left(\delta_{JK} \delta_{LM} + \delta_{JL} \delta_{KM}
+ \delta_{JM} \delta_{KL} \right) \Big] ~.
\label{skoubi4}
\ea

After these explanations, we are now in position to describe the structure of the
various terms that contribute to the gravitational anomaly of Lifshitz fermions
in the limit $M \rightarrow \infty$ and $x \rightarrow x^{\prime}$. We will
also provide some identities that are necessary to cast them in
familiar form. It is most convenient to split the terms in two groups as
\be
A(t, x) = A_1 (t, x) + A_2 (t, x) ~,
\ee
where $A_1 (t, x)$ denotes collectively the terms that contain no derivatives of
$\Delta_{\mu}$ and $A_2 (t, x)$ denotes all other terms that contain two derivatives of
$\Delta_{\mu}$ and which can be evaluated using the fourth--order Synge--DeWitt tensor.
As noted before, terms that contain only one derivative of $\Delta_{\mu}$ vanish in the
coincidence limit, whereas higher order derivative terms do not enter in the calculation.

{\bf (i). $A_1$ terms}: These terms are the easiest to describe since the coincidence
limit can be taken from the beginning. There are two different type of such terms,
so we write $A_1 (t, x) = A_1^{(1)} (t, x) + A_1^{(2)} (t, x)$, which can be read directly
from the power series expansion of ${\rm exp} (X+Y)$ (note in this respect that $[X, ~, Y]$
and all other higher order commutators of $X$ and $Y$ can only produce additional terms that
contain derivatives of $\Delta_{\mu}$ prior to the coincidence limit). First, we display
\ba
& & A_1^{(1)} (t, x) = {1 \over 256 \pi^{7/2}} {\rm Tr} \int d^3 k ~ e^{-k^6}
\gamma_5 ~ [\gamma^0, ~ \gamma^i] ~ [\gamma_a, ~ \gamma_b] \times \nonumber\\
& & \Big\{6 ~ \hat{k}_j k^6 \left({R^{ab}}_{0i} \hat{k}_l [D^j, ~ D^l]_+
+ {R^{ab}}_{0l} \hat{k}^l [D_i,  ~ D^j]_+
+ {R^{ab}}_{0l} \hat{k}_i [D^l,  ~ D^j]_+ \right) \nonumber\\
& & ~~~ + 3 ~ k^6 \left({R^{ab}}_{0i} + 2{R^{ab}}_{0l} \hat{k}_i \hat{k}^l \right)
\left(D^2 + 4\hat{k}_j \hat{k}_k D^j D^k \right)  \nonumber\\
& & ~~~ - 18 ~ \hat{k}_j \hat{k}_k k^{12} \left({R^{ab}}_{0i} + 2{R^{ab}}_{0l} \hat{k}_i
\hat{k}^l \right) D^j D^k \nonumber\\
& & ~~~ -  {R^{ab}}_{0i} D^2 - {R^{ab}}_{0l} [D^l, D_i]_+ \Big\} ~,
\label{figoo1}
\ea
which contains the cross product terms between \eqn{kasoulai} and \eqn{katra1} excluding all
derivatives of $\Delta_{\mu}$. In particular, the first line in \eqn{figoo1} includes the
product of terms of order $1/M$ with terms of order $1/M^5$, the second line the product of
terms of order $1/M^2$ with $1/M^4$, the third line the product of terms of order
$1/M$--squared with terms of order $1/M^4$, and, finally, the fourth line includes the terms
of order $1/M^6$ appearing in \eqn{katra1}. In writing \eqn{figoo1}, we introduced
unit vectors $\hat{k}_i$ and integrated over $k_0$, thus picking up a factor of $\pi^{1/2}$.
Then, we can easily perform the integration over three--momenta based on the relations
\eqn{skoubi1}--\eqn{skoubi3}, which are used here up to $n=2$, and find that the resulting
terms are either of the form ${R^{ab}}_{0i} D^2$ or ${R^{ab}}_{0l} [D_i , ~ D^l]_+$. These
terms cancel against each other separately before taking the trace of the gamma matrices,
and, therefore, we obtain
\be
A_1^{(1)} (t, x) = 0 ~.
\ee

Next, we display the remaining terms of this group that contain the cross product
terms of order $1/M^4$ and $1/M^2$ found in the expansions \eqn{katra1} and \eqn{katra2}.
They yield
\ba
& & A_1^{(2)} (t, x) = {1 \over 16 \cdot 512 \pi^{7/2}} {\rm Tr} \int d^3 k ~ e^{-k^6}
k^6 ~ \gamma_5 ~ [\gamma^0, ~ \gamma^i] ~ [\gamma_a, ~ \gamma_b] ~
[\gamma^j, ~ \gamma^k] ~ [\gamma_c, ~ \gamma_d] \times \nonumber\\
& & \left({R^{ab}}_{0i} {R^{cd}}_{jk}
+ 4 {R^{ab}}_{0i} {R^{cd}}_{jm} \hat{k}_k \hat{k}^m +
2 {R^{ab}}_{0l} {R^{cd}}_{jk} \hat{k}_i \hat{k}^l + 8 {R^{ab}}_{0l}
{R^{cd}}_{jm} \hat{k}_i \hat{k}^l \hat{k}_k \hat{k}^m \right)
\label{figoo2}
\ea
after introducing the unit vectors $\hat{k}_i$ and integrating over $k_0$.
These terms are analogous to the group of terms \eqn{kothon} that
contribute to the axial anomaly of a Lifshitz fermion coupled to gauge fields, but they
are more complicated now because they involve the product of $\gamma_5$ with eight
gamma matrices. As we will see shortly, they provide a non--vanishing contribution
to the gravitational form of the anomaly, which has no analogue in the relativistic case.
Integrating over the three--momenta, based on relations \eqn{skoubi1}--\eqn{skoubi3},
which are used again up to $n=2$, we arrive at the expression
\ba
& & A_1^{(2)}(t, x) = {1 \over 45 \cdot 512 \pi^2} ~ {\rm Tr} \left(\gamma_5 ~
{1 \over 2} [\gamma^{a^{\prime}} , ~ \gamma^{b^{\prime}}] {1 \over 2}
[\gamma_a , ~ \gamma_b] {1 \over 2} [\gamma^{c^{\prime}} , ~ \gamma^{d^{\prime}}]
{1 \over 2} [\gamma_c , ~ \gamma_d] \right) \times \nonumber\\
& & {E_{a^{\prime}}}^0
{E_{b^{\prime}}}^i {E_{c^{\prime}}}^j {E_{d^{\prime}}}^k
\left(53 {R^{ab}}_{0i} {R^{cd}}_{jk} +
8 {R^{ab}}_{0k} {R^{cd}}_{ji} + 8 G_{ik} G^{lm} {R^{ab}}_{0l}
{R^{cd}}_{jm} \right)
\ea
which is written here using the inverse vierbeins ${E_{a^{\prime}}}^{\mu}$ that trade
the space--time gamma--matrices $\gamma^{\mu}$ ($\mu = 0, i, j, k$) with their
tangent space--time counterparts $\gamma^{a^{\prime}}$.

At this point we employ the trace identities found in Appendix A in order to evaluate
$A_1^{(2)} (t, x)$. Using the trace of $\gamma_5$ with the product of eight gamma--matrices,
which is given by equation \eqn{eighttra} in terms of flat space--time indices and which
should be appropriately adapted to current notation, we obtain the result\footnote{In the
course of this calculation we also find terms of the form $\epsilon^{0ijk} R_{0ijk}
{R^{mn}}_{mn}$, but they vanish because $R_{0ijk} + R_{0jki} + R_{0kij} = 0$. Such terms
also arise in the computation of $A_2$. We also employ the identity
\be
{1 \over 2} \epsilon^{0ijk} \left({{R^n}_{i0}}^l R_{lkjn} + {R^{nl}}_{i0} R_{ljkn} +
{R^{ln}}_{in} R_{jl0k} \right) = \epsilon^{0ijk} R_{jk0l} {R^{ln}}_{in} \nonumber
\ee
to simplify the intermediate expressions and express everything in terms of
$R_{0i0l}{R^{0l}}_{jk}$ and $R_{jk0l} {R^{ln}}_{in}$.}
\be
A_1^{(2)} (t, x) = {1 \over 360 \pi^2} \epsilon^{0ijk} \left(R_{0l0i}
{R^{0l}}_{jk} - R_{jk0l} {R^{ln}}_{in} \right) .
\ee
A useful identity that can be used here and later to combine terms of the form
$R_{0l0i}{R^{0l}}_{jk}$ and $R_{jk0l} {R^{ln}}_{in}$ (these are the only terms that
can arise in the computation of the gravitational anomaly in non--relativistic theories)
is provided by
\be
{1 \over 2} \epsilon^{0ijk} R_{ab0i} {R^{ab}}_{jk} = \epsilon^{0ijk} \left(R_{0i0l}
{R^{0l}}_{jk} - R_{jk0l} {R^{ln}}_{in} \right) .
\ee
This look rather odd at first sight, but it is in fact true in general for all
four--dimensional geometries.
The proof of this identity follows by brute force, writing down all terms that arise
from summation over repeated indices. Thus, overall, we obtain
\be
A_1 (t, x) = {1 \over 720 \pi^2} \epsilon^{0ijk} R_{ab0i} {R^{ab}}_{jk} ~,
\ee
which has the right form but comes with positive sign.

{\bf (ii). $A_2$ terms}: This group contains all other terms that meet the four criteria
outlined before and they have derivatives of $\Delta_{\mu} = k_{\nu} \nabla_{\mu}
\nabla^{\nu} \sigma (x, x^{\prime})$. Such terms will arise from the power series
expansion of the exponential, which now apart from ${\rm exp} (X+Y)$ it also receives
contributions from the commutators $[X, ~ Y]$, $[X, ~ [X , ~ Y]]$, $[Y, ~ [X , ~ Y]]$
and $[Y , ~ [X , ~ [X , ~ Y]]]$. The relevant terms are too many to write them down
explicitly. We only mention here that they can also be divided in two subgroups:
$A_2^{(1)} (t, x)$ that encompasses the terms for which the scaling of $\Delta_0$ and
$\Delta_i$ is same as $k_0$ and $k_i$ and $A_2^{(2)} (t, x)$ that encompasses the terms
for which the scaling relation is anomalous (their naive order with respect to the
scaling of $\Delta_{\mu}$ is $1/M^6$ and $1/M^8$, respectively, as explained earlier).
In the first subgroup we may simply replace
$\nabla_{\mu} \nabla_{\nu} \Delta_0 (x, x^{\prime})$ by
$k_0 [\nabla_{\mu} \nabla_{\nu} \nabla_0 \nabla^0 \sigma]$ and
$\nabla_{\mu} \nabla_{\nu} \Delta_i (x, x^{\prime})$ by
$k_j [\nabla_{\mu} \nabla_{\nu} \nabla_i \nabla^j \sigma]$ in the coincidence limit
$x \rightarrow x^{\prime}$, whereas in the second subgroup the anomalous scaling term
$M^2 k_0 [\nabla_{\mu} \nabla_{\nu} \nabla_i \nabla^0 \sigma]$ shown in \eqn{scala9}
should be used instead to replace $\nabla_{\mu} \nabla_{\nu} \Delta_i (x, x^{\prime})$.
In either case, we employ expression \eqn{synd3} for the fourth order Synge--DeWitt
tensor,
\be
[\nabla_{\mu} \nabla_{\nu} \nabla_{\kappa} \nabla_{\lambda} \sigma] = {1 \over 3}
\left(R_{\mu \kappa \lambda \nu} + R_{\mu \lambda \kappa \nu} \right) ,
\ee
and obtain after a very long calculation that also involves the term--by--term
integration in momentum space (using all relations \eqn{skoubi1}--\eqn{skoubi4}
up to $n=3$) the following result
\ba
A_2 (t, x) & = & -{19 \over 1440 \pi^2} \epsilon^{0ijk} \left(R_{0i0l}
{R^{0l}}_{jk} - R_{jk0l} {R^{ln}}_{in} \right) \nonumber\\
& = & -{19 \over 2880 \pi^2} \epsilon^{0ijk} R_{ab0i} {R^{ab}}_{jk} ~.
\ea
Note that the contribution of $A_2 (t, x)$ comes with the opposite sign compared
to $A_1 (t, x)$.

Combining these expressions, it turns out that all terms that contribute to
the sum $A(t, x) = A_1 (t, x) + A_2 (t, x)$ yield a net result
\be
A(t, x) = - {1 \over 192 \pi^2} \epsilon^{0ijk} {R^{ab}}_{0i} R_{ab ~ jk} ~.
\ee
Thus, in conclusion, the axial anomaly of a Lifshitz fermion in the background of
a metric field, which equals $2A(t, x)$, turns out to be identical to that of a
Dirac fermion, i.e.,
\be
\nabla_{\mu} J_5^{\mu} = - {1 \over 96 \pi^2} \epsilon^{0ijk}
{R^{ab}}_{0i} R_{ab ~ jk}
\equiv -{1 \over 192 \pi^2} {\rm Tr} (R \wedge R)
\label{berlous}
\ee
with the same overall numerical coefficient, as advertised before.

Note that if we were computing the anomaly of a Weyl--Lifshitz fermion, as opposed to
the Dirac--Lifshitz fermion that we have been considering so far, the overall coefficient
would be half of \eqn{berlous}.

Finally, we end this section with some general remarks regarding alternative approaches
to the axial anomaly which can also be used in non--relativistic field theories.

Although the Fujikawa method for computing the anomaly is straightforward to implement
in principle, it turns out to be very cumbersome for higher order operators, as we have
just seen. Later, in section 4, we will provide an alternative method for determining
the overall coefficient of the topological density ${\rm Tr} (R \wedge R)$ based on the
computation of the $\eta$--invariant of the three dimensional Lifshitz operator on
$\Sigma_3$ and obtain the corresponding index theorem associated to the integrated form of
the gravitational anomaly for the $(3+1)$--dimensional Dirac--Lifshitz operator. In
effect, this will provide a faster though more mathematically oriented way to arrive
at the same result.

There is also an alternative physical
method based on supersymmetric quantum mechanics to compute efficiently and reliably
the axial anomaly, as described in the classic work \cite{gaume} for the cases of the Dirac
and the Rarita--Schwinger fermion operators. In that context, one has to find the appropriate
supersymmetric two--dimensional non--linear sigma and reduce it to $0+1$ dimensions so that
the index of the fermion operator can be obtained by path integral methods of the underlying
supersymmetric quantum mechanics model. The general philosophy of this procedure is to find
a one--dimensional quantum mechanical system defined on the Riemannian space--time manifold
$M_4$ such that its Hamiltonian is the square of the fermion operator. Then, the expression
for $A(t, x)$ is interpreted as the partition function for an ensemble with density
matrix $\rho = \gamma_5 ~ {\rm exp} [-\beta (i \gamma^{\mu} D_{\mu})^2]$ at temperature
$\beta^{-1}$ ($\beta^{-1} = M^2$ for relativistic Dirac fermions) and its evaluation is
equivalent to the high temperature expansion of the system under periodic boundary
conditions for both bosons and fermions. The details of the calculation are usually
much simpler in that approach than in Fujikawa's treatment of the same problem as they
circumvent the very many terms that can arise otherwise. This approach also allows to relate
the result to the computation of the fermion propagator in a constant uniform magnetic field
\cite{julian} (but see also \cite{zuber}, in particular p. 100), thus providing a more physical
interpretation to the origin of the topological density that enters into the anomaly (in effect,
Schwinger's computation provides a physical derivation of the so called A--roof genus of the
fermion operator on $M_4$). We have not
considered this possibility for Lifshitz fermions so far, but we think that the right framework
should be provided by sigma models with non--relativistic supersymmetry that generalize the
construction reported recently in \cite{xue} to theories with multi--component scalar
fields associated to a $z=3$ Lifshitz non--linear sigma model in $1 + 1$ dimensions.
If this step is successfully implemented in the future, the treatment of anomalies
in quantum field theories with non--relativistic fermions will be on par with
the relativistic theories. Work in this direction is in progress and will be reported
elsewhere.

\section{Instantons of Ho\v{r}ava--Lifshitz gravity}
\setcounter{equation}{0}

In this section we present an overview of the instanton solutions of
$z=3$ Ho\v{r}ava--Lifshitz gravity following our earlier work on the subject \cite{BBLP}.
After some general considerations that set up the notation and framework for the definition
of instantons in terms of geometric flows, we focus on the explicit construction and
classification of solutions with $SU(2)$ isometry. These will be used later as metric
backgrounds to evaluate the integrated form of the axial anomaly and provide examples
of chiral symmetry breaking in Lifshitz theories induced by gravity. It will turn out that
such phenomena become possible only in a unimodular phase of Ho\v{r}ava--Lifshitz gravity,
which is thought to govern the deep ultra--violet regime of the theory.
Our treatment of instantons is rather complementary to the results presented in \cite{BBLP}
giving a more qualitative picture that is also appropriate for
comparison with similar solutions that arise in Euclidean Einstein gravity.
The presentation is self--contained as several of these results are not widely known.
We also reveal the curvature characteristics of the solutions that make
them relevant for applications to chiral symmetry breaking.

\subsection{General considerations}

Ho\v{r}ava--Lifshitz gravity is a non--relativistic theory that is defined by assuming
that space--time is of the form $\mathbb{R} \times \Sigma_3$. It is naturally formulated
using the ADM decomposition of the space--time metric $G_{\mu \nu}$,
\be
ds^2 = -N^2 dt^2 +g_{ij}\left(dx^i +N^i dt\right) \left(dx^j +N^j dt\right)
\ee
with lapse and shift functions $N$ and $N_i$, respectively. The spatial slices $\Sigma_3$,
which are the leaves of a foliation, have induced metric $g_{ij}$ and extrinsic curvature
(second fundamental form) $K_{ij}$,
\be
K_{ij} = {1 \over 2N} \left(\partial_t g_{ij} - \nabla_i N_j - \nabla_j N_i \right) ,
\ee
which provides the conjugate momentum to the metric $g_{ij}$ in the canonical
formalism. The momenta are simply expressed using the superspace metric ${\cal G}^{ijkl}$
as
\be
\pi^{ij} = {2 \over \kappa^2} \sqrt{{\rm det} g} ~ {\cal G}^{ijkl} K_{kl} ~.
\ee

Recall that superspace is defined as the infinite dimensional space of all Riemannian
metrics on $\Sigma_3$, and, as such, it provides the arena for geometrodynamics.
It is naturally endowed with a metric
\be
{\cal G}^{ijkl} = {1 \over 2} \left(g^{ik} g^{jl} + g^{il} g^{jk} \right) -
\lambda g^{ij} g^{kl}
\ee
that generalizes the standard DeWitt metric \cite{witt2} using an arbitrary parameter
$\lambda$. There is also the inverse metric
\be
{\cal G}_{ijkl} = {1 \over 2} \left(g_{ik} g_{jl} + g_{il} g_{jk} \right) -
{\lambda \over 3\lambda -1} g_{ij} g_{kl}
\ee
which satisfies the relation
\be
{\cal G}^{ijkl} {\cal G}_{klmn} = {1 \over 2} \left(\delta_m^i \delta_n^j +
\delta_n^i \delta_m^j \right) .
\label{gadger1}
\ee
In general relativity $\lambda =1$, but in Ho\v{r}ava--Lifshitz gravity $\lambda$
can take more general values. It is important to note that the superspace metric is
positive definite when $\lambda < 1/3$ and it is indefinite when $\lambda > 1/3$;
in the latter case, which includes the value $\lambda =1$, the indefiniteness of the
DeWitt metric accounts for the conformal factor problem in all gravitational theories.
The value $\lambda = 1/3$ is special since the DeWitt metric becomes degenerate. It is
closely related to the limit $\lambda \rightarrow \pm \infty$ for which the inverse
DeWitt metric becomes degenerate. In these special cases, the conformal factor of the
three--dimensional geometries decouples and a unimodular phase of the theory
emerges; we will say more about this later as it provides a very important class
of models for the purposes of the present work.

With these definitions in mind, the action of Ho\v{r}ava--Lifshitz gravity takes
the following form written as a sum of kinetic and potential terms,
\cite{horava1, horava2},
\be
S = {2 \over \kappa^2} \int dt d^3x \sqrt{{\rm det} g} ~ N ~ K_{ij} {\cal G}^{ijkl} K_{kl}
- {\kappa^2 \over 2} \int dt d^3x \sqrt{{\rm det} g} ~ N ~ E^{ij} {\cal G}_{ijkl} E^{kl} ~,
\label{acthora}
\ee
where $\kappa^2 = 32 \pi G$ is the four--dimensional gravitational coupling and $E^{ij}$
is the gradient of a suitably chosen {\em local} functional $W[g]$,
\be
E^{ij} = -{1 \over 2 \sqrt{{\rm det} g}} {\delta W[g] \over \delta g_{ij}} ~.
\ee
Here and in the following we are only considering the form of the theory which is said to
satisfy the {\em detailed balance condition}, meaning, in particular, that the potential
term is derived for a superpotential functional $W[g]$. Other forms of the theory have
also been studied in the literature, but they will not be in focus in the present work.
Ordinary general relativity resembles the form of the action \eqn{acthora} but
the potential is given by minus the Ricci scalar curvature $R$ of the metric on
$\Sigma_3$. In that case, one may still derive the potential from a superpotential
functional, which is known as the Hamilton--Jacobi functional of general relativity,
but it is a non--local functional\footnote{Non--local functionals can also be used
formally to recast Ho\v{r}ava--Lifshitz theory without detailed balance into detailed
balance form, but these generalizations will not be pursued further.} of the metric
$g$ in general.

The resulting theory exhibits scaling anisotropy in space and
time depending on the choice of $W[g]$. We are going to consider models with anisotropic
scaling exponent $z=3$ by choosing as superpotential functional the action of
three--dimensional topologically massive gravity on $\Sigma_3$ which is defined as
follows, \cite{DJT},
\be
W [g] = {2 \over \kappa_{\rm w}^2} \int_{\Sigma_3} d^3x \sqrt{{\rm det} g} ~
(R-2 \Lambda_{\rm w}) + {1 \over \omega} W_{\rm CS} ~.
\ee
The first term is the usual Einstein--Hilbert action in three dimensions with
gravitational coupling $\kappa_{\rm w}$ and cosmological constant $\Lambda_{\rm w}$, whereas
the second term is the gravitational Chern--Simons action \cite{simons}, which is
written in terms of the Christoffel symbols of the metric $g$,
\ba
W_{\rm CS} [g] & = & \int_{\Sigma_3} d^3x \sqrt{g} ~ \epsilon^{ijk}
\Gamma_{im}^l \left(\partial_j \Gamma_{lk}^m + {2 \over 3} \Gamma_{jn}^m
\Gamma_{kl}^n \right)  \nonumber\\
& = & {1 \over 2} \int_{\Sigma_3} {\rm Tr} \left(\omega \wedge d \omega
+ {2 \over 3} \omega \wedge \omega \wedge \omega \right)
\ea
using the fully antisymmetric symbol in three dimensions with $\epsilon^{123} = 1$
or equivalently using the spin connection one--forms $\omega$ (they should not be
confused with the coupling constant $\omega$ of the Chern--Simons term).
Obviously, this theory is not invariant under parity, since orientation reversing
transformations on $\Sigma_3$ flip the sign of the Chern--Simons coupling $\omega$.
This affects the vacuum structure of topologically massive gravity, since it introduces
chirality in the geometry, and it also has implications for the instanton solutions of
Ho\v{r}ava--Lifshitz gravity.

The classical equations of motion that follow by varying $W[g]$ take the form
\be
R_{ij} - {1 \over 2} R g_{ij} + \Lambda_{\rm w} g_{ij} - {\kappa_{\rm w}^2 \over \omega}
C_{ij} = 0 ~,
\label{equatmg}
\ee
where $C_{ij}$ is the Cotton tensor of the metric $g$, which is defined as follows,
\be
C_{ij} = {1 \over \sqrt{{\rm det} g}} {\epsilon_i}^{kl} \nabla_k \left(R_{jl} -
{1 \over 4} R g_{jl} \right) .
\ee
The Cotton tensor is a covariantly conserved and traceless tensor, which is third order with
respect to the space derivatives of the metric, and it accounts for the anisotropic scaling
exponent $z=3$ of the associated Ho\v{r}ava--Lifshitz gravity. A special case arises
when $\kappa_{\rm w}$ becomes infinite and the Einstein--Hilbert term drops from the
action $W$; the resulting three--dimensional theory is conformal gravity whose vacua
are conformally flat metrics satisfying the field equations $C_{ij} = 0$ (see, for
instance, \cite{witten} for more details). Another special case arises
when $\omega$ becomes infinite and the gravitational Chern--Simons term drops out
from the action $W$; the resulting three--dimensional theory is ordinary Einstein
gravity and in this case the anisotropic scaling of the associated Ho\v{r}ava--Lifshitz
gravity reduces to $z=2$. In all cases, the vacua of three--dimensional gravity
provide static solutions of the $3+1$ non--relativistic theory \eqn{acthora}.

Ho\v{r}ava--Lifshitz gravity, in its original formulation that we are using here, does
not exhibit general coordinate invariance but only a restricted symmetry associated
to foliation preserving diffeomorphisms of space--time. Then, for general values of
$\lambda$, the physical viability of the theory becomes questionable because it contains a
spin--$0$ graviton mode that is unwanted in the infrared limit and it should
decouple. For certain values of $\lambda$ this problem does not arise; we will say more
about this later, since most of our present work will concentrate to those special values
of $\lambda$.
For now it suffices to say that we are adopting the projectable version of the theory, and,
furthermore, in view of the applications, we choose the lapse and shift functions
\be
N(t) = 1 ~, ~~~~~~ N_i (t, x) = 0
\label{ourchoicea}
\ee
exactly as in the case of Lifshitz fermions, unless stated otherwise. Setting $N(t) = 1$
amounts to using proper time.

Gravitational instanton solutions can be obtained by considering the analytic continuation
of Ho\v{r}ava--Lifshitz theory in the time coordinate \cite{BBLP}. The four--dimensional
Euclidean action that follows from \eqn{acthora} by inverting the potential term is
bounded from below provided that the superspace metric is positive definite. Thus,
assuming that $\lambda < 1/3$, we have the following relations
\ba
S_{\rm Eucl.} & = & {2 \over \kappa^2} \int dt d^3x \sqrt{{\rm det} g} ~ K_{ij}
{\cal G}^{ijkl} K_{kl} + {\kappa^2 \over 2} \int dt d^3x \sqrt{{\rm det} g} ~ E^{ij}
{\cal G}_{ijkl} E^{kl} \nonumber\\
& = & {2 \over \kappa^2} \int dt d^3x \sqrt{{\rm det} g} \left(K_{ij} \pm {\kappa^2 \over 2}
{\cal G}_{ijmn} E^{mn} \right) {\cal G}^{ijkl} \left(K_{kl} \pm {\kappa^2 \over 2}
{\cal G}_{klrs} E^{rs} \right) \nonumber\\
& & \mp 2 \int dt d^3x \sqrt{{\rm det} g} ~ K_{ij} E^{ij} ~.
\ea
We also implicitly assume that $\Sigma_3$ is a compact manifold
with no boundary in order to avoid unnecessary complications that may arise from
additional boundary terms in the Euclidean action (persistent boundary effects in gravitational
theories require more careful treatment as in the monopole sector of gauge theories). It is
then clear that the action is bounded from below as
\be
S_{\rm Eucl.} \geq \mp 2 \int dt d^3x \sqrt{{\rm det} g} ~ K_{ij} E^{ij} = \mp
\int dt d^3x \sqrt{{\rm det} g} ~ E^{ij} \partial_t g_{ij} = \pm {1 \over 2}
\int dt ~ {dW \over dt} ~.
\ee
The lower bound is saturated for special configurations satisfying the first order
equations in time
\be
\partial_t g_{ij} = \mp {\kappa^2 \over 2 \sqrt{{\rm det} g}} {\cal G}_{ijkl}
{\delta W [g] \over \delta g_{kl}} ~,
\label{loulakii}
\ee
which describe gradient flow equations for the metric on $\Sigma_3$. They provide
extrema of the four--dimensional action, and, as such, they also satisfy the classical
equations of motion which are second order in time.

The instantons of Ho\v{r}ava--Lifshitz gravity are defined to be {\em eternal} solutions
of the geometric flow equations \eqn{loulakii} that exist for all time
$-\infty < t < +\infty$. The $\pm$ sign refers to instantons and anti--instantons, which
are mutually related by $t \rightarrow -t$. These flow lines interpolate smoothly
between any two fixed points of \eqn{loulakii}, i.e., between different extrema of $W[g]$,
and they have finite Euclidean action
\be
S_{\rm instanton} = {1 \over 2} |\Delta W|
\label{abcgd}
\ee
given by the difference of $W$ at the two fixed points reached as $t \rightarrow \mp \infty$.
It can be easily seen that $W$ changes monotonically along the flow lines, $\pm dW/dt \geq 0$, 
provided that $\lambda < 1/3$ so that the superspace metric is positive definite. Therefore,
the instanton action is always positive definite, justifying the use of the absolute value in 
\eqn{abcgd}. This definition is analogous to the instanton solutions of point particle 
systems that interpolate smoothly between different degenerate vacua (minima
of the potential). In the present context, the absence of singularities along the
eternal solutions of the flow equation \eqn{loulakii} ensures that the corresponding
space--time metric is complete and regular. Other flow lines that may only exist for a finite
time interval before they become extinct are not physically acceptable as they are
inflicted with singularities.

Specializing $W[g]$ to the action of three--dimensional topologically massive gravity,
we obtain from \eqn{loulakii} the following third order non--linear equation for $g$,
\be
\partial_t g_{ij} = \mp {\kappa^2 \over \kappa_{\rm w}^2} \left(R_{ij} -
{2\lambda -1 \over 2(3\lambda -1)} R g_{ij} - {\Lambda_{\rm w} \over 3\lambda -1} g_{ij} \right)
\pm {\kappa^2 \over \omega} C_{ij} ~,
\label{basicoex}
\ee
which is called Ricci--Cotton flow \cite{BBLP}. Its eternal solutions are the instantons
of $z=3$ Ho\v{r}ava--Lifshitz gravity that will be studied later in detail focusing, in
particular, to a mini--superspace model that allows us to derive explicit results and
obtain complete classification of all such special configurations.

Although the flow lines of \eqn{basicoex} depend explicitly on $\lambda$,
the fixed points are in fact independent of it (they are vacua of topologically
massive gravity irrespective of $\lambda$). As such, they satisfy equation
\eqn{equatmg}, which is equivalently written as
\be
R_{ij} - {1 \over 3} R g_{ij} - {\kappa_{\rm w}^2 \over \omega} C_{ij} = 0 ~~~~
{\rm and} ~~~~ R = 6 \Lambda_{\rm w} ~.
\ee
The first relation is the traceless part of \eqn{equatmg} and the other is the trace
telling us, in particular, that the two end--points of the instantons are metrics on
$\Sigma_3$ with Ricci scalar curvature $R$ fixed by $\Lambda_{\rm w}$. The curvature
of $\Sigma_3$ changes along the flow lines. The flow equation
becomes independent of $\Lambda_{\rm w}$ only when $\lambda$ is infinite, in which case
the end--points of the instantons can have arbitrary $R$ (in this case, the fixed points
satisfy only the traceless part of \eqn{equatmg}).
We also note for completeness that the Ricci--Cotton flow \eqn{basicoex} becomes the
celebrated Ricci flow when $\omega \rightarrow \pm \infty$. Actually, in that case, we
obtain Hamilton's Ricci flow for $\lambda = 1/2$ (see, for instance, the collected papers
on the subject \cite{styau} and the textbooks \cite{topping, lni}). Finally,
when $\kappa_{\rm w} \rightarrow \infty$, equation \eqn{basicoex} specializes to the
Cotton flow that was introduced in the literature in more recent years \cite{cotton}.

In all cases, the instantons of Ho\v{r}ava--Lifshitz gravity are regular space--time
configurations of the form $\mathbb{R} \times \Sigma_3$ that extend for all time
$-\infty < t < +\infty$ and they are suspended from two fixed points. They can
be visualized as infinitely long cylinders which provide the envelope of the geometric
flow on $\Sigma_3$ as shown in Fig.1. The slices (which are depicted here by
suppressing two of the three spatial dimensions) are the portraits of the geometry
shown at different instances of proper time. Similar pictures can also be drawn for
the instantons of Einstein gravity in proper time, but the cylinder is semi--infinite in
that case, since the solutions are suspended from removable singularities (nuts or bolts)
at one end. We will say more about these differences in due course.

\vspace{1cm}

\begin{figure}[h]
\centering
\epsfxsize=8cm\epsfbox{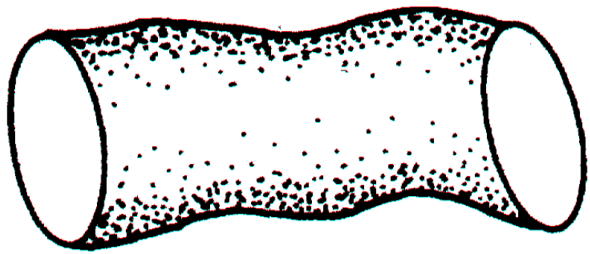}
\vspace{+0.6cm}
\caption{Schematic picture of an interpolating instanton solution $\mathbb{R} \times \Sigma_3$}
\end{figure}

\subsection{Unimodular phase at special values of $\lambda$}

Next, we examine the properties of Ho\v{r}ava--Lifshitz gravity at special values of
$\lambda$ that will play important role in the applications. First, we consider the case
$\lambda =1/3$ which amounts to
\be
{\cal G}^{ijkl} g_{kl} =0 ~, ~~~~~
\pi^{ij} = {2 \over \kappa^2} \sqrt{{\rm det} g} ~ {\cal G}^{ijkl} K_{kl} = {2 \over \kappa^2}
\sqrt{{\rm det} g} \left(K^{ij} - {1 \over 3} g^{ij} K\right) .
\ee
The first relation states that the superspace metric is degenerate as it develops a
null eigen--vector provided by $g$, whereas the second relation shows that the conjugate
momentum becomes traceless. More generally, ${\cal G}^{ijkl}$ projects any tensor to its
traceless part. At the same time, the inverse metric in superspace is not well defined (it
becomes infinite), and appropriate projection is required to make sense of the theory
at $\lambda = 1/3$. Thus, at $\lambda = 1/3$, we are led to define
\be
{\cal G}^{ijkl} = {1 \over 2} \left(g^{ik} g^{jl} + g^{il} g^{jk}\right) - {1 \over 3}
g^{ij} g^{kl}
\ee
and
\be
{\cal G}_{ijkl} = {1 \over 2} \left(g_{ik} g_{jl} + g_{il} g_{jk}\right) - {1 \over 3}
g_{ij} g_{kl} ~,
\ee
so that ${\cal G}_{ijkl}$ also projects any tensor to its traceless part. In this case
we have
\be
{\cal G}^{ijkl} {\cal G}_{klmn} = {1 \over 2} \left(\delta_m^i \delta_n^j +
\delta_n^i \delta_m^j \right) - {1 \over 3} g^{ij} g_{mn} ~,
\label{gadger2}
\ee
which differs from the corresponding relation \eqn{gadger1} by subtracting its trace
part. This modification is necessary for consistency of the projection.

The projection to traceless tensors is clearly related to the role that the conformal
factor (volume of space) has in the theory. The conformal factor of the metric $g$
is a propagating degree of freedom with positive kinetic energy in superspace when
$\lambda < 1/3$ and with negative kinetic energy when $\lambda > 1/3$. At the boarder value
$\lambda =1/3$ its kinetic energy vanishes, and, therefore, the conformal factor has no
dynamics at all. This does not necessarily imply that the theory is invariant under rescaling,
but it means that the volume of space can only appear as spectator. Then, the action of
Ho\v{r}ava--Lifshitz gravity \eqn{acthora} at $\lambda = 1/3$ takes the form (here we
reinstate momentarily $N$ and $N_i$ in order to make some general comments)
\ba
S & = & {2 \over \kappa^2} \int dt d^3x \sqrt{{\rm det} g} ~ N \left(K_{ij} - {1 \over 3}
g_{ij} K \right) \left(K^{ij} - {1 \over 3} g^{ij} K \right)
- {\kappa^2 \over 2} \int dt d^3x \sqrt{{\rm det} g} ~ N ~ \times \nonumber\\
& & \Big[{1 \over \kappa_{\rm w}^2} \left(R_{ij} -
{1 \over 3} R g_{ij} \right) - {1 \over \omega} C_{ij} \Big]
\Big[{1 \over \kappa_{\rm w}^2} \left(R^{ij} -
{1 \over 3} R g^{ij} \right) - {1 \over \omega} C^{ij} \Big] ~.
\label{tsiroto}
\ea
The underlying three--dimensional theory of topologically massive
gravity becomes {\em unimodular} in this case, since the volume of space $\Sigma_3$ is
held fixed. Its field equations are solely described by the traceless condition
\be
R_{ij} - {1 \over 3} R g_{ij} - {\kappa_{\rm w}^2 \over \omega} C_{ij} = 0
\label{kolitiro}
\ee
without imposing any restriction on $R$. Then, $\Lambda_{\rm w}$ can assume arbitrary
values as it has the interpretation of an integration constant arising from the contracted
Bianchi identity (this is the standard viewpoint for the cosmological constant in
all unimodular theories of gravity).
Similar reasoning applies to the limiting case $\lambda \rightarrow \pm \infty$ that
also leads to the action \eqn{tsiroto}.

The kinetic term of the action \eqn{tsiroto} is invariant under the following
transformation
\be
g_{ij} \rightarrow \Omega^2 (t, x) g_{ij} ~, ~~~~~
N \rightarrow \Omega^3 (t, x) N ~, ~~~~~ N_i \rightarrow \Omega^2 (t, x) N_i
\label{anisoweyl}
\ee
that changes the measure of integration as $\sqrt{{\rm det} g} ~ N \rightarrow \Omega^{6}
\sqrt{{\rm det} g} ~ N$. If $W[g]$ were the action of three--dimensional conformal gravity,
by letting $\kappa_{\rm w} \rightarrow \infty$, the potential term of Ho\v{r}ava--Lifshitz
gravity would also remain invariant under this transformation, since the Cotton tensor
transforms as $C_{ij} \rightarrow \Omega^{-1} C_{ij}$. In that case, the theory would
exhibit anisotropic Weyl invariance, as it was first noted in \cite{horava1, horava2},
thus turning the rigid anisotropic scaling of space and time coordinates $(t, x)
\rightarrow (a^3t, ax)$ into a larger local symmetry\footnote{Only in that case it is
appropriate to consider the non--projectable version of the theory, since the lapse
function necessarily depends on all space--time coordinates; even if we were choosing $N$
to depend only on $t$, the transformation \eqn{anisoweyl} would make it to depend on
both $t$ and $x$.}. However, the deformation that arises when $\kappa_{\rm w}$
is held finite breaks anisotropic Weyl invariance because the traceless part of the
Ricci tensor is invariant under conformal transformations and the potential term of the
action \eqn{tsiroto} depends explicitly upon $\Omega$. Thus, at $\lambda = 1/3$ (and
likewise at $\lambda = \pm \infty$) the action of Ho\v{r}ava--Lifshitz gravity
is not Weyl invariant, in general. It is only invariant under the restricted group
of foliation preserving diffeomorphisms that keep the volume of $\Sigma_3$ unchanged.
They account for the additional Hamiltonian constraint ${\pi^i}_i = 0$, which now follows
from the unimodular projection of topologically massive gravity on $\Sigma_3$.

It is also interesting to note that the scalar (spin--$0$) graviton mode which haunts
Ho\v{r}ava--Lifshitz gravity for generic values of $\lambda$ decouples in this case
and only the usual tensor graviton mode remains in the physical spectrum. This was
first noted in the literature by considering linear perturbations around the Minkowski
vacuum of the anisotropic Weyl invariant theory of gravity \cite{horava1, horava2},
but the proof was later generalized to the more general theory defined by \eqn{tsiroto}
\cite{park}. Thus, the phase of the theory arising at $\lambda = 1/3$ is not plagued
with the inconsistencies that otherwise haunt the non--relativistic theories of gravity.
We will stick with it in the applications that will be considered later for reason
that will become transparent in due course. Two important questions will remain
unanswered though, but they are both lying beyond the scope of the present work: one is
the extrapolation to large scale physics and the other is the fate of Weyl invariance
and unimodular symmetry upon quantization.

The instantons of the gravitational theory \eqn{tsiroto} arising at these special values
of $\lambda$ are eternal solutions of the flow equation
\be
\partial_t g_{ij} = \mp {\kappa^2 \over \kappa_{\rm w}^2} \left(R_{ij} -
{1 \over 3} R g_{ij} \right) \pm {\kappa^2 \over \omega} C_{ij} ~,
\label{basicoex2}
\ee
which can be obtained formally from \eqn{basicoex} by taking the limit $\lambda \rightarrow \pm
\infty$ (we opt the limit $\lambda \rightarrow -\infty$ as it fits in the range
$\lambda < 1/3$ that was considered earlier). We call equation \eqn{basicoex2}
{\em normalized} Ricci--Cotton flow because the driving terms of the metric deformations
are traceless tensors that preserve the volume of space, as consequence of unimodularity in
three dimensions. The fixed points of the flow lines satisfy equation \eqn{kolitiro} and
they depend on the volume of space, which, however, is appearing now as a spectator (as noted
before, the traceless Ricci and Cotton tensors scale differently under conformal
transformations of the metric). As a result, the end--points of these instantons can have
different Ricci scalar curvature. This should be contrasted to the curvature of the fixed points
of the unnormalized Ricci--Cotton flow \eqn{basicoex}, which is the same.

Next, we obtain explicit solutions by considering instantons with sufficiently large
group of isometries so that the equations become manageable. Instantons with less or no
isometries may also exist but we are lacking the mathematical tools to investigate them
properly. A few general remarks about this will only be made at the end of section 3.4.

\subsection{Instantons with $SU(2) \times U(1)$ isometry}

Instantons with $SU(2)$ isometry arise by choosing $\Sigma_3 \simeq S^3$ and introducing
homogeneous metrics of Bianchi IX type
\be
ds^2 =  \gamma_1 (t) (\sigma^1)^2 + \gamma_2 (t) (\sigma^2)^2 + \gamma_3 (t) (\sigma^3)^2
\label{spatoulio}
\ee
in terms of the left--invariant one--forms $\sigma^I$ of the group $SU(2)$ that satisfy
the defining relations
\be
d\sigma^I + {1 \over 2} {\epsilon^I}_{JK} \sigma^J \wedge \sigma^K = 0 ~.
\ee
Remarkably, this ansatz leads to consistent reduction of the Ricci--Cotton flow to a
system of ordinary differential equations for the metric coefficients $\gamma_I$ which
are taken to depend only on time. The same applies to all other homogeneous model
geometries on $\Sigma_3$, but we are only considering here the mini--superspace model
of the flow equations with $SU(2)$ symmetry. The resulting equations are highly non--linear
and unfortunately they can not be integrated when the geometry of $S^3$ is totally
anisotropic with $\gamma_1 \neq \gamma_2 \neq \gamma_3$. It is possible to investigate
these equations numerically and study some of their qualitative features, which turn out to
be sufficient for establishing the existence of $SU(2)$ gravitational instantons, in
general, following \cite{BBLP}. We will not repeat the same line of presentation
here, but rather give a complementary account of the solutions and then expand on
their properties that are most relevant to the present work.

In view of the applications that will be considered in the next section, it suffices to
restrict attention to axially symmetric configurations (also known as Berger spheres) by
letting
\be
\gamma_1 = \gamma_2 = {xL^2 \over 4} ~, ~~~~~~ \gamma_3 = {L^2 \over 4x^2}
\label{bergersphe}
\ee
and present all gravitational instantons with
enhanced isometry $SU(2) \times U(1)$. Here, $L$ is a characteristic length scale so that
the volume of space equals $2 \pi^2 L^3$. We also confine ourselves to the normalized
Ricci--Cotton flow, in which case the volume of space is preserved by the evolution (a few
remarks about the unnormalized flow associated to more general values of $\lambda$ will
be made later). Then, in this case, the system \eqn{basicoex2}
reduces consistently to a single differential equation for $x(t)$ in proper time $t$,
\be
{dx \over dt} = \pm {4 \kappa^2 \over \omega L^3} \left({1 \over x^3} - 1 \right)
\left({1 \over x^2} + {\omega L \over 3 \kappa_{\rm w}^2} \right) ,
\label{odey}
\ee
whose solutions will provide instantons in the unimodular phase of
Ho\v{r}ava--Lifshitz gravity. The variable $x$ is positive definite and it
parametrizes the shape modulus of space. The value $x=1$ corresponds to a round sphere,
where $x>1$ (respectively $x<1$) describes a geometrically deformed sphere which is squashed
(respectively stretched) along the axis of symmetry. Of course, the choice of axis is
arbitrary because one can permute the three principal directions of $S^3$, and, as a result, our
construction of instantons is unique up to a $\mathbb{Z}_3$ permutation symmetry.

The instanton solutions are by definition eternal
solutions of the flow equation that exist for all time $-\infty < t < + \infty$ and
interpolate smoothly
between any two fixed points, provided that there are more than one degenerate vacua in the
system. It is clear, by inspecting equation \eqn{odey}, that the number of available fixed
points with $SU(2) \times U(1)$ isometry depends crucially on the sign of the gravitational
Chern--Simons coupling $\omega$. For $\omega > 0$, there is only one fixed point located at
$x=1$, whereas for $\omega < 0$ there is an additional fixed point with
\be
\gamma_1 = \gamma_2 = {L^2 \over 4a} ~, ~~~~~~ \gamma_3 = {a^2 L^2 \over 4} ~,
\label{tsoutsouli}
\ee
which is axially symmetric with shape modulus specified in terms of the couplings by
\be
a = \sqrt{- {\omega L \over 3 \kappa_{\rm w}^2}} ~.
\label{tsulai}
\ee
Thus, for $\omega > 0$, there is no instanton, whereas for $\omega < 0$ there is a single
instanton solution with Euclidean action that turns out to be
\be
S_{\rm instanton} = {4 \pi^2 \over |\omega|} (a-1)^2 \left(4a^4 + 8a^3 + 12a^2
+ 2a +1 \right) .
\label{instacti}
\ee
Note that the two fixed points coalesce for special values of the couplings associated
to the choice $a =1$, in which case the instanton collapses to a point and its action
\eqn{instacti} vanishes, as required on general grounds. For $a < 1$ (respectively
$a > 1)$ the anisotropic fixed point is located to the right (respective left) of the
isotropic fixed point on the $x$--axis and the corresponding instanton interpolates between a
squashed (respectively stretched) sphere and the round one. The reduced equation
\eqn{odey} can be easily integrated to express $t (x)$ in closed form, as in \cite{BBLP},
but the particular expression will not be needed in the present work.

This construction provides the prime example of an instanton in Ho\v{r}ava--Lifshitz gravity,
which has the important property of being {\em chiral} for general couplings of the theory.
The reason is that the Cotton tensor is odd under parity, as it involves the totally
antisymmetric tensor $\epsilon_{ijk}$ in its definition, and, therefore, it flips sign
under orientation reversing transformations on $S^3$. Instanton and, likewise,
anti--instanton configurations only exist for one orientation (sign of $\omega$) but not
for the other. This occurrence can be intuitively explained
by comparing the effect of the driving curvature terms of the Ricci--Cotton flow.
For $\omega > 0$ all curvature terms work in the same direction to form the totally
isotropic fixed point, but when $\omega < 0$ the Cotton tensor competes against the
other curvature terms so that more than one fixed points can arise. Then, interpolating
flow lines between the two become possible. Chiral instantons are not commonly used in
physics\footnote{We note in passing that analogous configurations are known to
exist in some simpler non--relativistic integrable field theory models in $1+1$
dimensions, as in a derivative variant of the non--linear Schr\"odinger equation,
which possesses chiral solitons (for a brief account see, for instance, \cite{japi}
and references therein). The chiral nature of such solitons is also attributed to the
presence of Chern--Simons terms in a $(2+1)$--dimensional gauge theory that
undergoes dimensional reduction and gives rise to these integrable systems
(see also \cite{seminara} for more technical details on this subject).}, but we have to
work with them now as they arise naturally in $z=3$ Ho\v{r}ava--Lifshitz gravity.

It is instructive at this point to provide an effective point particle description
of the instanton solutions associated to the normalized Ricci--Cotton flow in the
mini--superspace sector of axially symmetric Bianchi IX metrics. In this case, it can
be easily verified that the action \eqn{tsiroto} of the unimodular Ho\v{r}ava--Lifshitz
gravity consistently reduces to the action
\be
S_{\rm eff.} = {3 \pi^2 L^3 \over \kappa^2} \int dt \Big[\left({d \beta \over dt}
\right)^2 - {16 \kappa^4 \over \omega^2 L^6} \left(e^{-6\beta} +
{\omega L \over 3 \kappa_{\rm w}^2} e^{-4\beta} - e^{-3\beta} -
{\omega L \over 3 \kappa_{\rm w}^2} e^{-\beta} \right)^2 \Big]
\label{partimodel}
\ee
for a single degree of freedom $\beta = {\rm log} x$ that ranges in the entire real line.
This variable is commonly used in the mixmaster model of the universe (see, for instance,
\cite{mixmast1, mixmast2}) and it is required to bring the kinetic energy in canonical form.
The effective potential follows from a superpotential $W$ which is the action
of topologically massive gravity evaluated for the class of Bianchi IX metrics.
Upon analytic continuation in time, the instantons of the point particle
model provide our gravitational instanton solutions satisfying equation \eqn{odey}, which
now reads in terms of the new variable $\beta$
\be
{d \beta \over dt} = \pm {4 \kappa^2 \over \omega L^3} \left(e^{-6\beta} +
{\omega L \over 3 \kappa_{\rm w}^2} e^{-4\beta} - e^{-3\beta} -
{\omega L \over 3 \kappa_{\rm w}^2} e^{-\beta} \right) .
\label{odey2}
\ee
Furthermore, the gravitational instanton action is equal to
the Euclidean point particle action evaluated for the corresponding instanton solution,
thus providing a simple derivation of \eqn{instacti}.

Next, we plot the effective point particle potential $V(\beta)$ appearing in the action
\eqn{partimodel} for different values of the coupling $\omega$ (positive or negative) to
get a picture of the resulting vacuum structure in mini--superspace. The results are shown
in Fig.2 where the origin of the axis $\beta = 0$ represents the round sphere. When
$\omega \rightarrow \pm \infty$, the anisotropic scaling exponent of Ho\v{r}ava--Lifshitz
gravity reduces to $z=2$ since the Chern--Simons term drops out from the action.
In all other cases the theory exhibits anisotropic scaling $z=3$.
The value $\omega = 0$ is also special in that the action is given by the
gravitational Chern--Simons term alone, whereas
the Einstein--Hilbert term drops out. This limit is best described by letting
$\kappa_{\rm w} \rightarrow \infty$ so that the effective potential remains finite.

Fig.2a depicts the potential $V(\beta)$ for infinitely large $|\omega|$, in which case the
flow equation we are considering becomes the normalized Ricci flow that was studied long time
ago in all generality (we also refer the reader to the work \cite{jackson} for its reduction
to homogeneous model geometries). In our case, the effective potential has a hill peaked
at $\beta = ({\rm log} 4)/3$ where $\gamma_1 = \gamma_2 = 4 \gamma_3$.
Fig.2b depicts the potential for $\omega = 0$ (or better to say $\kappa_{\rm w}$ infinite),
in which case the flow equation becomes the Cotton flow \cite{cotton}. Here, the plot of the
effective potential resembles Fig.2a with the difference that the hill is now peaked at
$\beta = ({\rm log} 2)/3$ where $\gamma_1 = \gamma_2 = 2 \gamma_3$. For all other positive
values of $\omega$ the potential looks alike with the hill now being peaked somewhere between
the two extreme values $({\rm log} 2)/3$ and $({\rm log} 4)/3$; for this reason $V(\beta)$
is not drawn separately.

Next, we consider the case of negative $\omega$ (with respect to a given orientation of $S^3$),
which is different because the
potential exhibits two vacua located at $\beta = 1$ and $\beta = - {\rm log} a$, using the
parameter $a$ given by equation \eqn{tsulai}. Fig.2c represents the typical form of the potential
for $a < 1$, whereas Fig.2d depicts the potential for $a > 1$. As the parameter $a$
approaches $1$, the two valleys (and, likewise, the two hills) are beginning to merge before
they cross each other. For $a=1$ the potential has a single hill and its plot resembles Fig.2a.
Finally, when $a$ becomes larger and larger the left hill is pushed further and further
away to the left until it is completely taken over by the infinitely steep wall of the
potential arising as $\beta \rightarrow -\infty$. The fact that the anisotropic ground state is
not accessible anymore as $a \rightarrow \infty$ is also apparent from the instanton action
\eqn{instacti} which becomes infinite; in that case one returns back to the
Ricci flow potential represented by Fig.2a, as noted before.

\begin{figure}[h]
\centering
\epsfysize=4.75cm\epsfbox{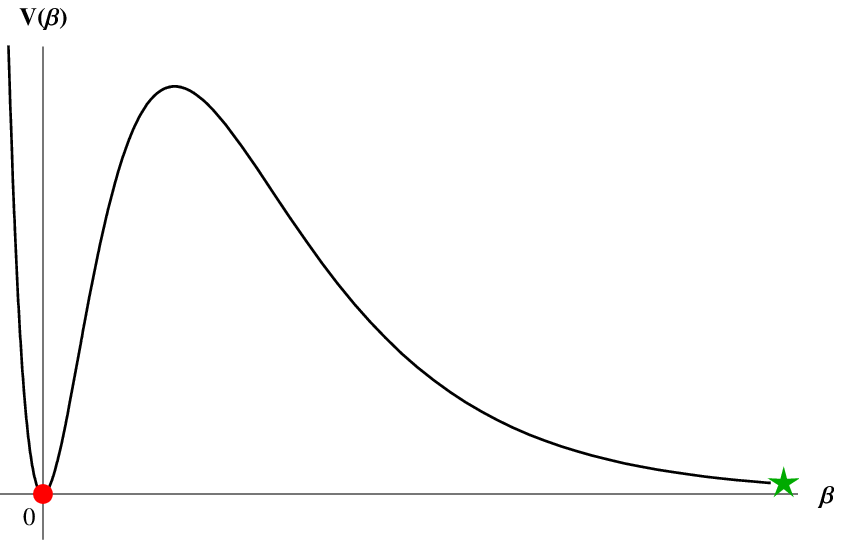}\nolinebreak
\qquad\epsfysize=4.75cm \epsfbox{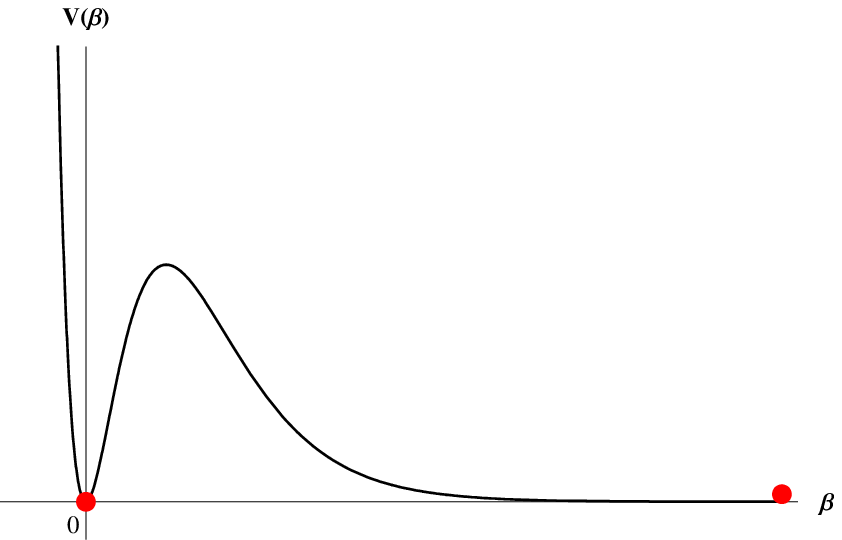}
\put(-350,-20){(a)}\put(-106,-20){(b)}
\vspace{+2cm}\\
\epsfysize=4.75cm \epsfbox{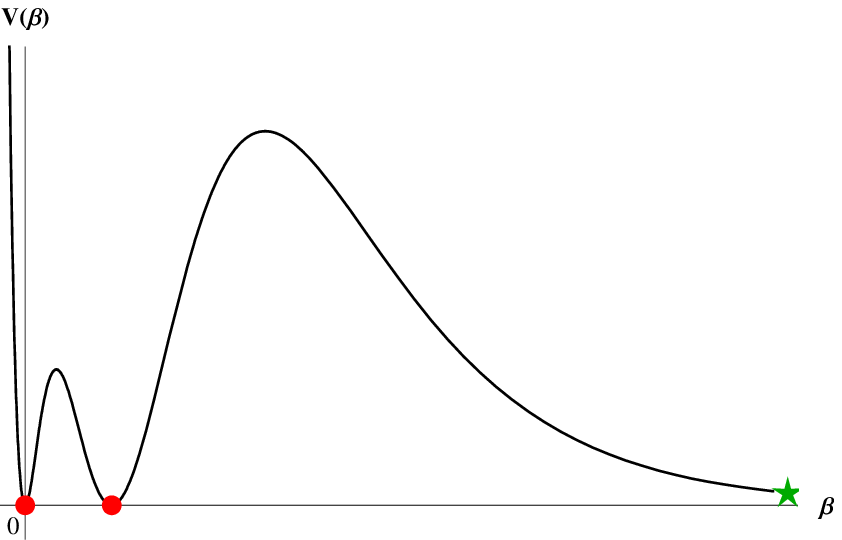}
\qquad\epsfysize=4.75cm\epsfbox{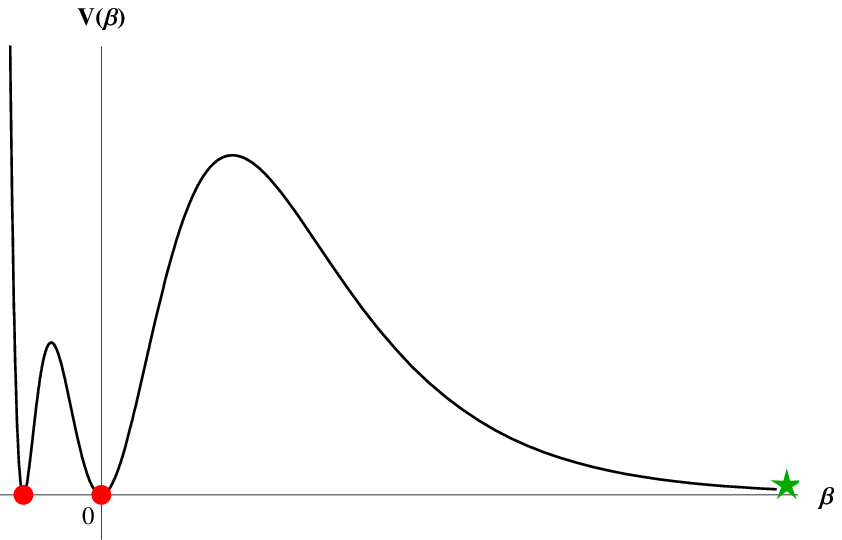}
\put(-350,-20){(c)}\put(-106,-20){(d)}
\vspace{+0.2cm}
\caption{Effective potential barriers for instanton tunneling with varying $\omega$}
\end{figure}

In Fig.2, the red bullets stand for the fixed points of the flow equation \eqn{odey}, whereas
the green star represents the additional fixed point that has been added at infinity by
formulating the same equation in terms of the variable $\beta$ given by \eqn{odey2}.
We note in this respect that when $x \rightarrow \infty$ the speed $dx/dt$ approaches
a limiting constant value $\mp 4\kappa^2 / 3 \kappa_{\rm w}^2 L^2$ as seen by equation
\eqn{odey}, whereas in terms of the variable $\beta$ we have $d\beta /dt = 0$ at
infinity. In either case, the potential looks flat at infinity.
Its value is simply lowered to $0$ in the $\beta$--parametrization, thus putting the
asymptotic configuration on par with the other degenerate vacua. Only when $\kappa_{\rm w}$
becomes infinitely large there is no height difference between the two
parametrizations as illustrated by the red bullet placed at infinity in Fig.2b for the
effective potential of pure Cotton flow.
The configuration associated to the point $\beta = +\infty$ is a completely squashed
sphere, which is nevertheless regular as it does not exhibit curvature singularities.
It arises in the correlated limit keeping the volume of $S^3$ fixed. The components of
its Ricci curvature tensor are $R_{11} = 1 = R_{22}$,  $R^{11} = 0 = R^{22}$ and
$R_{33} = 0 = R^{33}$, whereas all components of the Cotton tensor vanish. This degenerate
configuration also plays important role in general relativity as it describes a bolt in the
zero volume limit.

It is customary to think of instantons as describing the motion of a point particle
in the inverted potential $-V(\beta)$ that follows by analytic continuation of time.
At $t = -\infty$ the particle is located at the top of a hill (location of red bullet
in the inverted potential) and then starts rolling down until it comes to a complete stop at
the top of a nearby hill at $t = +\infty$. If the potential develops asymptotically a flat
direction, the point particle will also roll down the hill and reach the plateau after
infinitely long time. The limiting speed of the particle depends of the height difference
between the hill and the plateau, which it is zero for $V(\beta)$. The height difference
is not zero when the potential is written in terms of the variable $x$, in which case the
particle will reach infinity with finite speed. This solution can also be regarded as a
bona fide instanton solution with finite action. In all cases, an anti-instanton corresponds
to a particle that evolves reversely in time and has the right initial velocity to reach the
top of a nearby hill without overshooting or falling short of speed after sufficiently long time.
Based on this picture, it is then natural to count as instantons not only the flow lines of
the Ricci--Cotton flow that interpolate between vacua marked with red bullets but also
between configurations marked with a red bullet and a green star, as shown in Fig.2.

In our previous work \cite{BBLP} we have not included the latter solutions in the list of
instantons with $SU(2)$ isometries, because we were only considering flow lines that
interpolate between two different fixed points. But we may also add to list those configurations
that extend from the fully squashed sphere (green star) to a nearby fixed point (first red
bullet appearing to the left), since they also describe complete and regular metrics on
$\mathbb{R} \times S^3$ with $-\infty < t < +\infty$ (although the
extrinsic curvature of their slices does not vanish at both ends). The construction
of instanton solutions in the unimodular phase of Ho\v{r}ava--Lifshitz gravity is now
complete and can be used further for general purposes. Note, however, that these
additional instantons will not play substantial role in the present work, because the Ricci
scalar curvature of their spherical sections is everywhere non--negative. But they have
some common elements to the instantons of ordinary Einstein gravity,
which also admit an effective point particle description and reach asymptotically a flat
direction. These remarks will become more transparent in section 5.

\subsection{Sectional Ricci curvature of instantons}

For later use, let us consider in detail the variation of the Ricci scalar curvature $R$
of the three--dimensional spatial slices $\Sigma_3$ of the gravitational instanton solutions.
First, restricting attention to axially symmetric spheres with coefficients \eqn{bergersphe},
we find that
\be
R = {2 \over x^4 L^2} (4x^3 - 1) ~.
\ee
Fixing the volume of $S^3$, which is determined by $L$, we observe that $R$ as function
of the shape modulus $x$ attains its maximum value in the fully isotropic case $x=1$.
Squashing of the sphere corresponds to $x>1$, in which case $R(x)$ decreases monotonically
and it becomes asymptotically zero in the degenerate limit of a completely flat sphere.
Stretching of the sphere corresponds to $x<1$, in which case $R(x)$ also decreases monotonically
but without a lower bound. There is a critical value $x^3 = 1/4$ where the curvature
vanishes and then turns negative by stretching the sphere more and more. The
infinitely stretched sphere that arises as $x \rightarrow 0$ has infinitely negative
curvature and in this sense it is not a regular configuration, unlike the infinitely
squashed sphere that is regular. The range of allowed values of $R(x)$ is depicted
in Fig.3.

\begin{figure}[h]
\centering
\epsfxsize=9cm\epsfbox{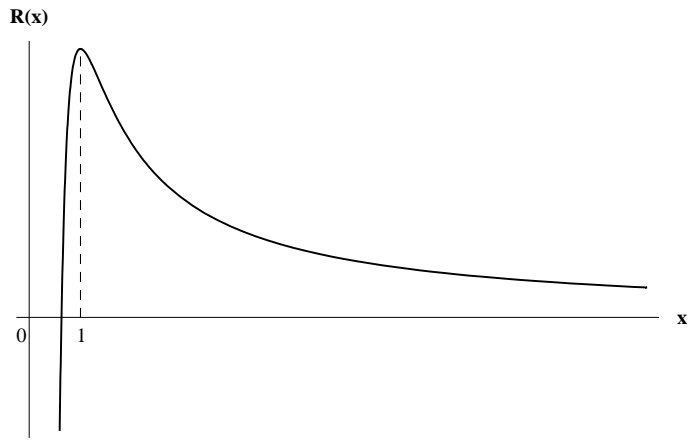}
\vspace{+0.2cm}
\caption{Dependence of curvature of Berger spheres upon squashing and stretching}
\end{figure}

Clearly, $R$ changes monotonically along the flow lines \eqn{odey}, since the instantons
have either $x\geq 1$ or $x\leq 1$. For this it suffices to recall that $x=1$ is a fixed point and,
as such, it can never arise as intermediate configuration. Also, $x(t)$ is a monotonic
function of time (the same also applies to $\beta (t)$) as can be easily seen by following
the point particle motion in the effective potential. The same conclusion is also
reached by computing $dR / dt$ for the special trajectories \eqn{odey}.

It is important to realize at this point that the chiral
instantons of $z=3$ Ho\v{r}ava--Lifshitz gravity, which arise for $\omega < 0$,
will develop spherical sections with negative curvature after some time, if the bound
$a^3 > 4$ is satisfied. On the other hand, chiral instantons with relatively
small coupling $\omega$ satisfying\footnote{When $a^3 < 4$, all $SU(2)$ instantons of the
theory have enhanced $SU(2) \times U(1)$ isometry. When $a^3 > 4$,
there are more general $SU(2)$ instantons that interpolate between the two fixed
points without necessarily having an extra $U(1)$ symmetry. These instantons have a modulus
which can be adjusted to yield the axially symmetric instanton solution we are
considering here. We also note that when $a^3 > 4$ there are two additional totally
anisotropic fixed points with $R=0$ which coexist with the axially symmetric point
\eqn{tsoutsouli}; then, there can also be $SU(2)$ instantons that interpolate between
any of these totally anisotropic fixed points and the axially symmetric or the totally
isotropic ones, but none of these has enhanced symmetry. More details on this and related
matters can be found in \cite{BBLP}. None of these more exotic possibilities will be further
discussed here as we are only considering instantons with
$SU(2) \times U(1)$ isometries.} $a^3 < 4$ have positive sectional Ricci
scalar curvature for all time $t$. Then, by Lichnerowicz's theorem \cite{lichn},
there can be no harmonic spinors on the spherical slices unless negative curvature
is allowed to occur by tuning the parameters of the theory. As will be seen later, this
suffices to show that the index of the four--dimensional fermion operator vanishes on all
backgrounds with positive sectional curvature. Ho\v{r}ava--Lifshitz gravity also
admits instantons with non-vanishing fermion index because there can be
solutions which are suspended from very stretched spheres with sufficiently
negative Ricci scalar curvature; in fact, it will turn out, that this can only happen
provided that $a^3 > 16$. Such extreme
geometric deformations of space are not allowed to occur in the instanton slices of
Einstein gravity, which always have positive sectional curvature.

Ho\v{r}ava--Lifshitz gravity with more general values of the superspace parameter
$\lambda$ also admit instantons. These solutions, however, do not permit
extreme geometric deformations of their spatial slices. In that case, the fixed points of the
unnormalized Ricci--Cotton flow have Ricci scalar curvature $R = 6 \Lambda_{\rm w}$ which is
determined by the size of the three--dimensional cosmological constant $\Lambda_{\rm w}$.
Then, all $SU(2) \times U(1)$ instantons that arise for $\Lambda_{\rm w} \geq 0$
(assuming that $\lambda < 1/3$ so that the superspace metric is positive definite)
have everywhere non--negative sectional Ricci scalar curvature. This can also be inferred
from Fig.3 (but see also \cite{BBLP}
which explains why we are not considering such instanton solutions when
$\Lambda_{\rm w} < 0$). The result should be contrasted to the normalized
Ricci--Cotton flow whose fixed points are three--geometries with different Ricci scalar
curvature, in general, and, as a result, the interpolating instantons have slices that
may extend from positive to negative values of $R$. This also explains why we are only
considering the case of normalized flow while searching for models of chiral symmetry
breaking induced by gravity.

The above considerations were based on the large isometry group of the specific
instanton solutions. In order to make more general statements for the variation of the
Ricci scalar curvature on $\Sigma_3$, it is necessary to
compute the time derivative of $R$ under the flow equation
\be
\partial_t g_{ij} = X_{ij} ~,
\ee
where $X_{ij}$ is the driving term of the deformations.
For example, for $X_{ij} = - 2 R_{ij}$ we have the Ricci flow, for
$X_{ij} = C_{ij}$ we have the pure Cotton flow, and for an appropriate combination
of Ricci and Cotton curvature terms we have the more general flow equation \eqn{basicoex}.
Since $\partial_t g^{ij} = - X^{ij}$, we find that the Christoffel symbols of the
metric evolve as
\be
\partial_t \Gamma_{ij}^k = {1 \over 2} g^{kl} \left(\nabla_i X_{jl} +
\nabla_j X_{il} - \nabla_l X_{ij} \right) ,
\ee
which in turn yield the time evolution of the Ricci scalar curvature,
\be
\partial_t R = \nabla_i \nabla_j X^{ij} - \Delta X - R_{ij} X^{ij} ~,
\label{maxapri}
\ee
where $X= g^{ij} X_{ij}$ is the trace of the driving curvature terms and
$\Delta = \nabla_i \nabla^i$ is the Laplace operator with respect to $g$.

Ideally, we would like to use \eqn{maxapri} to put bounds on $R$ and examine
its behavior against the possible formation of singularities. For the Ricci flow
equation we have $X_{ij} = - 2 R_{ij}$, and, therefore, $(\partial_t - \Delta)R =
2 R_{ij} R^{ij}$ which is manifestly positive definite. Then, in this case, if $R$ is positive
at some initial time, it will stay positive throughout the evolution. Also, by
appealing to the maximum principle, we have more generally that if $R \geq c$ at
$t =0$ (for any $c \in \mathbb{R}$), then the curvature will stay bounded for all subsequent
times, as $R \geq c/[1- (ct/3)]$. Also, for $c>0$, it turns out that the
Ricci flow becomes extinct since the three--geometry develops a curvature
singularity at finite time $T \leq 3/c$ (for this and related issues see, for instance,
the textbooks \cite{topping} and \cite{lni} and references therein). The normalized Ricci
flow is straightforward to study too. Similar estimates can not be presently made for
higher order flow equations (Cotton or Ricci--Cotton flows) because the maximum principle
does not apply to them directly. Thus, without imposing any isometries, it is not at all
clear what are the general criteria for having eternal solutions of the Ricci--Cotton flow
and how $R$ varies along the flow lines without hitting a singularity.

These last remarks show our current mathematical limitations to study the Ricci--Cotton
flow equation on general grounds. They also make us appreciate the existence of simple
instanton backgrounds for the propagation of fermions coupled to gravity.

\newpage

\section{The index of Lifshitz fermion operator}
\setcounter{equation}{0}

We are now in position to combine the results of the previous sections and consider
the index of the Dirac--Lifshitz operator associated to the integrated form on the
anomaly on gravitational instanton backgrounds of $z=3$ Ho\v{r}ava--Lifshitz theory.
The computation can be easily done based on the spectral flow in Euclidean time $t$
and the result turns out to be identical to the index of the Dirac operator on these
backgrounds, as suggested by the local form of the anomaly. We make some general remarks
first and then specialize the discussion to fermions propagating on backgrounds
$\mathbb{R} \times S^3$ with axially symmetric homogeneous geometries on $S^3$. This
restriction allows us to illustrate some special features of the non--relativistic
fermion--gravity models, based on explicit mathematical results for harmonic spinors
\cite{lichn} on Berger spheres \cite{hitchin, bar} that are neatly summarized in
Appendix D. It also allows to compare them directly to known results for relativistic
field theories. The implications will be discussed towards the end of this section
together with some more general remarks about non--relativistic theories of gravity
in the deep ultra--violet regime.

We consider the coupling of massless Lifshitz fermions \eqn{startipoint} to gravity given
by the following non--relativistic action, which is  written in ADM form,
\be
S = \int dt d^3x \sqrt{{\rm det} g} ~ \Big[i \bar{\psi} \left(\gamma^0 D_0 +
\gamma^i \mathbb{D}_i \right) \psi + {2 \over \kappa^2} K_{ij} {\cal G}^{ijkl} K_{kl}
- {\kappa^2 \over 8 {\rm det} g} {\delta W \over \delta g_{ij}} {\cal G}_{ijkl}
{\delta W \over \delta g_{kl}} \Big] ,
\label{couplees}
\ee
using the action of three-dimensional topologically massive gravity on $\Sigma_3$ with
metric $g_{ij}$ as superpotential functional $W$ and the extrinsic curvature $K_{ij}$ of
the spatial slices $\Sigma_3$ in space--time. As such, it exhibits anisotropic scaling
$z=3$ in both fermion and boson sectors. For the extended theory \eqn{couplees}, a
systematic coupling constant expansion may be given in which the Fermi fields enter only
at the first correction to the lowest approximation. Thus, according to the standard lore
(see, for instance, \cite{rama1}, as well as \cite{rama2} section II.G), the backreaction
of the Fermi fields can be consistently ignored in the Bose field equations and solve
the Fermi equation in the background of an external Bose configuration. This is the
approximation that will be adopted through out this section while studying fermions in
gravitational instanton backgrounds.

The chiral nature of instantons in Ho\v{r}ava--Lifshitz gravity should be contrasted
to the behavior of gravitational instanton solutions in Euclidean Einstein
theory under parity. In ordinary gravity, which preserves parity, a change of
orientation in space simply exchanges instantons with anti--instantons, and, as a result,
the background geometry shows no chirality. Based on this difference, it
is then natural to inquire whether there is violation of chiral symmetry for Lifshitz
fermions in the background of such chiral gravitational instanton solutions in sharp
contrast with the negative result obtained for Dirac fermions in the background of
Einstein gravitational instantons (with boundaries) \cite{gibbons, eguchi2, pope, eguchi3}.
This question is addressed in detail in this section and the answer depends crucially
on the geometry of the three--dimensional spherical slices that the instanton transverses
as $t$ varies from $-\infty$ to $+\infty$. For Berger spheres, in particular, for which
explicit results can be obtained relatively easy, the index of the four--dimensional
fermion operator is non--zero provided that the instanton slices can become sufficiently
stretched. In fact, as it turns out, there is a critical value in the couplings of the
theory beyond which the background geometry becomes sufficiently chiral to allow level
crossing in the spectrum and lead to chiral symmetry breaking by the axial anomaly.

\subsection{Atiyah--Patodi--Singer index theorem}

Let us consider the fermion operator ${\cal D} = i \gamma^{\mu} {\cal D}_{\mu}$ for either
Dirac fields (${\cal D}_{\mu} = D_{\mu}$) or Lifshitz fields (${\cal D}_{\mu} = \mathbb{D}_{\mu}$),
which are treated here together in a unified way. They act on four--component spinors $\Psi (t, x)$
which we split into two subspaces of chirality $\pm 1$ with respect to the operator $\gamma_5$,
i.e., $\gamma_5 \Psi_{\pm} = \pm \Psi_{\pm}$. Introducing the pair of corresponding spin bundles
$\Delta_{\pm}$ over the space--time $M_4$ and assuming that the first and second Stiefel--Whitney
classes vanish so that the manifold is orientable and it admits a spin structure, we define
the spin complex by
\ba
& & {\cal D}: ~~~~~~ C^{\infty} (\Delta_+) \rightarrow C^{\infty} (\Delta_-) ~, \nonumber\\
& & {\cal D}^{\dagger}: ~~~~~ C^{\infty} (\Delta_-) \rightarrow C^{\infty} (\Delta_+) ~.
\ea
Then, the index of the spin complex (also called index of the operator ${\cal D}$) is
\be
{\rm Ind}({\cal D}) = {\rm dim ~ Ker} ~ {\cal D} - {\rm dim ~ Ker} ~ {\cal D}^{\dagger}
= n_+ - n_- ~,
\ee
where $n_{\pm}$ is the number of normalizable solutions of the equation
$i \gamma^{\mu} {\cal D}_{\mu} \Psi = 0$ with chirality $\pm 1$, respectively.
These definitions are analogous to the familiar spin complex associated to the ordinary
Dirac operator. In either case, the operator ${\cal D}^{\dagger} {\cal D}$ is elliptic
on spaces with Euclidean signature which will be considered throughout this section (though
the order is different for the Dirac and Lifshitz operators), and, thus, the general mathematical
framework for computing the index by analytic means is applicable.

The Atiyah--Singer index theorem on compact manifolds without boundary \cite{AS} states that
$n_+ - n_-$ equals the integral of the A--roof genus on $M_4$. In physical
terms, this is provided by the local form of the axial anomaly, which is the same for both
Dirac and Lifshitz fermions, and the index theorem states
\be
{\rm Ind}({\cal D}) = - {1 \over 8 \pi^2} \int_{M_4} {\rm Tr} (F \wedge F)
- {1 \over 192 \pi^2} \int_{M_4} {\rm Tr} (R \wedge R)
\ee
when both the gauge and metric field contributions are combined together (it is called the
twisted spin complex); here we are only considering the simplest case of a $U(1)$ gauge field,
but, apparently, the result easily generalizes to non--Abelian gauge fields.
Thus, when $\partial M_4$ is empty, the integrated form of the anomaly yields the relation
${\rm Ind}(D) = {\rm Ind}(\mathbb{D})$. This is an interesting result in its own right, but it
is not directly applicable to the cases we have in mind. First of all, the Lifshitz models
of gauge theories that are available in the literature\footnote{The Lifshitz gauge field theory
(Abelian or non--Abelian) is naturally defined by the following action, which is written here
in flat space--time in $3+1$ dimensions, \cite{horava4},
\be
S = \int dt ~ d^3x ~ \Big[ {\rm Tr} (E_i E^i) - {1 \over g^2} {\rm Tr} \left((D_i F^{ik})
(D_j {F^j}_k) \right) \Big] ~, \nonumber
\ee
where $E_i = \partial_t A_i$ and $F_{ij} = \partial_i A_j - \partial_j A_i - i[A_i, ~ A_j]$
provide the electric and the magnetic fields, respectively, in the axial gauge $A_0 (t, x) = 0$.
$S$ satisfies the detailed balance condition using as superpotential functional $W$
the action of three--dimensional Yang--Mills theory with coupling constant $g$. Augmenting $W$
by the Chern--Simons term for the gauge fields $A_i$, so that the three-dimensional action
is that of topologically massive gauge theory \cite{DJT}, simply shifts $D_i F^{ik}$ by
a term proportional to $\epsilon^{ijk} F_{ij}$ in the potential term of $S$. In either case,
the anisotropic scaling parameter can not be higher than $2$ contrary to the gravitational
case for which the addition of a Chern--Simons term increases $z$ from $2$ to $3$.} exhibit
anisotropic scaling $z=2$ and not $z=3$. But even if this mismatch is not a major concern in the
coupled spinor--vector Lifshitz theory, the problem is that the space--time has boundaries as the
theory is naturally
defined on $M_4 \simeq \mathbb{R} \times \Sigma_3$. Furthermore, we have no instanton solutions
in our disposal for Lifshitz gauge theories\footnote{The instantons equations for Lifshitz gauge
theories can be derived by completing the square of the Euclidean action $S$, as we did for the
gravitational case. They correspond to eternal solutions of the first order equations in time
\be
\partial_t A_i (t, x) = \pm \left(D_j {F^j}_i + \mu ~ \epsilon_{ijk} F^{jk} \right) \nonumber
\ee
taking also into account the presence of a Chern--Simons term in $W$ with coupling constant
$\mu$. This equation is a variant of the gauge field flow in three dimensions, which, however,
has not been studied in detail so far. We hope to return to it elsewhere with explicit results.}
in order to investigate the relation between the index of the fermion operator and the instanton
number of the background configurations, as in relativistic Yang--Mills theories. For all
these reasons we will not consider the gauge field sector in the remaining part of this paper.
As for the gravitational sector, we need to work with the generalization of the Atiyah--Singer
index theorem to manifolds with boundaries in order to account correctly for the end--points of
our instanton solutions.

The index of the fermion operator ${\cal D}$ in the background of a four--dimensional metric
on a manifold $M_4$ with boundary $\partial M_4$ takes the following form, according to the
Atiyah--Patodi--Singer index theorem \cite{APS} (but see also \cite{gilka, melrose}),
\be
{\rm Ind}({\cal D}) = - {1 \over 192 \pi^2} \int_{M_4} {\rm Tr} (R \wedge R) +
{1 \over 192 \pi^2} \int_{\partial M_{4}} {\rm Tr} (\theta \wedge R) - {1 \over 2}
\eta_{\cal D} (\partial M_4) ~.
\label{rotames}
\ee
In writing this formula it is implicitly assume that we are computing the $L^2$--index
of the fermion operator acting on square integrable spinors on the manifold $M_4$ with
respect to the inner product
\be
(\Psi , ~ \Psi^{\prime}) = \int_{M_4} d^4 x \sqrt{{\rm det} G} ~ \Psi^{\dagger} (x)
\Psi^{\prime} (x) ~.
\ee
For manifolds of the form $\mathbb{R} \times \Sigma_3$, in particular, with compact slices
without boundaries, this means that we are considering spinors with asymptotic behavior
\ba
& & \Psi \sim e^{- k_+ t} ~~~~~~ {\rm as} ~~ t \rightarrow + \infty ~~~~ {\rm with} ~~
k_+ > 0 ~,
\label{apsb1} \\
& & \Psi \sim e^{- k_- t} ~~~~~~ {\rm as} ~~ t \rightarrow - \infty ~~~~ {\rm with} ~~
k_- < 0 ~,
\label{apsb2}
\ea
which is characteristic of bound states. These are often called Atiyah--Patodi--Singer
boundary conditions, which are assumed from now on. When the slices $\Sigma_3$ are
compactified due to the physical boundary conditions imposed on the fields at infinity, the
index theorem remains valid as it stands. However, for non--compact spaces that account for
permanent boundary effects, one has to device a variant of the index theorem that is
applicable to open spaces \cite{callias}, as in the case of monopole configurations. None
of these generalizations will be further pursued in this paper as we restrict attention to
$\Sigma_3 \simeq S^3$ exclusively.

The bulk contribution to the index \eqn{rotames} is given by the integral over $M_4$ of the
characteristic class ${\rm Tr}(R \wedge R)$, which is appropriately normalized, as before.
In the presence of boundaries one adds two surface terms.
The first surface term is the integral over $\partial M_4$ of the secondary characteristic class
${\rm Tr}(\theta \wedge R)$ that is computable from the second fundamental form $\theta$
determined by the choice of the normal to the boundary, and, as such, it accounts for the
difference of the metric from a cross--product form at the boundary. Introducing a cross--product
metric $G_0$ on $M_4$, which agrees with the original metric $G$ on $\partial M_4$ and serves
as reference frame in the calculations, we then define the second fundamental form as the difference
of the corresponding connection one--forms $\theta = \omega - \omega^0$. By definition $\omega^0$
has only tangential components to $\partial M_4$. Furthermore, we have ${\rm Tr}(R \wedge R) =
dQ (\omega, \omega^0)$, where
\be
Q (\omega, \omega^0) = {\rm Tr} \left(2 \theta \wedge R + {2 \over 3} \theta \wedge \theta
\wedge \theta - 2 \theta \wedge \omega \wedge \theta - \theta \wedge d \theta \right)
\ee
is the Chern--Simons form \cite{simons} and ${\rm Tr} (R_0 \wedge R_0) = 0$ for the cross--product
reference metric. A simple calculation yields
\be
Q (\omega, \omega^0)|_{\rm boundary} = {\rm Tr} (\theta \wedge R) ~.
\ee
Then, the surface correction term has to be subtracted from the bulk term with the same
normalization in order to have a well--defined quantity when the metric is not of a
cross--product form. Finally, the last surface term is the so called $\eta$--invariant
determined by the eigen--values of the tangential part of the operator restricted to the
boundary. It is a non--local term that accounts for the asymmetry between the positive and
negative modes on $\partial M_4$ and it is commonly defined using the spectral Riemann zeta
function. It is also corrected by the dimension of the zero modes if they are present at the
boundary. In all examples that will be considered later the $\eta$--invariant will be computed
explicitly by algebraic methods.

When the index is computed on a given metric background, the bulk term and the first
boundary correction are identical for the Dirac and Lifshitz operators on $M_4$
because the normalization constants are the same. We claim that a much stronger statement
is actually true, namely that the indices of the two operators are equal
\be
{\rm Ind}(D) = {\rm Ind}(\mathbb{D})
\ee
on spaces with boundaries. It means that the $\eta$--invariants of the tangential part of
the Dirac and Lifshitz operators are the same even though their eigen--values differ. We
will prove this relation by direct computation only for space--times whose slices have
homogeneous and axially symmetric metrics. This is sufficient to
demonstrate that the index of the Dirac operator can still be used to determine whether
there is chiral symmetry breaking on a given background with $SU(2) \times U(1)$ isometry
in the non--relativistic theory of Ho\v{r}ava--Lifshitz gravity. Note, however, that this
is also sufficient to establish the general validity of the relation $\eta_{\rm D} =
\eta_{\mathbb{D}}$, beyond mini--superspace models, without putting extra effort.
According to the general theory (see, in particular, \cite{APS}), the $\eta$--invariant
should differ by a constant from the Chern--Simons invariant (appropriately
normalized). Since the local form of the anomaly yields the same Chern--Simons invariant in
both cases, it suffices to show that the constants are also the same. This can be easily
established using the standard metric on $S^3$, as will be seen later, and, therefore, the
$\eta$--invariants of the Dirac and Lifshitz operators ought to be equal, as they are, for
all metrics\footnote{Care should be taken when the operators at the boundary admit zero
modes, in which case one has to add the necessary harmonic corrections to $\eta$.}. Verifying
this general result for all Berger spheres seems superfluous, but we will go through it anyway.
It serves as an independent consistency check based on spectral methods that the local
form of the gravitational anomaly is the same for Dirac and Lifshitz
fermions\footnote{We intend to revisit this problem elsewhere by different methods based
on the integral representation of the $\eta$--invariant of an operator ${\cal D}$,
\be
\eta_{\cal D} = {1 \over \Gamma ((s+1) / 2)} \int_0^{\infty} dy ~ y^{(s-1)/2} ~ {\rm tr}
\left({\cal D} ~ e^{-y {\cal D}^2} \right) |_{s=0} ~. \nonumber
\ee}.

The computation of the index can be alternatively described as spectral flow of the
three--dimensional Dirac operator on the slices $\Sigma_3$. This approach is very important
for understanding intuitively the meaning of the index. It also puts in better context the
results that will be given later. Let us assume first, for simplicity, that the metric on
$M_4$ has the cross--product form
\be
ds^2 = N(t_0) dt^2 + g_{ij} (t_0 , x) dx^i dx^j
\ee
around $t=t_0$ (we may also set $N=1$ without loss of generality). Then, the fermion operator
takes the block-diagonal form (see Appendix A for the fermion operator in Euclidean space, which
can be either Dirac or Lifshitz)
\be
\gamma^{\mu} {\cal D}_{\mu} = \left(\begin{array}{ccc}
0          &  & \partial / \partial t -i \sigma_I {E_I}^i {\cal D}_i\\
           &  &            \\
\partial / \partial t + i \sigma_I {E_I}^i {\cal D}_i &  & 0
\end{array} \right) ,
\ee
where $\sigma_I$ are the Pauli matrices and ${E_I}^i$ are the inverse dreibeins associated
to the metric $g_{ij}$ (summation over the space indices $i$ and the tangent space indices $I$
are implicitly assumed). The $2 \times 2$ blocks $\partial / \partial t \pm i \sigma_I {E_I}^i
{\cal D}_i$ are operators mapping the two--component Weyl spinors $\Psi_{\pm}$ to $\Psi_{\mp}$
and they are mutually related by conjugation as $(\partial / \partial t + i \sigma_I {E_I}^i
{\cal D}_i)^{\dagger} = -(\partial / \partial t - i \sigma_I {E_I}^i {\cal D}_i)$.
Equation $i\gamma^{\mu} {\cal D}_{\mu} \Psi (t, x) = 0$ acting on four--component
spinors reduces to the following system of Weyl equations on $M_4$,
\be
\left({\partial \over \partial t} \pm i \sigma_I {E_I}^i {\cal D}_i\right) \Psi_{\pm} (t, x) =
0 ~.
\ee
The cross--product form of the metric on $M_4$ allows to introduce the separation of variables
$\Psi_{\pm} (t, x) = {\rm exp} (-Et) \Psi_{\pm} (x)$ and obtain
\be
\pm i \sigma_I {E_I}^i {\cal D}_i  \Psi_{\pm} (x) = E \Psi_{\pm} (x)
\ee
in the vicinity of $t = t_0$. Thus, the solutions of the four--dimensional fermion equation
$i\gamma^{\mu} {\cal D}_{\mu} \Psi (t, x) = 0$ reduce to the eigen--value problem of the
three--dimensional fermion operators $\pm i \sigma_I {E_I}^i {\cal D}_i$, which is
analogous to the familiar reduction of the time--dependent Schr\"odinger equation to the time
independent one (though, here, the time is Euclidean).

In reality, however, $M_4$ may not have a cross--product metric for all $t$ and the
three--dimensional operators ${\cal D}_i$ may also depend upon $t$. To remedy the situation we
make the simplifying assumption that the time evolution is very slow so that in the
{\em adiabatic} approximation we may write
\be
\Psi_{\pm} (t, x) = {\rm exp} \left(- \int^t dt^{\prime} E(t^{\prime}) \right) \Psi_{\pm} (x)
\ee
with
\be
\pm i \sigma_I {E_I}^i {\cal D}_i (t) \Psi_{\pm} (x) = E (t) \Psi_{\pm} (x)
\ee
replacing the naive relations given before. In this more realistic (but not yet ultimate)
picture, the solutions of the four--dimensional fermion equation still reduce to the
eigenvalue problem of the corresponding three--dimensional fermion operators but with
time dependent eigenvalues $E(t)$ that are slowly varying with time. Normalizable solutions
$\Psi_{\pm} (t, x)$ on $\mathbb{R} \times S^3$ correspond to those cases that
$E(t)$ is positive for $t \rightarrow + \infty$ and negative for $t \rightarrow - \infty$
(these are the Atiyah--Patodi--Singer boundary conditions \eqn{apsb1} and \eqn{apsb2} with
$k_{\pm} = E(t= \pm \infty)$). This, in turn, implies that the three--dimensional fermion
operators $\pm i \sigma_I {E_I}^i {\cal D}_i$ should admit zero modes for some intermediate
values of $t$ in order for the spectral flow $E(t)$ to be able to produce the necessary level
crossing from negative to positive eigen--values. Otherwise, there will be no normalizable
solutions of the four--dimensional fermion operator $i \gamma^{\mu} {\cal D}_{\mu}$ and its
index will be zero. In either case, the eigen--value problem
$i \sigma_I {E_I}^i {\cal D}_i \Psi_{\pm} (x) = \pm E (t) \Psi_{\pm} (x)$ is central to
the subject and level crossing from negative (resp. positive) eigen--values
to positive (resp. negative) eigen--values is the dominant effect as $t$ varies continuously
from $-\infty$ to $+\infty$. Every normalizable solution contributes one unit to the
index (either to $n_+$ or $n_-$ depending on the chirality), as required on general grounds.
Letting $t \rightarrow -t$ in the applications, so that instantons and anti--instantons are
interchanged, simply amounts to replacing $n_{\pm}$ by $n_{\mp}$.

Finally, it can be shown that the index does not depend on the adiabatic approximation
used to establish its relation to the difference of the normalizable zero modes of the
the operators $A = \partial / \partial t + i \sigma_I {E_I}^i
{\cal D}_i$ and  $A^{\dagger} = - \partial / \partial t + i \sigma_I {E_I}^i {\cal D}_i$.
It is obvious that the zero modes of $A$ are also zero modes of $A^{\dagger} A$, and,
likewise, the zero modes of $A^{\dagger}$ are also zero modes of $A A^{\dagger}$. These
two sets are in general different, whereas the non--zero modes of $A^{\dagger} A$ and
$A A^{\dagger}$ coincide (for this note that if $\Psi$ is an eigen--state of
$A A^{\dagger}$ with eigen--value $E \neq 0$, the state $E^{-1} A^{\dagger} \Psi$ will
be eigen--state of $A^{\dagger} A$ with the same eigen--value $E$). Then, the difference
of the number of zero modes can be equivalently expressed by
\be
n_+ - n_- = {\rm dim ~ Ker} ~ A - {\rm dim ~ Ker} ~ A^{\dagger} =
{\rm dim ~ Ker} ~ (A^{\dagger} A) - {\rm dim ~ Ker} ~ (A A^{\dagger}) ~.
\ee
The operators $A^{\dagger} A$ and $A A^{\dagger}$ are better behaved than $A$ and
$A^{\dagger}$, since they elliptic for either Dirac or Lifshitz cases, and they are more
suited for the argument we are about to make. Smooth variations
of the background metric that keep the boundaries invariant deform the entire spectrum
of the operators used in the adiabatic approximation, but they can not affect the index.
Indeed, if a non--zero eigen--value of $A^{\dagger} A$ deforms to zero, the same eigen--value
of $AA^{\dagger}$ will also be forced to become zero. Then, both $n_{\pm}$ will increase
by $1$ and their difference with stay the same. By the same token, any non--zero mode of
$A^{\dagger} A$ that crosses from negative (resp. positive) to positive (resp. negative)
values will be accompanied by the same level crossing of the operator $A A^{\dagger}$.
Hence, the result of the adiabatic approximation extends without modification to the index
of the fermion operator ${\cal D}$ in the background of instanton metrics, which interpolate
continuously between the two end--points of time, irrespective of the intermediate details.

For a more detailed account of the index theorem and related subjects we refer the reader
to the original papers \cite{AS, APS} and also to the mathematics textbooks
\cite{gilka, ezra, melrose} and the physics report \cite{eguchi3}. An excellent account of
the physical applications of the index theorem to fermions in topologically non--trivial
background fields is provided in the modern textbook \cite{rubakov} (see, in particular,
part III) with many references to the original works. It is beyond our purpose to list separately
all important contributions made in the literature in this area of research over the years.

\subsection{Lifshitz operator on Berger spheres}

We will determine the spectrum of the three--dimensional Lifshitz operator and compute
its $\eta$--invariant for the class of axially symmetric Bianchi IX metrics. Our presentation
is parallel to that for the Dirac operator. Thus, we consider the third order operator
\be
i \gamma^i \mathbb{D}_i = {1 \over 2} \Big[(i \gamma^i D_i) (-D^2) + (-D^2)(i \gamma^i D_i)
\Big]
\label{routhina}
\ee
with $D^2 = D_i D^i$, which is the tangential part of $i \gamma^{\mu} \mathbb{D}_{\mu}$
restricted to any given slice of space--time (eventually it will be taken at the boundary),
and examine the eigen--value problem
\be
i \gamma^i \mathbb{D}_i \Psi (x) \equiv i \sigma_I {E_I}^i \mathbb{D}_i \Psi (x) = Z ~ \Psi (x)
\ee
for two--component spinors $\Psi (x)$. This equation admits zero mode solutions (the analogue
of harmonic spinors for the three--dimensional Dirac operator) when the geometry has sufficient
negative curvature. These modes are very important in our discussion as they provide the critical
points for level crossing under geometric flows, which, here, will have the interpretation of
spectral flow. Note in this respect that the number of zero modes is not a topological invariant
as their existence solely depends on the geometry of space.

Using the three--dimensional analogue of Lichnerowicz's formula \eqn{finalrel} for the square of
the Dirac operator, we may substitute
\be
-D^2 = (i \gamma^i D_i)^2 - {1 \over 4} R
\ee
into the Lifshitz operator \eqn{routhina}, which now takes the following form
\be
i \gamma^i \mathbb{D}_i = (i \gamma^i D_i)^3 - {1 \over 4} R (i \gamma^i D_i) -
{1 \over 8} i \gamma^i (\nabla_i R) ~.
\ee
This equation is valid in general for all metrics on a given manifold $\Sigma_3$ and forms
the basis of our construction. Note that the
Lifshitz operator does not necessarily commute with the Dirac operator since there is a
geometric obstruction
\be
[i \gamma^i D_i , ~ i \gamma^j \mathbb{D}_j] = {1 \over 4} (\nabla^i R) D_i +
{1 \over 8} \nabla^2 R
\label{obstraq}
\ee
that involves the derivatives of the Ricci scalar curvature $R$ on $\Sigma_3$. Here,
$\nabla^2$ is the three--dimensional Laplacian acting on scalars as opposed to $D^2$ that acts
on two--component spinors.

When the obstruction term \eqn{obstraq} vanishes
the eigen--value problem of the Lifshitz operator reduces to that of the Dirac operator.
A class of metrics that realize this possibility is provided by homogeneous model geometries
because $R$ becomes purely algebraic: in these cases the curvature depends on the metric
coefficients, since the geometry is not isotropic in general, but it is independent of the
coordinates on $\Sigma_3$ (it is the same at all points) as simple consequence of
homogeneity. Then, the three--dimensional Lifshitz operator simplifies to
\be
i \gamma^i \mathbb{D}_i = (i \gamma^i D_i)^3 - {1 \over 4} R (i \gamma^i D_i)
\ee
and the problem of its diagonalization reduces to that of the three--dimensional Dirac
operator. In fact, any class of metrics arising in Bianchi's classification of homogeneous
three--geometries will serve this purpose, but we will not delve into the general discussion.
It is also appropriate to note at this point that the use of homogeneous geometries also
removes the factor ambiguity in the definition of the Lifshitz fermion operator, which is
now uniquely defined (recall that different factor orderings differ by total derivative terms
of the curvature tensor which vanish for all homogeneous metrics).

Here, we restrict attention to Bianchi IX metrics on $S^3$ with isometry group $SU(2)$.
Still the spectrum of the fermion operator is not easy to determine in closed form unless
there is an additional symmetry that makes the problem tractable. We consider axially symmetric
metrics with isometry $SU(2) \times U(1)$, as in section 3, which we write in the form
\be
ds^2 = \gamma \Big[(\sigma^1)^2 + (\sigma^2)^2 + \delta^2 (\sigma^3)^2 \Big]
\ee
setting $\gamma_1 = \gamma_2 \equiv \gamma$ and $\gamma_3 = \delta^2 \gamma$. The Dirac and
Lifshitz operators scale uniformly by $\sqrt{\gamma}$ and $(\sqrt{\gamma})^3$, respectively,
under conformal transformations of the metric. Thus, we may set without loss of generality
$\gamma = 1$. The zero modes, in particular, are independent of $\gamma$ and they only
depend on $\delta$. The entire spectrum can also be determined as function of $\delta$. The
parameter $\delta$ varies from $0$ to $\infty$, depending on the degree of spatial anisotropy,
and this induces a spectral flow that will also be studied in the following.

Based on these observations, we immediately see that both the Dirac and Lifshitz operators
have the same eigen--states $\Psi$ on Berger spheres, whereas the eigen--values $Z$ are
simply expressed as
\be
Z= \zeta \Big[\zeta^2 + {1 \over 8} (\delta^2 -4) \Big]
\label{spelifz}
\ee
in terms of the eigen--values $\zeta$ of the Dirac operator.
Recall at this point that the spectrum of the Dirac operator on Berger spheres is given by
\cite{hitchin} (but see also Appendix D for the notation and essential details)
\be
\zeta_{\pm} = {\delta \over 4} \pm {1 \over 2 \delta} \sqrt{4 \delta^2 pq + (p-q)^2} ~;
~~~~~~ (p, q) \in \mathbb{N} ~.
\ee
The multiplicity of these eigen--values is $p+q$ for each
pair $(p, q)$. The positive integers $p$ and $q$ are not ordered which means that
the conjugate pair $(q, p)$ also yield the same eigen--values $\zeta_{\pm}$ with the same
multiplicity. Of course, $p$ and $q$ can also be equal to each other. Also, there are
additional eigen--values arising for $q = 0$,
\be
\zeta_0 = {\delta \over 4} + {p \over 2 \delta} ~; ~~~~~~ p \in \mathbb{N}
\ee
with multiplicity $2p$. Equally well we could have set $p=0$ and let $q \in \mathbb{N}$,
but there is no solution when both integers are equal to zero ($\zeta = \delta /4$ is not
allowed to occur). These formulae provide the
complete spectrum of the Lifshitz operator on Berger spheres together with the multiplicities.
We will denote the eigen--values by $Z_{\pm}$ and $Z_0$ depending on the choice $\zeta_{\pm}$
and $\zeta_0$ made, respectively, in \eqn{spelifz}.

The general results simplify in three special cases that are worth commenting as for
the Dirac case. First, we consider $\delta =1$ that corresponds to the homogeneous
and isotropic constant curvature metric on $S^3$. The spectrum of the Lifshitz operator
becomes
\be
\delta = 1: ~~~~ Z = \pm {1 \over 64} (2n+1) (4n^2 + 4n -5) ~, ~~~~ n \in \mathbb{N} ~~~
{\rm with ~ multiplicity} ~~ n(n+1)
\ee
and it is equally distributed to positive and negative values. Next, we consider $\delta = 0$
that corresponds to a fully squashed three--sphere along its third principal axis. In this
case, the eigen--values $Z_0$ become infinite. The eigen--values $Z_{\pm}$ also tend to
$\pm \infty$, respectively, as long as $p \neq q$, whereas for $p=q \equiv n$  finite
eigen--values arise
\be
\delta = 0: ~~~~ Z = \pm {1 \over 2} n(2n^2 - 1) ~, ~~~~ n \in \mathbb{N} ~~~
{\rm with ~ multiplicity} ~~ 2n
\label{stalousr}
\ee
and they coincide with the spectrum of the Lifshitz operator on the round $S^2$. Finally, we
consider the limiting case $\delta \rightarrow \infty$ that corresponds to a fully stretched
three--sphere along its third principal axis. All eigen--values become infinite and exhibit
the following universal behavior,
\be
\delta \rightarrow \infty: ~~~~ Z \simeq {3 \delta^3 \over 64} ~~~~~
{\rm with ~ infinite ~ multiplicity} ~.
\ee
The spectra in these cases should be compared with the corresponding expressions \eqn{cameen1},
\eqn{cameen2} and \eqn{asymptf} found in Appendix D for the Dirac operator.

Zero modes of the Lifshitz operator arise for those special values of $\delta$ that $Z=0$ and,
as such, they are invariant under rescaling of the metric.
Note at this end that the eigen--values $Z_0$ and $Z_+$ are positive definite for all $\delta$,
whereas $Z_-$ are negative for $\delta < 4$. This follows from the fact that
$\zeta^2 + (\delta^2 -4)/8$ in \eqn{spelifz} stays positive definite\footnote{The validity of this
statement, which is very important in the following, can be easily shown case
by case. First, we consider $\zeta_0$ and find
\be
\zeta_0^2 + {1 \over 8} (\delta^2 - 4) = {1 \over 16 \delta^2} \left(3 \delta^4 + 4 (p-2) \delta^2
+ 4p^2 \right) , \nonumber\\
\ee
which is manifestly positive definite for all $p \geq 2$. For $p=1$ the terms in the parenthesis
take the form $3 \delta^4 - 4 \delta^2 + 4 = 2 \delta^4 + (\delta^2 -2)^2$ which is also positive
definite for all $\delta$. Next, we consider $\zeta_{\pm}$ and note that
$\zeta_{\pm}^2 + (\delta^2 - 4)/8 > 0$ is equivalent to the inequality
\be
9\delta^8 + 16 (2pq -3) \delta^6 + 8 \left((p-q)^2 + 8 (2pq -1)^2 \right) \delta^4 +
64 (p-q)^2 (2pq-1) \delta^2 + 16 (p-q)^4 > 0 ~. \nonumber\\
\ee
This is manifestly true for all integer values of $p$ and $q$ apart from $p=q=1$ that should be
examined separately. In the latter case, $\zeta_{\pm} = \delta /4 \pm 1$ and, thus,
$\zeta_{\pm}^2 + (\delta^2 - 4)/8 = (3 \delta^2 \pm 8\delta +8)/16$. Since the discriminant is
negative, the sign is positive for all values of $\delta$, which completes the proof.} for all
values $\zeta_0$ and $\zeta_{\pm}$ irrespective of $\delta$. Then, the sign of the eigen--values
$Z$ is completely determined by the sign of $\zeta$. Zero modes arise when $\zeta_- = 0$, i.e.,
when there are positive integer solutions $(p, q)$ to the equation
\be
\delta^2 = 2 \sqrt{4 pq \delta^2 + (p-q)^2} ~,
\ee
as for the Dirac operator (see equation \eqn{spevalh}). Thus, the necessary condition for the
existence of Lifshitz zero modes is provided by the bound $\delta \geq 4$ on the anisotropy
parameter of Berger spheres. The first zero modes arise at $\delta = 4$ with multiplicity $2$.
When the Berger sphere is stretched more and more, zero modes occur at other special values of
$\delta$ with the same multiplicities as for the zero modes of the Dirac operator described in
Appendix D.2 (where we refer the reader for details and examples).

The presence of zero modes implies level crossing under spectral flow when $\delta$ varies.
As soon as an eigen--value $Z_-$ crosses zero, it will stay positive for all higher values of
$\delta$ and never cross back to negative values. Fig.4 provides the spectral flow of $Z_-$
as functions of $\delta$ for the lowest values of $p$ and $q$. We plot the
results for $(p, q) = (1, 1)$, $(1, 2)$, $(1,3)$, $(2, 2)$, $(1, 4)$ and $(2, 3)$ for which
level crossing occurs at $\delta = 4$, $5.67$, $6.95$, $8$, $8.03$ and $9.80$, respectively.
The eigen--values $Z_-$ with $p=q$ pass from the values $-1/2, -7, \cdots$ that correspond
to the negative modes of the Lifshitz operator on $S^2$, whereas all other ones tend to
$-\infty$ as $\delta \rightarrow 0$.
Direct comparison can be made with the corresponding eigen--values $\zeta_-$ of the Dirac
operator shown in Fig.6 in Appendix D.2. Of course, the dependence of $Z_-$ upon $\delta$ is
more complicated now. All eigen--values behave asymptotically as $Z = 3 \delta^3 / 64$
when $\delta \rightarrow \infty$ (compare to the asymptotic linear slope lines
$\zeta = \delta / 4$ in Fig.6).

\begin{figure}[h]
\centering
\epsfig{file=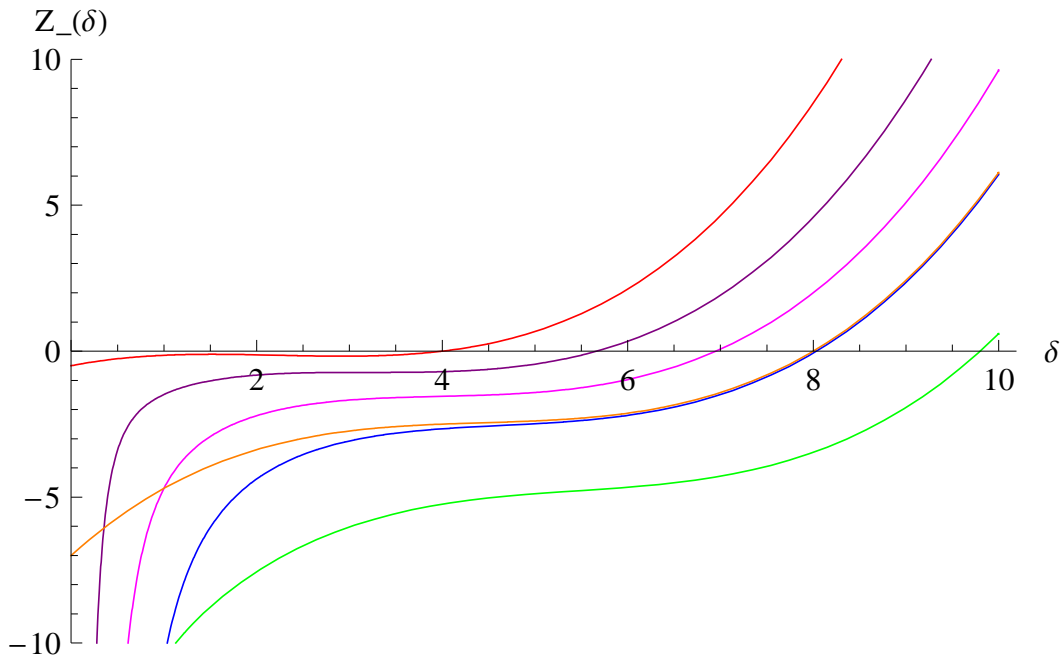,width=14cm}
\caption{Level crossing of the modes $Z_-(\delta)$ from negative to positive values}
\end{figure}

In anticipation of the general results for the $\eta$--invariant of the Lifshitz operator on
Berger spheres, we examine the spectral asymmetry between positive and negative eigen--values
in the special cases $\delta = 1$ and $\delta = 0$. This computation can be done by elementary
means and can be used for comparison with the more general formulae that will be derived
next. Since there is no spectral asymmetry for $\delta = 1$, $\eta_{\mathbb{D}}$ vanishes in
this case, as $\eta_{\rm D}$ also does. For $\delta = 0$ the finite eigen--values \eqn{stalousr} are
equally distributed to positive and negative values and the same thing applies to all other
eigen--values $Z_{\pm}$ that tend to $\pm \infty$. In this case, however, there is an excess
of positive modes provided by $Z_0$. They have multiplicity $2p$ and depend upon $p$ as $p^3$
up to an overall constant $1 / (8\delta^3)$. Although this normalization constant is infinite at
$\delta = 0$, the $\eta$--invariant does not depend upon it, since $\eta_{\mathbb{D}}$ is
inert to uniform rescaling of the spectrum. We immediately find $\eta_{\mathbb{D}} (s) =
2 \zeta (3s-1)$, and, therefore, $\eta_{\mathbb{D}} = \eta_{\mathbb{D}} (s = 0) =
2 \zeta (-1) = -1/6$. The final result is identical to the $\eta$--invariant of the
Dirac operator for $\delta = 0$ although in that case the spectrum depends linearly upon
$p$ and, thus, $\eta_{{\rm D}} (s) = 2 \zeta (s-1)$ (see the relevant footnote in Appendix
D.3); setting $s=0$ yields the same answer $2 \zeta (-1)$ for both operators.

These preliminary results make us suspect that $\eta_{\mathbb{D}} = \eta_{{\rm D}}$ holds
for all values of $\delta$ and not only for $0$ and $1$. We will confirm this relation by direct
calculation and also explain its occurrence in the context of Atiyah--Patodi--Singer theory
that was outlined before.

\subsection{Computation of the $\eta$--invariant}

In analogy with the Dirac operator discussed in Appendix D.3, we employ the spectral
Riemann zeta--function and define the regularized difference between positive and negative
eigen--values of the Lifshitz operator by the general formula
\be
\eta_{\mathbb{D}} (s) = \sum_{{\rm eigenvalues}} ({\rm sign} ~ Z) ~ |Z|^{-s}
\ee
excluding zero modes. Then, the $\eta$--invariant of the three--dimensional Lifshitz
operator is naturally defined by appropriated analytic continuation at $s=0$, as
$\eta_{\mathbb{D}} = \eta_{\mathbb{D}} (s=0)$, accounting for its spectral asymmetry.

First, we consider Berger spheres with $\delta < 4$ so that $Z_0$ and $Z_+$
are all positive definite and $Z_-$ are all negative. Then, up to uniform rescaling
of the spectrum that does not affect the final result for $\eta_{\mathbb{D}}$ (recall
that when the spectrum scales uniformly the $\eta$--invariant remains the same because
it is evaluated at $s=0$), we have
\ba
\eta_{\mathbb{D}} (s) & = & \sum_{p, q >0} (p+q) \left(\left({\delta^2 \over 2} + X
\right) \Big[\left({\delta^2 \over 2} + X\right)^2 + {\delta^2 \over 2} (\delta^2 - 4)
\Big] \right)^{-s} - \nonumber\\
& & \sum_{p, q >0} (p+q) \left(\left(-{\delta^2 \over 2} + X \right)
\Big[\left(-{\delta^2 \over 2} + X \right)^2 + {\delta^2 \over 2} (\delta^2 - 4)\Big]
\right)^{-s} + \nonumber\\
& & \sum_{p>0} 2p \left(\left({\delta^2 \over 2} + p \right)
\Big[\left({\delta^2 \over 2} + p \right)^2 + {\delta^2 \over 2} (\delta^2 - 4)\Big]
\right)^{-s}
\label{horibh}
\ea
setting $X = \sqrt{4 \delta^2 pq + (p-q)^2}$ for notational convenience.
The first line refers to the contribution of $Z_+$, the second to $Z_-$ and the third
to $Z_0$. We will compute the individual sums by expanding all fractions in powers of
$\delta$ (up to the appropriate order) and then set $s=0$. The steps we follow are
analogous to those outlined in Appendix D.3 for $\eta_{\rm D}$, but, of course, the
intermediate expressions are much more involved now.

The contribution of the modes $Z_0(\delta)$ is the easiest to evaluate. Using the
power series expansion
\ba
& & \left(\left({\delta^2 \over 2} + p \right)
\Big[\left({\delta^2 \over 2} + p \right)^2 + {\delta^2 \over 2} (\delta^2 - 4)\Big]
\right)^{-s} = \nonumber\\
& & {1 \over p^{3s}} \Big[1 + {s \delta^2 \over 2} \left({4 \over p^2} - {3 \over p} \right)
+ {s(9s-1) \delta^4 \over 8p^2} + \cdots \Big]
\label{muzch}
\ea
we find that the last term in \eqn{horibh} assumes the following expansion in terms of
Riemann zeta--functions,
\be
\tilde{I}_0 (s) = 2 \zeta (3s-1) + s \delta^2 \left(4 \zeta (3s+1) - 3 \zeta (3s) \right) +
{s(9s-1) \delta^4 \over 4} \zeta (3s+1) + \cdots ~.
\ee
The terms that are omitted vanish at $s=0$ since they contain the factor $s\zeta (3s+n)$
with integer $n \geq 2$ and $\zeta (s)$ is absolutely convergent for ${\rm Re} s > 1$;
they include a term of the form $\delta^4 /p^3$ in \eqn{muzch}
as well as all terms of order $\delta^6$ and higher which come multiplied with
$1/p^{n+1}$ with $n \geq 2$. Then, taking into account that $\zeta (-1) = -1/12$,
$\zeta(0) = -1/2$ and that $\zeta (s)$ has a simple pole at $s=1$ with residue $1$ (and,
thus, $3s \zeta (3s +1)$ equals $1$ at $s=0$), we obtain
\be
\tilde{I}_0 (0) = -{1 \over 6} + {4 \delta^2 \over 3} - {\delta^4 \over 12} ~.
\ee

The contribution of the modes $Z_{\pm} (\delta)$ to the $\eta$--invariant is more difficult
to extract because of the double sums that are involved. Again, we expand the fractions in
power series of $\delta$
\ba
& & \left(\left(\pm {\delta^2 \over 2} + X \right) \Big[\left(\pm {\delta^2 \over 2} +
X \right)^2 + {\delta^2 \over 2} (\delta^2 - 4) \Big] \right)^{-s} =
{1 \over X^{3s}} \Big[1 + {s \delta^2 \over 2} \left(\pm {3 \over X} + {4 \over X^2}
\right) + \nonumber\\
& & ~~~~~~~~~~~~~~~ {s \delta^4 \over 8} \left({9s-1 \over X^2} \mp {8(3s+2) \over X^3}
\right) \mp {3s(s-1)(3s+2) \delta^6 \over 16 X^3} + \cdots \Big]
\ea
omitting all terms of the form $1/X^{n+2}$ with $n \geq 2$ that do not contribute to the
final result when $s=0$, as will be seen shortly. Then, the first two sums in the general
expression \eqn{horibh} for $\eta_{\mathbb{D}}(s)$, which we denote respectively by
$\tilde{I}_{\pm} (s)$, are expanded as follows,
\be
\tilde{I}_+ (s) - \tilde{I}_- (s) = -3s \delta^2 f\left({3s+1 \over 2}\right) -
{s(3s+2) \delta^4 \over 8} \left(16 + 3 (s-1) \delta^2 \right) f\left({3s+3 \over 2}
\right) + \cdots ~,
\ee
when they are combined together, setting for convenience
\be
f(s) = \sum_{p, q >0} {p+q \over [4 \delta^2 pq + (p-q)^2]^s} ~.
\ee
The function $f(s)$ also arises in the computation of the $\eta$--invariant of the
Dirac operator. It is absolutely convergent on the complex $s$--plane for ${\rm Re} s > 3/2$,
which justifies the suppression of the higher order terms in the power series expansion
(all such terms are multiplied with $s$ and vanish at $s=0$). We also recall from Appendix
D.3 that $f(s)$ has simple poles at $s=1/2$ and $s=3/2$ with residues given by equation
\eqn{resqar}. These are the values of $(3s/2) f((3s+1)/2)$ and $(3s/2) f((3s+3)/2)$ at $s=0$
and they are equal to $(\delta^2 -1)/6$ and $1/(2\delta^2)$ respectively. Taking these into
account, we obtain
\be
\tilde{I}_+ (0) - \tilde{I}_- (0) = - \delta^2 - {\delta^4 \over 12} ~.
\ee

The final result for the $\eta$--invariant of the Lifshitz operator on Berger spheres
with $\delta < 4$ is identical to the result for the Dirac operator, i.e.,
\be
\eta_{\mathbb{D}} = - {1 \over 6} (\delta^2 - 1)^2 ~.
\ee
This is not surprising in retrospect because the axial anomaly is identical in both
cases including the overall normalization constant. Then, by the Atiyah--Patodi--Singer
index theorem, the gravitational Chern--Simons action should be related to the
$\eta$--invariant of the Lifshitz operator exactly as the Chern--Simons action is related
to $\eta$--invariant of the Dirac operator by equation \eqn{relaspr}.
From this point of view, the computation of $\eta_{\mathbb{D}}$ provides
an alternative derivation/verification of the axial anomaly for Lifshitz fermions without
path integral methods. Hence our interest in pursuing this calculation to the end.
It can also be seen here that there are more terms contributing to $\eta_{\mathbb{D}} (s)$,
compared to $\eta_{\rm D}(s)$, but they are weighted with different factors. The two
expressions are clearly different when $s \neq 0$ (for comparison see the corresponding
steps taken in Appendix D.3 for the Dirac operator). Nevertheless, the individual terms add
up to produce the same final result at $s=0$, as in the path integral computation of the
local form of the axial anomaly.

Next, we consider Berger spheres with $\delta \geq 4$. In analogy with the Dirac operator
discussed in Appendix D, let us denote by
\be
{\cal C}(\delta_{\rm c}) = \{(p, q) \in \mathbb{N}^2 ; ~~ \delta_{\rm c}^2 = 2 \sqrt{4pq
\delta_{\rm c}^2 + (p-q)^2} \}
\ee
the set of all positive integers $(p, q)$ associated to zero modes of the Lifshitz operator
at each one of the special values $\delta_{\rm c} < \delta$. Since the multiplicity of
the zero modes at $\delta_{\rm c}$ is $p+q$ for each $(p, q)$, the quantity
\be
S(\delta) = \sum_{\delta_{\rm c} < \delta} \Big[\sum_{(p, q) \in {\cal C}(\delta_{\rm c})}
(p+q) \Big]
\label{oumarta}
\ee
provides the total number of modes that cross from negative to positive values as $\delta$
varies from $\delta < 4$ to any given value $\delta \geq 4$. Taking into account the multiplicity
of eigen--values, it follows that $S(\delta)$ is an even integer number.
Thus, for $\delta \geq 4$, the $\eta$--invariant of the Dirac operator for Berger spheres is
shifted by twice the number of these modes and equals
\be
\eta_{\mathbb{D}} = - {1 \over 6} (\delta^2 - 1)^2 + 2 S(\delta) ~.
\ee
$S(\delta)$ is the same number for the Dirac and Lifshitz operators, since level crossing
occurs at the same values of $\delta$ and the mode multiplicities are also the same.
$S(\delta)$ simply vanishes when $\delta < 4$. Thus, we have
\be
\eta_{\mathbb{D}} = \eta_{{\rm D}}
\ee
as advertised above. This formula is actually valid for all $0 \leq \delta < \infty$,
including the special values $\delta_{\rm c}$, where zero modes appear, and it will be used
in the following.

\subsection{Chiral symmetry breaking by instantons}

We apply the index theorem to the Dirac--Lifshitz operator defined on Ho\v{r}ava--Lifshitz
instanton backgrounds with $SU(2) \times U(1)$ isometry. Our treatment is quite
general, with no reference to the details of the particular gravitational solutions, in
order to identify the sector of the theory that leads to chiral symmetry breaking.
As will be seen in the following only the gross qualitative features of the geometry matter
for the analytic derivation of the index formula.

Using the Euclidean four--dimensional metric on $M_4 \simeq \mathbb{R} \times S^3$ written
in terms of the left--invariant one--forms\footnote{The left--invariant one--forms $\sigma^I$ of
$SU(2)$ should be distinguished from the Pauli matrices $\sigma_I$ used earlier to express
the three--dimensional fermion operator as $i \gamma^i {\cal D}_i = i \sigma_I {E_I}^i
{\cal D}_i$ in terms of tangent space indices.} $\sigma^I$ of $SU(2)$ with coefficients
$\gamma_1 = \gamma_2 \equiv \gamma$,
\be
ds^2 = dt^2 + \gamma (t) (\sigma^1)^2 + \gamma (t) (\sigma^2)^2 +
\gamma_3 (t) (\sigma^3)^2 ~,
\ee
we introduce the orthonormal coframe $e^a$ of vierbeins given by the one--forms
\be
e^0 = dt ~,  ~~~~~ e^1 = \sqrt{\gamma} ~ \sigma^1 ~, ~~~~~ e^2 =
\sqrt{\gamma} ~ \sigma^2 ~, ~~~~~ e^3 = \sqrt{\gamma_3} ~ \sigma^3
\ee
and obtain the corresponding connection one--forms computed by $de^a +
{\omega^a}_b \wedge e^b = 0$,
\ba
& & {\omega^0}_1 = -{1 \over 2 \gamma} {d \gamma \over dt} ~ e^1 ~, ~~~~~
{\omega^0}_2 = - {1 \over 2 \gamma} {d \gamma \over dt} ~ e^2 ~, ~~~~~
{\omega^0}_3 = - {1 \over 2 \gamma_3} {d \gamma_3 \over dt} ~ e^3 ~,\nonumber\\
& & {\omega^2}_3 = - {\sqrt{\gamma_3} \over 2 \gamma} ~ e^1 ~, ~~~~~~
{\omega^3}_1 = - {\sqrt{\gamma_3} \over 2 \gamma} ~ e^2 ~, ~~~~~~
{\omega^1}_2 = - {2 \gamma - \gamma_3 \over 2 \gamma \sqrt{\gamma_3}} ~ e^3 ~.
\ea
Then, the curvature two--forms computed by ${R^a}_b = d{\omega^a}_b + {\omega^a}_c \wedge
{\omega^c}_b$ turn out to be
\ba
& & {R^0}_1 = -{1 \over \sqrt{\gamma}} \left({d^2 \sqrt{\gamma} \over dt^2} \right) e^0
\wedge e^1 - {1 \over 4 \sqrt{\gamma_3}} {d \over dt} \left({\gamma_3 \over \gamma} \right)
e^2 \wedge e^3 ~, \\
& & {R^0}_2 = -{1 \over \sqrt{\gamma}} \left({d^2 \sqrt{\gamma} \over dt^2} \right) e^0
\wedge e^2 - {1 \over 4 \sqrt{\gamma_3}} {d \over dt} \left({\gamma_3 \over \gamma} \right)
e^3 \wedge e^1 ~, \\
& & {R^0}_3 = -{1 \over \sqrt{\gamma_3}} \left({d^2 \sqrt{\gamma_3} \over dt^2} \right) e^0
\wedge e^3 + {1 \over 2 \sqrt{\gamma_3}} {d \over dt} \left({\gamma_3 \over \gamma} \right)
e^1 \wedge e^2 ~, \\
& & {R^2}_3 = - {1 \over 2 \sqrt{\gamma}} \left({d \over dt} \sqrt{{\gamma_3 \over \gamma}}
\right) e^0 \wedge e^1 + {1 \over 4 \gamma \gamma_3} \left({\gamma_3^2 \over \gamma} -
{d \gamma \over dt} {d \gamma_3 \over dt} \right) e^2 \wedge e^3 ~, \\
& & {R^3}_1 = - {1 \over 2 \sqrt{\gamma}} \left({d \over dt} \sqrt{{\gamma_3 \over \gamma}}
\right) e^0 \wedge e^2 + {1 \over 4 \gamma \gamma_3} \left({\gamma_3^2 \over \gamma} -
{d \gamma \over dt} {d \gamma_3 \over dt} \right) e^3 \wedge e^1 ~, \\
& & {R^1}_2 = {1 \over 2 \sqrt{\gamma_3}} {d \over dt} \left({\gamma_3 \over \gamma}\right)
e^0 \wedge e^3 - {1 \over 4 \gamma^2} \left(3 \gamma_3 - 4 \gamma +
\left({d \gamma \over dt} \right)^2 \right) e^1 \wedge e^2
\ea
and after some calculation the topological density takes the form of a total derivative term,
without using any equations of motion,
\be
{\rm Tr}(R \wedge R) = {d \over dt} ~ \Big[~{1 \over 2} \gamma_3 \left({d \over dt} {\rm log}
{\gamma_3 \over \gamma} \right)^2 + \left({\gamma_3 \over \gamma} - 1 \right)^2 \Big] ~
dt \wedge \sigma^1 \wedge \sigma^2 \wedge \sigma^3 .
\label{klarota1}
\ee

We also introduce a variant of the four--dimensional metric that has the cross--product form
\be
ds_{(0)}^2 = dt^2 + \gamma (t_0) (\sigma^1)^2 + \gamma (t_0) (\sigma^2)^2 +
\gamma_3 (t_0) (\sigma^3)^2 ~,
\ee
using a fixed parameter $t_0$ instead of $t$ in the metric components. It serves as reference
frame for the computations in the vicinity of $t=t_0$. Since the connection one--forms of the
cross--product metric have components ${\omega^0}_i = 0$ and ${\omega^i}_j$ are the same as
above with $t$ simply replaced by $t_0$, we infer that the second fundamental form
$\theta = \omega - \omega_0$ has the following non--vanishing components at $t = t_0$,
\be
{\theta^0}_1 = - {d \sqrt{\gamma} \over dt} |_{t=t_0} ~ \sigma^1 ~, ~~~~~~
{\theta^0}_2 = - {d \sqrt{\gamma} \over dt} |_{t=t_0} ~ \sigma^2 ~, ~~~~~~
{\theta^0}_3 = - {d \sqrt{\gamma_3} \over dt} |_{t=t_0} ~ \sigma^3 ~.
\ee
Then, ${\rm Tr} (\theta \wedge R)$ at any given slice $t=t_0$ is given by
\be
{\rm Tr} (\theta \wedge R) |_{t=t_0} = {1 \over 2} \gamma \left({d \over dt} {\rm log}
{\gamma_3 \over \gamma} \right) {d \over dt} \left(
{\gamma_3 \over \gamma} \right) |_{t=t_0} ~ \sigma^1 \wedge \sigma^2 \wedge \sigma^3 ~.
\label{klarota2}
\ee

Applying the Atiyah--Patodi--Singer index theorem \eqn{rotames} to the fermion operator
${\cal D}$ (it can be either $D$ or $\mathbb{D}$) we obtain
\ba
{\rm Ind} ({\cal D}) & = & -{1 \over 12} ~ \Big[~{1 \over 2} \gamma_3 \left({d \over dt}
{\rm log} {\gamma_3 \over \gamma} \right)^2 + \left({\gamma_3 \over \gamma} - 1 \right)^2
\Big]_{t = -\infty}^{t= +\infty} + \nonumber\\
& & {1 \over 24} \Big[~ \gamma \left({d \over dt} {\rm log} {\gamma_3 \over \gamma} \right)
{d \over dt} \left({\gamma_3 \over \gamma} \right) \Big]_{t = -\infty}^{t= +\infty}
- {1 \over 2} \Big[\eta_{\cal D} \Big]_{t = -\infty}^{t= +\infty}
\label{mnlka}
\ea
after performing the integral in the time variable $t$ and
the angular coordinates of $S^3$, using the volume element $\sigma^1 \wedge \sigma^2 \wedge
\sigma^2 = {\rm sin} \theta ~ d\theta \wedge d \varphi \wedge d \psi$, which yields a factor
of $16 \pi^2$. This result is quite general, as it stands, because it does not
depend on the specific form of the metric coefficients $\gamma (t)$ and $\gamma_3 (t)$ for
intermediate times, but it relies on their existence as $t$ varies from $-\infty$ to $+\infty$.

Furthermore, we observe that the time derivative terms cancel against each other without
reference to any field equations, and, thus, equation \eqn{mnlka} simplifies to
\be
{\rm Ind} ({\cal D}) = -{1 \over 12} ~ \Big[\left({\gamma_3 \over \gamma} - 1 \right)^2
\Big]_{t = -\infty}^{t= +\infty} - {1 \over 2} \Big[\eta_{\cal D}
\Big]_{t = -\infty}^{t= +\infty} ~.
\ee
The only information needed for computing the index are the boundary values of the metric
coefficients, whereas the boundary values of their time derivatives are irrelevant. Thus,
this formula is equally valid for the instantons of Ho\v{r}ava--Lifshitz gravity that
interpolate between any two fixed points of the Ricci--Cotton flow with $\delta = 1$ and
$0< \delta \neq 1$ as well as for the instantons that extend all the way to the fully
squashed sphere with $\delta =0$. Setting $\gamma_3 / \gamma = \delta^2$, as we did
before, and using the general expression for the $\eta$--invariant on Berger spheres,
$\eta_{\cal D} = - (\delta^2 -1)^2 /6 + 2S(\delta)$, we arrive at the final expression
\be
{\rm Ind} ({\cal D}) = - \Delta S
\ee
given by the number of level crossings (including their multiplicities) that have occurred on
the spatial slices, via deformation of $\delta$, in the entire history of time (for the
definition of $S(\delta)$ see equation \eqn{oumarta} or \eqn{oumarta2}). This is precisely
what was expected by the spectral flow argument. Instantons
and anti--instantons will have opposite signs of $\Delta S$, if it is not zero, since they
are mutually related by $t \rightarrow -t$. It is also obvious now that the index vanishes
on all instanton backgrounds whose end--points have non--negative sectional Ricci scalar
curvature.

The only instantons that can give rise to level crossing, and, thus, to non--vanishing index
occur in the unimodular version of Ho\v{r}ava--Lifshitz gravity with superspace parameter
$\lambda = 1/3$. They are eternal solutions of the normalized Ricci--Cotton flow whose
end--points are not constrained to have the same Ricci scalar curvature. When, in particular,
the Chern--Simons parameter $\omega$ is negative (for a given choice of orientation on
$S^3$), there are flow lines that interpolate between the totally isotropic configuration
and a Berger metric with anisotropy parameter
\be
\delta^2 \equiv a^3 = \left(-{\omega L \over 3 \kappa_{\rm w}^2} \right)^{3/2} ~.
\ee
Recall from section 3.3 (to which we refer for the notation) that such chiral instantons are
associated to the tunneling of an effective point--particle between the red bullet vacua of the
potential well shown in Fig.2c and Fig.2d; as such, they correspond to continuous deformations
of the round sphere to configurations with $\delta < 1$ and $\delta > 1$, respectively.

The {\em necessary and sufficient condition} to have non--vanishing index is provided by
$\delta > 4$, which singles out Fig.2d with the left red bullet pushed sufficiently far away
from the origin. This, in turn, amounts to the following inequality among the various
parameters of the theory,
\be
-{\omega \over \kappa_{\rm w}^2} > 24 \left({\pi^2 \over {\rm Vol}(S^3)}\right)^{1/3} .
\label{inamorat}
\ee
Here, we have expressed the characteristic length scale (size) $L$ of space in terms of
the volume of $S^3$ which equals to $2 \pi^2 L^3$ and it remains constant in time. Then,
the index is given by the total number of modes that changed sign as $\delta$ was varying
from $1$ to $a^{3/2}$ and equals
\be
{\rm Ind} ({\cal D}) = \pm S(a^{3/2})
\ee
with plus or minus signs referring to the way that the cylinder $\mathbb{R} \times S^3$ is
transversed (backward or forward in Euclidean time). This is in agreement with the spectral
flow interpretation of the index of the spin complex.

In more physical terms, a non--vanishing index implies that the axial charge associated
to the (anomalous) conservation law of the axial current $J_5^{\mu}$,
\be
Q_5 = \int_{\Sigma_3} d^3x \sqrt{{\rm det} g} ~ J_5^0 (t, x) ~,
\ee
is not necessarily conserved but it rather changes as
\be
\Delta Q_5 = {\rm Ind} ({\cal D})
\ee
in all processes mediated by the corresponding gravitational instanton backgrounds
and it leads to lepton and baryon number non--conservation. The index is an integer
number\footnote{$S(\delta)$ is an even integer and thus ${\rm Ind} ({\cal D})$ that
counts the difference of positive and negative chirality zero modes of the four--dimensional
fermion operator acting on four--component spinors is even. This means, in particular, that
Weyl fermions can be consistently coupled to Ho\v{r}ava--Lifshitz gravity without suffering
from global gravitational anomalies. Global anomalies can only arise when the number of level
crossings is odd, in which case the sign of $\sqrt{{\rm det}(i \gamma^{\mu} {\cal D}_{\mu})}$ becomes
ambiguous \cite{gaume}. Apparently, there are no large diffeomorphisms of $S^3$ (embedded in the
foliation preserving diffeomorphisms of $\mathbb{R} \times S^3$) that can flip the sign of
$\sqrt{{\rm det}(i \gamma^{\mu} {\cal D}_{\mu})}$ and trigger an inconsistency. Global gravitational
anomalies may nevertheless be present in higher dimensional generalizations of the theory, as
in ordinary gravity in $8k$ or $8k+1$ dimensions, where disconnected diffeomorphisms can arise
and play important role \cite{hitchin}.} (this is consistent
with the fact that $M_4$ admits spin structure), which can become arbitrarily large when the
parameter $a$ is sufficiently large. In terms of the inequality \eqn{inamorat}, it implies
that Ho\v{r}ava--Lifshitz gravity can exhibit chiral symmetry breaking when the relative
coupling of the Cotton term ($\sim \kappa_{\rm w} / \omega$) is sufficiently small. Said
differently, chiral symmetry breaking becomes possible when there is a lower bound on the
volume of space for fixed couplings $\kappa_{\rm w}$ and $\omega$,
\be
{\rm Vol}(S^3) > 13824 \pi^2 \left(-{\kappa_{\rm w}^2 \over \omega} \right)^3 .
\label{morat21}
\ee
A Lifshitz universe that exhibits a bounce \cite{calca, kiritsis, brande} may help to
realize this novel possibility in practice. We intend to study the implications of this
scenario elsewhere in more detail.

Another interesting question is the behavior of
inequality \eqn{morat21} for chiral symmetry breaking under the renormalization group
equations of topologically massive gravity. The one--loop beta functions for the three
dimensional Newton coupling $\kappa_{\rm w}^2$ and the cosmological constant $\Lambda$ have
been computed explicitly in the recent paper \cite{ergin}, where it was also found that the
Chern--Simons coupling $\omega$ does not change with the energy scale. One may try to
recast these results into a form that is directly applicable to the unimodular variant of
topologically massive gravity, in which $\Lambda$ assumes the role of an integration
constant and the volume of three--space appears as a running coupling, and inquire whether
both sides of the inequality \eqn{morat21} are renormalization group invariant (at least to
one--loop level). It seems that the answer is negative\footnote{Other important relations
in the three--dimensional world, such as the defining relation of the chiral point of
topologically massive gravity, also fail to be renormalization group invariant, as noted
in \cite{ergin}.}. A weaker
(and perhaps more appropriate) question that can be posed in this context is whether the
inequality \eqn{morat21} is not overturned by the renormalization group flow once it is
satisfied at a given energy scale; and if so, what will be the deeper meaning of this? We
are primarily interested in studying this question in the vicinity of a non--trivial fixed
point, which is UV attractive in all directions and has negative $\Lambda$, but there are
still some issues left open in the analysis given in \cite{ergin} that prevent us from
drawing definite conclusions at this moment. Intertwining our results with renormalization
group ideas and exploring their meaning in the four--dimensional theory of
Ho\v{r}ava--Lifshitz gravity are some important problems to which we also hope to return
elsewhere.

\subsection{Topological invariants of the instantons}

The index of the Dirac--Lifshitz operator on our gravitational instanton backgrounds is not
a topological invariant of space--time. This is mere reflection of the fact that the
dimension of the space of harmonic spinors on $S^3$ is not bounded by the topology. It
only depends on the geometry and it can grow without bound for metrics with very negative
Ricci scalar curvature. Thus, to complete our discussion, we will compute explicitly the
topological numbers of the instanton spaces, namely their signature and Euler numbers,
using the Atiyah--Patodi--Singer theorem for the Hirzebruch and de Rham complexes,
respectively. They both turn out to be zero, as expected by the cylindrical form of $M_4$
(but see also the exact arguments below).

{\bf Hirzebruch signature complex:} The signature of a Riemannian four--manifold with
boundaries is closely related to the formula for the index of the fermion operator.
It reads \cite{APS}
\be
\tau(M_4) = - {1 \over 24 \pi^2} \int_{M_4} {\rm Tr} (R \wedge R) +
{1 \over 24 \pi^2} \int_{\partial M_4} {\rm Tr} (\theta \wedge R) -
\eta_{\rm S} (\partial M_4) ~.
\label{lokobik}
\ee
The first two terms are identical to those appearing in \eqn{rotames} for
${\rm Ind}({\cal D})$, but they are now multiplied with $1 / 24 \pi^2$ that differs from
$1 / 192 \pi^2$ by a factor of $8$ (this is also consistent with the fact that a compact
four--dimensional spin manifold without boundaries has signature that is an integer multiple
of $8$). Thus, their particular form is known from before (see equations \eqn{klarota1} and
\eqn{klarota2}). The last term is the associated $\eta$--invariant for the signature
complex\footnote{To define the $\eta_{\rm S}$ one considers the self--adjoint operator $B$
acting on forms $\omega$ on the manifold $\partial M_4$ as $B(\omega) = (-1)^p (\epsilon
\star d - d ~ \star ) ~ \omega$, choosing $\epsilon = 1$ for $2p$--forms and $\epsilon = -1$
for $(2p-1)$--forms. $B$ preserves the parity of forms on $\partial M_4$ and commutes with
$\omega \rightarrow (-1)^p \star \omega$, so that $B = B^{\rm even} \oplus B^{\rm odd}$
and $B^{\rm even}$ is isomorphic to  $B^{\rm odd}$. In particular, the $\eta$--invariant
of the operator $B$ is twice the $\eta$--invariant of $B^{\rm even}$. Normally, the index
theorem \eqn{lokobik} would involve one--half the $\eta$--invariant of $B$, but this is
finally expressed in terms of $\eta_{\rm S}$ associated to $B^{\rm even}$ alone, explaining
the factor of $1/2$ that is missing.},
which has been computed by Hitchin on Berger spheres\footnote{The computation is straightforward
and it relies (once more) on the relation between the $\eta$--invariant and the Chern--Simons
action implied by the Atiyah--Patodi--Singer index theorem \cite{APS}. Note, however,
that there is a discrepancy by a factor of $2$ in the final expression for $\eta_{\rm S}$
compared to the result derived in \cite{nigel} (it is apparently a typo).} \cite{nigel} and
takes the following form,
\be
\eta_{\rm S} (\delta) = -{2 \over 3} (\delta^2 - 1)^2
\label{signaeta}
\ee
for all values of the anisotropy parameter $\delta$. Since it is also $8$ times larger than
$\eta_{\cal D}/2 = -(\delta^2 - 1)^2 /12$, without counting the effect of level crossing, as it
does not occur here, we find immediately that $\tau (M_4) = 0$.

{\bf Euler--de Rham complex:} The Euler number of a Riemannian four--manifold with
boundaries is given by
\be
\chi(M_4) = {1 \over 32 \pi^2} \int_{M_4} \epsilon_{abcd} ~ {R^a}_b \wedge {R^c}_d -
{1 \over 16 \pi^2} \int_{\partial M_4} \epsilon_{abcd} \left({\theta^a}_b \wedge {R^c}_d -
{2 \over 3} {\theta^a}_b \wedge {\theta^c}_e \wedge {\theta^e}_d \right) .
\ee
The bulk term is the integral of the Gauss--Bonnet topological density and the boundary
term is the integral of the corresponding Chern--Simons secondary class which is required
for well definiteness. In this case there is no analogue of the $\eta$--invariant boundary
term. Explicit computation shows that all our gravitational instanton backgrounds obey the
following relation,
\ba
& & {1 \over 2} \epsilon_{abcd} ~ {R^a}_b \wedge {R^c}_d = \epsilon_{abcd} ~ dt \wedge
{d \over dt} \left({\theta^a}_b \wedge {R^c}_d - {2 \over 3} {\theta^a}_b \wedge {\theta^c}_e
\wedge {\theta^e}_d \right) = \\
& & {d \over dt} ~ \Big[ ~ {1 \over 2 \gamma \sqrt{\gamma_3}} \left(\left({d \gamma \over dt}
\right)^2 - 4 \gamma + 3 \gamma_3 \right) \left({d \gamma_3 \over dt} \right) -
{(\sqrt{\gamma_3})^3 \over \gamma^2} \left({d \gamma \over dt} \right) \Big] ~
dt \wedge \sigma^1 \wedge \sigma^2 \wedge \sigma^3 , \nonumber
\ea
and, therefore, we easily obtain that $\chi(M_4) = 0$. This relation follows from the
particular form of the four--dimensional metric $ds^2 = dt^2 + \gamma(t) (\sigma^1)^2 +
\gamma(t) (\sigma^2)^2 + \gamma_3 (t) (\sigma^3)^2$ without using any equations of motion.

Another way to view these results is to think of $M_4 \simeq \mathbb{R} \times S^3$ as being
topologically equivalent to a compact manifold $M_{\rm c} \simeq S^4$ with two points removed,
say the "north" and the "south" poles. This is best seen by considering the canonical de
Sitter metric on $S^4$ with radius $L$ and $r$ running from $0$ to $\infty$,
\be
ds^2 = {1 \over (1 + (r/2L)^2)^2} \Big[~ dr^2 + 4r^2 \left((\sigma^1)^2 + (\sigma^1)^2 +
(\sigma^1)^2 \right) \Big] ~.
\ee
Passing to the proper coordinate $t$ given by the change of variables ${\rm tan}(t/2L) = r/2L$,
the metric takes the form
\be
ds^2 = dt^2 + {\rm sin}^2 \left({t \over L} \right) \left((\sigma^1)^2 + (\sigma^1)^2 +
(\sigma^1)^2 \right)
\ee
with $t$ running from $0$ to $\pi L$. It resembles the form of the four--dimensional
metrics that we have been considering all along having homogeneous and isotropic slices
with $\gamma_1 = \gamma_2 = \gamma_3 = {\rm sin}^2 (t/L)$. In this frame, $S^4$ is
described as $S^1$ fibration over $S^3$. At the two poles which are located at
$t = 0$ and $\pi L$, respectively, the three--dimensional slices shrink to a point.
Then, removing the two poles yields a compact space with boundaries that has the
topology of $\mathbb{R} \times S^3$.
Next, using the following general relations $\tau (M_4) = \tau (M_{\rm c})$ and
$\chi (M_4) = \chi (M_{\rm c}) - 2$, and knowing, in particular, that $\tau (S^4) = 0$
and $\chi (S^4) = 2$, we obtain immediately that $\tau (M_4) = 0 = \chi (M_4)$.

It is now appropriate to compare these results to the topological invariants of instanton
solutions of Einstein gravity, focusing, in particular, to the Taub--NUT and Eguchi--Hanson
spaces. These metrics take exactly the same form in proper time as the instantons
of Ho\v{r}ava--Lifshitz gravity, and, therefore, one may (erroneously) think that their
signature and Euler numbers also vanish. Here, we focus to the topological aspects of these
spaces (their analytic properties are discussed in the next section to which we refer for
details) and explain in brief how the presence of removable singularities affects the
computation of $\chi$ and $\tau$. Such singularities are not present in the instanton metrics
of Ho\v{r}ava--Lifshitz gravity, by construction, and, naturally both $\chi$ and $\tau$
vanish there.

The removable singularities that arise at one end of the instanton solutions of Einstein
gravity form the fixed point set of a Killing vector field. Then,
one has to employ the Lefschetz fixed--point theorem, which is a special case of the more
general $G$--index theorem, to account for the presence of the fixed points in the
computation of the Euler number (see, for instance, \cite{eguchi3} for a rather
elementary account). According to this theorem, an isolated fixed point (nut)
adds one unit to the Euler number of the manifold and a surface of fixed points that
form an invariant $S^2$ submanifold (bolt) adds to it two units. As a result, the
Euler number of Taub--NUT space is $1$ and that of Eguchi--Hanson space is $2$.
The computation of the signature requires some additional ingredients in
the case of  Eguchi--Hanson space: the $\eta$-invariant of the signature complex \eqn{signaeta}
should be made equivariant with respect to a discrete $\mathbb{Z}_2$ symmetry that identifies
antipodal points on $S^3$ (see, for instance, \cite{nigel}) and integration over $S^3$ yields
a factor of $8\pi^2$ instead of $16\pi^2$. Taking these into account one finds that the
signature of Eguchi--Hanson space is $\pm 1$ (the sign refers to the choice of instanton or
anti--instanton metric), whereas the signature of Taub--NUT space vanishes. In either case
we have $\chi (M_4) = | \tau (M_4) | + 1$, as required on general grounds for this class of
Einstein spaces.

\section{Comparison to relativistic theories}
\setcounter{equation}{0}
This section is devoted to instanton solutions
of Einstein gravity and the associated index formulae for the Dirac operator. It will make
more transparent the issues we addressed earlier but in a more familiar context.

The index of the Dirac operator $D$ in the background of a four--dimensional metric on a
manifold $M_4$ with boundary $\partial M_4$ takes the same form \eqn{rotames} as for the
Lifshitz operator, according to Atiyah--Patodi--Singer index theorem \cite{APS}, where
$\eta_{\rm D}$ is the $\eta$--invariant of the three--dimensional Dirac operator at the boundary,
which is discussed extensively in Appendix D. ${\rm Ind}(D)$ counts the difference in the number
of normalizable positive and negative chirality spinors obeying the appropriate boundary conditions,
as usual. Our main objective is to compute the index of the Dirac operator in the background
of Taub--NUT \cite{hawking} and Eguchi--Hanson \cite{extra1} gravitational instanton metrics so
that direct comparison can be made to the results derived in the non--relativistic case. Our
exposition is somewhat extensive containing all relevant details and it includes a relatively
new (alternative) derivation of the index in the case of Eguchi--Hanson space.

\subsection{Gravitational instantons with $SU(2) \times U(1)$ isometry}

The Taub--NUT and Eguchi--Hanson metrics provide the only complete and regular gravitational
instanton solutions of Einstein gravity with $SU(2) \times U(1)$ group of isometries. They
also admit an effective point particle description in the context of mixmaster dynamics
\cite{mixmast1} that parallels the description of $SU(2) \times U(1)$ instantons of
Ho\v{r}ava--Lifshitz gravity.

Recall that the potential of general relativity is minus the
Ricci scalar curvature of the spatial slices $\Sigma_3$ in the ADM formulation of the
theory on $\mathbb{R} \times \Sigma_3$ \cite{adm}. Using the Bianchi IX ansatz, we find
that the gravitational action with metric coefficients \eqn{bergersphe} takes the
following form\footnote{Integration over $S^3$ yields a volume factor $16 \pi^2$ that
fixes the normalization used in $S_{\rm eff.}$. This is correct for Taub--Nut space but
not for Eguchi--Hanson space whose slices are topologically $S^3/\mathbb{Z}_2$.
In the latter case the volume factor is $8 \pi^2$ and the overall normalization of
$S_{\rm eff.}$ should be $3 \pi^2 / \kappa^2$. We overlook this difference here in order to
make the presentation uniform, but we will come back to it later as it affects the
action of the instantons.} in proper time (with
$N(t) = 1$),
\be
S_{\rm eff.} = {6 \pi^2 \over \kappa^2} \int dt ~ e^{3 \Omega} ~
\Big[\left({d \beta \over dt}\right)^2 - 4 \left({d \Omega \over dt}\right)^2
- {16 \over 3} e^{- 2 \Omega} \left(e^{-4 \beta} - 4 e^{-\beta} \right) \Big] ~,
\label{mithrid}
\ee
setting for convenience $\kappa^2 = 32 \pi G$ and
\be
\beta = {\rm log} x ~, ~~~~~~ \Omega = {\rm log} L ~.
\ee
Note that the superspace metric is not positive definite for Einstein gravity, since
one has to take $\lambda = 1$, \cite{witt2}, and this is reflected in the negative kinetic
energy of the volume modulus found in the effective action. Because of this difference,
the instantons of Einstein gravity admit a slightly different description (compared to the
instantons of Ho\v{r}ava--Lifshitz theory), which is not directly based on Euclidean action
bounds. Now the action can in principle be made arbitrarily negative because of the
indefiniteness of the DeWitt metric associated to the conformal factor. Yet, the Euclidean
action of Einstein instantons is finite and it is given by a boundary term, since the bulk
term vanishes. Also, the instantons of Einstein gravity have self--dual Riemann curvature
tensor, whereas in Ho\v{r}ava--Lifshitz theory self--duality is not present.

Fig.5 is a plot of the effective potential $V(\beta)$ appearing in the action \eqn{mithrid}
as function of the anisotropy parameter $\beta$ with fixed $\Omega$.

\begin{figure}[h]
\centering
\epsfxsize=11cm\epsfbox{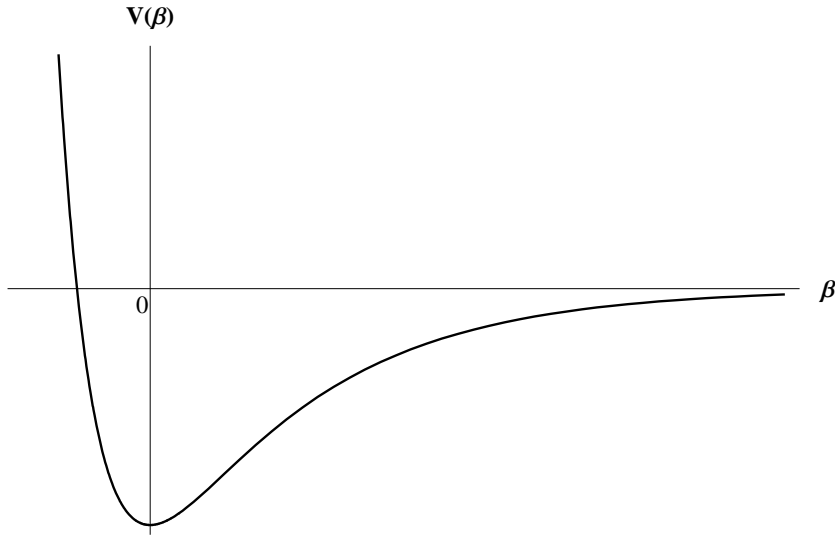}
\vspace{+0.2cm}
\caption{Effective potential for $SU(2) \times U(1)$ mixmaster dynamics in Einstein gravity}
\end{figure}

\noindent
In this case, however, the volume of space changes with time and $V(\beta)$ scales appropriately.
The plot is closely related to the one shown in Fig.3 for $R(x)$, since $V= -R$, but the potential 
is now expressed as function of $\beta$ rather than $x$. 
It should also be compared to the form of the potential in Ho\v{r}ava--Lifshitz gravity.
All these differences will be explained next from a particular point of view.

Following \cite{gibbons}, we
invert the potential $V(\beta)$ and represent an $SU(2) \times U(1)$ instanton by a
point particle which rolls from the top of the hill located at $\beta = 0$ towards the plateau
that is reached asymptotically as $\beta \rightarrow +\infty$. But the potential also varies
with time through its scaling factor, thus adjusting continuously the height difference in
the course of the motion. More importantly, the two end--points of the particle trajectory have
$\Omega = \pm \infty$. The top of the hill will correspond to a removable nut singularity in
space--time if $\Omega = -\infty$ and to a Euclidean (or conical infinity) if
$\Omega = +\infty$. Likewise, the plateau reached at the other end of the motion
will correspond to a removable bolt singularity in space--time if $\Omega = -\infty$ and
to a Taubian infinity if $\Omega = +\infty$. As a matter of fact, the self--dual Taub--NUT
metric corresponds to a trajectory which starts with zero velocity at the top of the hill
as $\Omega = -\infty$ and reaches the plateau at infinity as $\Omega = + \infty$. On
the other hand, the Eguchi--Hanson instanton corresponds to a trajectory that starts from
the plateau at infinity as $\Omega = -\infty$ and reaches the top of the hill
as $\Omega = +\infty$. These are the only two possible instanton solutions of Einstein
gravity with $SU(2) \times U(1)$ isometry that yield complete non--singular manifolds.
In both these cases, the proper time variable $t$ extends from a finite value (say $0$) to
$+\infty$ since there is a nut or a bolt forming at $t=0$ and space--time ends there.
The anti--instantons follow by letting $t \rightarrow -t$.

This qualitative picture can be made precise by writing down the first order equations
that $\beta$ and $\Omega$ satisfy in time. They are gradient flow equations derived from
a superpotential which is the Hamilton--Jacobi functional of general relativity computed
for Bianchi IX metrics. In our case, the potential takes the form
\be
{16 \over 3} e^{-2 \Omega} \left(e^{-4 \beta} - 4 e^{-\beta} \right) = e^{-6 \Omega} \Big[
\left({\partial {\cal W} \over \partial \beta} \right)^2 - {1 \over 4} \left({\partial
{\cal W} \over \partial \Omega} \right)^2 \Big]
\ee
and the first order equations satisfied by the instantons are
\be
{d \beta \over dt} = \pm e^{-3 \Omega} ~ {\partial {\cal W} \over \partial \beta} ~, ~~~~~
4 {d \Omega \over dt} = \mp e^{-3 \Omega} ~ {\partial {\cal W} \over \partial \Omega} ~.
\label{pousmin}
\ee
Note, however, that ${\cal W}$ is not uniquely determined because of the indefiniteness of
the DeWitt metric. There are two possible choices of superpotential given by
\be
{\cal W}_{\rm TN} = {4 \over 3} ~ e^{2 \Omega} \left(e^{-2 \beta} - 4 e^{-\beta/2} \right)
~~~~~~~ {\rm or} ~~~~ {\cal W}_{\rm EH} = {4 \over 3} ~ e^{2 \Omega} \left(e^{-2 \beta} +
2 e^{\beta} \right) ,
\ee
which correspond to the celebrated Taub--NUT and Eguchi--Hanson instantons, respectively.
In the first case, ${\rm exp}(-2 \Omega + 2 \beta) - {\rm exp}(-2 \Omega + \beta /2) =
1/(16 m^2)$ is a first integral of the system \eqn{pousmin} and the integration constant
$m$ is called nut parameter. In the second case, the integral is provided by
${\rm exp}(4 \Omega + 2 \beta) - {\rm exp}(4 \Omega - \beta) = a^4$ with $a$ being the
modulus of Eguchi--Hanson metric. These integrals allow to determine the metric of
gravitational instantons in closed form.

It can be easily verified that the first order equations \eqn{pousmin} solve the
second order equations of motion following from the action \eqn{mithrid} upon analytic
continuation in time. The relative minus sign between $(\partial {\cal W} /\partial \beta)^2$
and $(\partial {\cal W} / \partial \Omega)^2$ in the potential is also attributed to the
indefiniteness of the DeWitt metric.
The plus or minus signs appearing in the defining relations \eqn{pousmin} refer to the
instanton or anti--instanton configurations, which are related to each other by time reversal.
These equations also follow from the self--duality relations
by integrating once in time and making the appropriate choice of integration constants (we
refer the reader to the original work \cite{gibbons} for these technical details).
It so happens that the first order equations for the metric
coefficients of the self--dual Taub--NUT space are identical to the Ricci flow
equations on $S^3$ with metric \eqn{spatoulio} provided that $\gamma_i$ are traded with
$\gamma_i^2$. There is no analogous interpretation of the equations that determine the
Eguchi--Hanson metric.

The Euclidean point particle action for these trajectories is evaluated by completing
the squares and using equation \eqn{pousmin}. We obtain
\be
S_{\rm instanton}= \pm {12 \pi^2 \over \kappa^2} \int dt
\left({\partial {\cal W} \over \partial \beta}{d \beta \over dt} +
{\partial {\cal W} \over \partial \Omega}{d \Omega \over dt} \right) =
\pm {12 \pi^2 \over \kappa^2} \int dt ~ {d {\cal W} \over dt} =
\pm {12 \pi^2 \over \kappa^2} \Delta {\cal W} ~,
\label{mourmat}
\ee
which is solely determined by the difference of ${\cal W}$ at the end points of proper
time, $t = 0$ and $\pm \infty$. As such, it resembles the form of the action for
Ho\v{r}ava--Lifshitz instantons. However,  ${\cal W}$ blows up in the asymptotic region of
both instanton solutions (Taubian or conical infinity). This can be easily seen by
employing the first integrals and evaluating ${\cal W}$ at $t = \pm \infty$. At the
other end of the instanton, which corresponds to  a removable singularity (nut or bolt)
reached at $t=0$, ${\cal W}$ is finite. Thus, $S_{\rm instanton}$, as evaluated above,
turns out to be infinite and looks problematic.
To remedy the situation one has to subtract the infinite contribution of the flat space
metric which serves as reference frame in the calculations.

Notice in this respect that the integral of the mean curvature of
any given slice $\Sigma_3$ is the derivative of its volume, which, in turn yields the
following relation using the first order equations \eqn{pousmin},
\be
\int_{\Sigma_3} d^3 x \sqrt{{\rm det} g} ~ K = {1 \over 2} \int_{\Sigma_3} d^3 x
\sqrt{{\rm det} g} ~ g^{ij} {d \over dt} g_{ij} = {d \over dt} {\rm Vol}(\Sigma_3) =
2 \pi^2 {d \over dt} e^{3 \Omega} = \mp 3 \pi^2 {\cal W} ~.
\ee
The mean curvature is the trace of the extrinsic curvature, $K = g^{ij} K_{ij}$, with
$2K_{ij} = dg_{ij} / dt$ since we have made the choice of lapse and shift functions
$N=1$ and $N_i = 0$. The volume of space is ${\rm Vol}(\Sigma_3) = 2\pi^2 L^3$ with
$L= {\rm exp} \Omega$, and, finally, we have $\partial {\cal W} / \partial \Omega =
2 {\cal W}$ due to the particular dependence of ${\cal W}$ upon $\Omega$ for both
Taub--NUT and Eguchi--Hanson spaces. Based on this observation, we recast the Euclidean
instanton action \eqn{mourmat} in the form
\be
S_{\rm instanton}= - {4 \over \kappa^2} \int_{\partial M_4} d^3 x \sqrt{{\rm det} g} ~ K ~.
\ee
To make the action finite one simply has to subtract the contribution of a reference round
sphere with volume $16 \pi^2 t^3$ at both ends of space--time which we write as
\be
S_{\rm instanton}^{\prime}= - {4 \over \kappa^2} \int_{\partial M_4} d^3 x \sqrt{{\rm det} g}
~ (K - K_0) ~.
\ee
The integral of $K-K_0$ over $\partial M_4$ is precisely the so called
Gibbons--Hawking boundary term and it is rewarding to derive it here in the canonical ADM
formalism by a different line of arguments. Taking this into account, the regulated instanton
action of the Eguchi--Hanson metric turns out to be zero and the difference in the
normalization of $S_{\rm eff.}$ (being $8 \pi^2$ instead of $16 \pi^2$) becomes irrelevant.
The regulated action of Taub--NUT instanton is not zero but finite.

Recall for completeness that in the conventional approach to the problem one adds the
Gibbons--Hawking boundary term to the bulk Einstein--Hilbert action \cite{giha} (but see 
also \cite{york}).
This is provided by the difference between the trace of the
second fundamental form of the boundary in a given metric, $K$, and its value in the flat
space metric, $K_0$. More precisely, we have the Euclidean action (setting $\kappa^2 = 32 \pi G$, 
as before)
\be
S_{{\rm gravity}} = - {2 \over \kappa^2} \int_{M_4} d^4 x \sqrt{{\rm det} G} ~ R[G]
- {4 \over \kappa^2} \int_{\partial M_4} d^3 x \sqrt{{\rm det} g} ~ (K - K_0) ~.
\ee
This does not affect the classical equations of motion but it makes the variational problem
well--posed in the presence of boundaries. Then, the Euclidean action of the instanton metrics
is provided by the Gibbons--Hawking boundary term alone, since the bulk term vanishes, and it
is finite, as required on general grounds.

In summary, the instantons of Einstein gravity with $SU(2) \times U(1)$ isometry
resemble the instantons of Ho\v{r}ava--Lifshitz gravity that interpolate between
the round and a fully squashed sphere. The volume of their spherical slices is modulated
with time. In the non--relativistic case the extrinsic curvature vanishes at the end--points
of space--time, since these are the fixed points of the Ricci--Cotton flow, and the
Gibbons--Hawking term becomes obsolete. Nevertheless, it will be interesting to investigate
the form of the boundary terms in Ho\v{r}ava--Lifshitz gravity in more general situations.

\subsection{Index of Dirac operator for Taub--NUT metric}

The Taub--NUT metric is conveniently written in terms of a radial
coordinate $r \in [m, \infty)$ ($m$ is the nut parameter) as
\be
ds_{\rm TN}^2 = {r+m \over r-m} dr^2 + (r^2 -m^2) \Big[(\sigma^1)^2 + (\sigma^2)^2 +
{4m^2 \over (r+m)^2} (\sigma^3)^2 \Big] ~.
\label{tanurr}
\ee
The end--points of $r$ determine the two boundaries of Taub--NUT space: the removable
nut singularity and the asymptotic infinity. Proper time can be alternatively used, setting
\be
\pm t = \int_m^r dx \sqrt{{x+m \over x-m}} = m ~ {\rm arcosh} ~ {r \over m} +
\sqrt{r^2 - m^2} ~,
\ee
and it runs from $0$ to $\pm \infty$ (the sign depends on whether
one imposes self--duality or anti--self--duality condition on the Riemann tensor).
As for the three Euler angles that are used to represent $\sigma^i$, they range
as $0 \leq \theta \leq \pi$, $0 \leq \varphi \leq 2\pi$, $0 \leq \psi \leq 4\pi$, and,
therefore, the slices that arise at fixed $r$ have the topology of $S^3$ ($\psi$ is
extended to the double covering of the rotation group $SO(3)$). Thus, integrating the
volume form $\sigma^1 \wedge \sigma^2 \wedge \sigma^3 = {\rm sin} \theta ~ d\theta
\wedge d\varphi \wedge d\psi$ yields a factor $16\pi^2$.

The computation of the index of the Dirac operator $D$ on the Taub--NUT space relies on
the general formula that was obtained in section 4 without reference to any equations
of motion for the background geometry. Using equations \eqn{klarota1} and \eqn{klarota2},
we have, in particular,
\be
{\rm Ind} (D) = -{1 \over 12} ~ \Big[\left({\gamma_3 \over \gamma} - 1 \right)^2
\Big]_{r = m}^{r = +\infty} - {1 \over 2} \Big[\eta_{\rm D}
\Big]_{r = m}^{r = +\infty} ~.
\label{routhoca}
\ee
The derivation of equations \eqn{klarota1} and \eqn{klarota2} was done before using the proper
(Euclidean) time $t$, but it can also be done using the radial coordinate $r$.
The only difference is that the end--points of the instanton are now located at $r=m$ and
$r = \infty$ (equivalently at $t=0$ and $t = \infty$) as shown above.

Here, the slices arising at fixed $r$ are Berger spheres with anisotropy parameter 
\be
{\gamma_3 \over \gamma} = \delta^2 = {4m^2 \over (r + m)^2}
\ee
and the $\eta$--invariant takes the following form, according to formula \eqn{myetai},
\be
\eta_{\rm D} |_r = - {(r +3m)^2 (r -m)^2 \over 6 (r +m)^4} ~.
\ee
Note that $\eta_{\rm D} |_{r = m} = 0$, since the nut is conformally related to an $S^3$
endowed with the constant curvature metric (i.e., $\delta = 1$). We also note that
$\eta_{\rm D} |_{r = \infty} = -1/6$, since the Taubian infinity is conformally related
to the fully squashed $S^3$ (i.e., $\delta = 0$). These special cases are discussed in
detail in Appendix D and they provide the two extreme values of $\delta$ in Taub--NUT
space.

We have, in particular, $\delta \leq 1$ so that the spherical slices have positive Ricci
scalar curvature for all $r \geq m$. Thus, there is no level crossing from negative to
positive eigen--values of the three--dimensional Dirac operator as the slices are
deforming geometrically by varying $r$ and the index vanishes. This is also apparent
from equation \eqn{routhoca}.
Thus, there is no chiral symmetry breaking induced by this gravitational instanton
background reconfirming the results that were originally obtained in the literature
\cite{romer, eguchi2} long time ago. The Hirzebruch signature of Taub--NUT space is
eight times the index of the Dirac operator -- there are no caveats in this case --
and it vanishes, as noted before.

\subsection{Index of Dirac operator for Eguchi--Hanson metric}

Another example is provided by the Eguchi--Hanson
metric which is conveniently written using a radial coordinate $r \in [a, \infty)$ as
\be
ds_{\rm EH}^2 = {dr^2 \over 1 - a^4/r^4} + {r^2 \over 4} ~ \Big[(\sigma^1)^2 +
(\sigma^2)^2 + \left(1 - {a^4 \over r^4} \right) (\sigma^3)^2 \Big] ~.
\ee
The end--points of $r$ determine the two boundaries of Eguchi--Hanson space: the removable
bolt singularity and the asymptotic locally Euclidean infinity. Proper time can also be
used in this case, setting
\be
\pm t = {}_2 F_1 \left(-{1 \over 4} , ~
{1 \over 2} ; ~ {3 \over 4} ; ~ {a^4 \over r^4} \right) r - a \sqrt{\pi} ~
{\Gamma (3/4) \over \Gamma (1/4)} ~,
\ee
which is given in terms of the hypergeometric function ${}_2 F_1$ and runs
from $0$ to $\pm \infty$, as for the Taub--NUT
space. However, the three Euler angles range differently here, as $0 \leq \theta \leq \pi$,
$0 \leq \varphi \leq 2\pi$, $0 \leq \psi \leq 2\pi$, and, therefore, they cover only once
the rotation group $SO(3)$ ($SO(3) \simeq SU(2)/\mathbb{Z}_2$). Letting $\psi$ range up to
$2\pi$ instead of $4\pi$ is necessary to remove the bolt singularity from the space--time
metric. Then, in this case, the slices that arise at fixed $r$ have the topology of the
real projective space $\mathbb{R}P^3 \simeq S^3/\mathbb{Z}_2$ and angular integration
yields a factor $8 \pi^2$ instead of $16\pi^2$.

Repeating the steps that were taken before, we find that the index of the Dirac operator for
the Eguchi--Hanson space takes the form
\be
{\rm Ind} (D) = -{1 \over 24} ~ \Big[\left({\gamma_3 \over \gamma} - 1 \right)^2
\Big]_{r = a}^{r = +\infty} - {1 \over 2} \Big[\eta_{\rm D}
\Big]_{r = a}^{r = +\infty} ~.
\label{routhocb}
\ee
The numerical factor of the first term is $1/24$ instead of $1/12$ that was encountered
before because of the difference in the angular integration. The slices at constant $r$
are Berger spheres with anisotropy parameter
\be
{\gamma_3 \over \gamma} = \delta^2 = 1 - {a^4 \over r^4} ~,
\ee
which varies monotonically from $0$ to $1$ as $r$ runs from $r=a$ to $\infty$.

The $\eta$--invariant of the three--dimensional Dirac operator at any given
slice is more tricky to find as it is now defined on $\mathbb{R}P^3$. One has to use
the spectrum of the Dirac operator on Berger spheres given in Appendix D and
implement the appropriate projection by $\mathbb{Z}_2$ while computing $\eta_{\rm D}$.
$\mathbb{R}P^3$ has two inequivalent spin structures provided by the embedding
of $\mathbb{Z}_2$ into $SU(2)$. For the canonical spin structure, the spectrum of
eigen--values and their multiplicities are given by equations \eqn{sasha1} and
\eqn{sasha2} with $\zeta_{\pm}$ restricted to pairs of positive integers quantum
numbers $(p, q)$ with $p+q$ even and $\zeta_0$ restricted to even positive integers
$p$. This is so because the states $|j, m>$ used in Appendix D.1 to solve the
eigen--value problem flip sign under rotations by $2 \pi$ when $j$ assumes
half--integer values, i.e., when $2j+1$ is even (the canonical choice). For the other
spin structure, the eigen--values $\zeta_{\pm}$ are restricted to pairs of positive
integers $(p, q)$ with $p+q$ odd and $\zeta_0$ are restricted to odd positive integers
$p$ so that $2j+1$ is odd. Overall, the spectrum of the Dirac operator on $S^3$
consists of the disjoint union of the spectra on $\mathbb{R}P^3$ associated to the
two inequivalent spin structures (further details as well as generalizations to lens
spaces can be found in \cite{bar}).

Using the canonical spin structure on $\mathbb{R}P^3$, we can easily compute $\eta_{\rm D}$
from first principles at the end points of the Eguchi--Hanson instanton (there is no
need to know the exact equivariant expression for $\eta_{\rm D}$ for generic values of
the anisotropy parameter $\delta$). At one end, $r = a$, $\mathbb{R}P^3$ is fully
collapsed to a two--sphere (it is the removable bolt singularity of the instanton) and
$\delta = 0$. At the other end, $r = \infty$, the space becomes asymptotically locally
Euclidean and $\delta = 1$. The results differ from $\eta_{\rm}$ computed at the two end
points of Taub--NUT space (for which $\delta = 0$ and $1$) because of the $\mathbb{Z}_2$
moding that affects the spectral asymmetry.

First, we consider the case $\delta = 0$. Only the eigen--values $\zeta_{\pm}$ with
quantum numbers $p=q \equiv n \in \mathbb{N}$ tend to finite values and yield the spectrum
$\pm n$ of the Dirac operator on $S^2$ (see the special cases discussed in Appendix D.1).
They are the limiting finite eigen--values on $\mathbb{R}P^3$ endowed with the canonical spin
structure, since $p+q = 2n$ is even. The other eigen--values $\zeta_{\pm}$ tend $\pm \infty$.
All these eigen--values are equally distributed showing no asymmetry. The remaining
eigen--values $\zeta_0$ with even quantum numbers $p \in 2 \mathbb{N}$ also become infinite
at $\delta = 0$. These are the relevant ones here as they provide the excess modes on the
fully squashed $\mathbb{R}P^3$ endowed with the canonical spin structure. Thus, the spectral
asymmetry is determined by computing their contribution to the $\eta$--invariant,
\be
\eta_{\rm D} (s) |_{r=a} = \sum_{p ~ {\rm even}} {2p \over p^s} = 2 \sum_{n=1}^{\infty}
{1 \over (2n)^{s-1}} = 2^{2-s} \zeta (s-1) ~,
\ee
which is expressed in terms of the Riemann zeta--function $\zeta (s)$. Setting $s=0$, we
obtain the $\eta$--invariant on the fully squashed $\mathbb{R}P^3$
\be
\eta_{\rm D} |_{r=a} = 4 \zeta(-1) = -{1 \over 3} ~.
\ee
This limiting case also shows why the second spin structure on $\mathbb{R}P^3$ should be
excluded from the discussion. All eigen--values $\zeta_{\pm}$ and $\zeta_0$ associated to
it tend to $\pm \infty$ at $\delta = 0$ leaving no finite modes behind to account
for the spectrum of the Dirac operator on $S^2$. This is not surprising in retrospect
as $S^2$ admits only one spin structure that follows from the canonical spin
structure on $\mathbb{R}P^3$; it also explains the choice of spin structure that was
made on the slices of Eguchi--Hanson metric for all other values of $\delta$ that
connect continuously to $0$.

Next, we consider the case $\delta = 1$. The spectrum turns out to be
$\pm (2n \pm 1)/4$ with multiplicities $n(n \pm 1)$ for all even positive integers $n$
(this follows easily by projecting the results found in Appendix D.1 to $\mathbb{R}P^3$
endowed with the canonical spin structure and the canonical round metric). This projection
creates an asymmetry in the spectrum compared to the round $S^3$. The allowed eigen--values
are now restricted to the subset $\cdots, -11/4, -7/4, -3/4, ~ 5/4, ~ 9/4, ~ 13/4, \cdots$
instead of the more extended list of eigen--values $\cdots, -7/4, -5/4, -3/4, ~ 3/4, ~ 5/4,
~ 7/4, \cdots$ that arise without the projection. Then, taking into account the
multiplicities of these eigen--values, we obtain the corresponding spectral Riemann
zeta--function
\be
\eta_{\rm D} (s) |_{r= \infty} = \sum_{n ~ {\rm even}} {n(n+1) \over (2n+1)^s} -
\sum_{n ~ {\rm even}} {n(n-1) \over (2n-1)^s} = {1 \over 4} \left(\beta(s-2) - \beta (s)
\right)
\ee
where $\beta (s)$ is the so called Dirichlet (or Catalan) beta function, which is defined by the
following infinite sum over all integers $n \geq 0$
\be
\beta (s) = \sum_{n=0}^{\infty} {(-1)^n \over (2n+1)^s} ~.
\ee
Then, the $\eta$--invariant at infinity follows by setting $s=0$ and it is found to be
\be
\eta_{\rm D} |_{r= \infty} = {1 \over 4} \left(\beta (-2) - \beta (0) \right) =
{1 \over 8} \left(E_2 - E_0 \right) = - {1 \over 4} ~,
\ee
where $\beta (-k) = E_k /2$ are given by the Euler numbers $E_k$ for all integers
$k \geq 0$; we have, in particular, $E_0 = 1 = -E_2$.

Combining these results, we find that the index of the Dirac operator for the Eguchi--Hanson
gravitational instanton background is zero. This follows from \eqn{routhocb}, term by term, as
\be
{\rm Ind}(D) = - {a^8 \over 24 r^8} \Big|_{r=a}^{r=\infty} -
{1 \over 2} \left(-{1 \over 4} + {1 \over 3} \right) = {1 \over 24} - {1 \over 24} = 0 ~.
\ee
Thus, chiral symmetry
breaking is not allowed to occur in this case either, reconfirming the results derived
in the literature \cite{extra4} (but see also the review \cite{eguchi3}).

An alternative derivation of the $\eta$--invariant on $\mathbb{R}P^3$ endowed with the
canonical metric is provided in the literature by employing the $G$--index theorem
(for a physicists description see, for instance, \cite{eguchi3} and references therein).
This method was originally used in \cite{extra4} to compute the index of the
Dirac operator on Eguchi--Hanson space (but see also \cite{extra2} for the computation of
boundary terms in all A--D--E gravitational instanton backgrounds by similar methods).
Our derivation here is different, as it relies on spectral methods, and we have included it
since it is not discussed that way elsewhere to the best of our knowledge.

\subsection{Miscellaneous remarks}

The index of the four--dimensional Dirac operator on gravitational instanton backgrounds
$M_4$ with boundaries ought to be zero as simple consequence of Lichnerowicz's theorem
\cite{lichn}. Using $(i \gamma^{\mu} D_{\mu})^2 = - D_{\mu} D^{\mu}$, which is the
specialization of formula \eqn{finalrel} to four--metrics with zero Ricci scalar curvature,
one immediately sees that the zero modes of the Dirac operator should necessarily
be covariantly constant spinors. Such solutions exist in general, but they can not be
normalizable on spaces that extend to infinity. Thus, the index theorem that counts the
difference between the positive and negative chirality normalizable zero modes (with
respect to the Atiyah--Patodi--Singer boundary conditions) vanishes, as noted in
\cite{gibbons}. On the other hand, if $M_4$ is compact without boundaries, there exist
covariantly constant spinors and the index is non--zero; in particular, one finds that
there are two covariantly constant modes of one helicity and no modes of the opposite
helicity on $K3$. Normalizability of the zero modes is not an issue in the latter case
because of the compactness of space.

The calculations presented in this section provide a complementary viewpoint based on
the ADM decomposition of the four--metric that is most appropriate for comparison with
Lifshitz theories (there is no simple analogue of Lichnerowicz's formula for the
Dirac--Lifshitz operator that can lead to similar arguments for the index). In the
relativistic case things work out in such a way so that the three--dimensional sectional
Ricci scalar curvature on $\Sigma_3$ remains positive definite as one transverses $M_4$
from one end to the other ($0 \leq \delta \leq 1$ in both examples consider above).
Then, by Lichnerowicz's formula, which is now applied to the three--dimensional Dirac
operator, there can be no zero modes (harmonic spinors) on $\Sigma_3$. This,
in turn, implies the absence of level crossing under the spectral flow induced by the
deformation of the three--metric along the radial direction $r$ and the index of the
four--dimensional Dirac operator vanishes, as required on general grounds, but through
the application of a different argument akin to the canonical ADM formalism.

A novel possibility arises when the Taub--NUT space is taken to another region $-m < r < m$.
In this case, one has to flip the overall sign of the metric to maintain Euclidean
signature while going beyond the coordinate singularity $r=m$. Then, the anisotropy
parameter $\delta = 2m / (r+m)$ is allowed to become arbitrarily large as $r$ comes
closer to the curvature singularity located at $r=-m$. In fact, the Ricci scalar curvature
of the spherical slices becomes zero at $r = 0$ and then turns negative. Harmonic spinors
first arise at $r=-m/2$, where $\delta =4$, and level crossing becomes possible for all
$-m < r < -m/2$ according to the analysis given in Appendix D. This possibility was
recognized in literature quite early \cite{eguchi2}. It seems interesting as it allows for
non--vanishing index of the Dirac operator. The computation proceeds as for $r>m$ with the
difference that the $\eta$--invariant has to be replaced with the more general formula
\eqn{geneta}. Thus, by spectral flow, the index equals to the number of modes that have
crossed zero, $S(\delta)$. There is a problem, however, in that the space can not be
truncated to any given value of $r$ other than $r = -m$ for otherwise the metric will be
incomplete. This is also seen
pictorially by considering the point particle interpretation of the gravitational
instanton in the effective potential shown in Fig.5. As $r$ varies from $m$ to $-m$, the
shape modulus $\beta$ varies from $0$ to $-\infty$. Then, the point particle rolls
down the infinitely deep cliff of the inverted potential $-V(\beta)$ and there is no
way to stop it before hitting the curvature singularity. In this case, the singularity is
inevitable for otherwise the metric will be incomplete. Thus, formally, the index of the
Dirac operator in this region of Taub--NUT space is infinite. This example illustrates
the fact that the index of the Dirac operator is always zero in Einstein gravity provided
that no curvature singularities are present in the geometry. Certain variations of
this theme have also been considered in the physics and mathematics literature in recent
years (see, for instance, \cite{roman1, roman2}).

Finally, we note that the results obtained for Lifshitz fermions resemble more the
behavior of relativistic Rarita--Schwinger spin $3/2$ fields coupled to gravity rather
than spin $1/2$ fermions. In that case, the axial current also exhibits
an anomaly which is given by ${\rm Tr} (R \wedge R)$, since it is a total derivative term,
but its coefficient differs by a factor of $-21$ compared to the spin $1/2$ anomalous term
\cite{mike, vann, malcom}; consistency of the Rarita--Schwinger equations of motion also
requires that the metric background should be Ricci flat, which is certainly true for all
instanton solutions of Einstein gravity.
Then, the integrated form of the anomaly for a relativistic spin $3/2$ field reads
\be
I_{3/2} = {7 \over 64 \pi^2} \int_{M_4} {\rm Tr} (R \wedge R) - {7 \over 64 \pi^2}
\int_{\partial M_{4}} {\rm Tr} (\theta \wedge R) - {1 \over 2} \eta_{3/2} (\partial M_4) ~,
\ee
using the corresponding $\eta$--invariant at the boundary of space--time, and it provides
the index of the Rarita--Schwinger operator on $M_4$. Explicit computations have shown
that $I_{3/2}$ equals $2 \tau$ (twice the Hirzebruch signature of $M_4$) on all
asymptotically locally Euclidean gravitational
instanton backgrounds of Einstein gravity \cite{extra3, extra4, extra2, exerm} (but see
also \cite{eguchi3} for an overview). This follows by applying the $G$--index theorem to
these special spaces and not by computing $\eta_{3/2}$ directly by spectral methods (this
computation has not been carried out in the literature to the best of our knowledge). For
instance, one finds $I_{3/2} = 2$ for the Eguchi--Hanson space. As a result, chiral symmetry
breaking can be induced by gravitational instanton effects in supergravity, which contains spin
$3/2$ fields, and it may give rise to helicity changing amplitudes. These results should be
contrasted to the index of the Dirac operator that vanishes on all gravitational instanton
backgrounds with boundaries; only on $K3$, which is the unique compact gravitational
instanton without boundary, one obtains a non--vanishing result $I_{1/2} = \tau/8 = -2$,
whereas $I_{3/2} = -21 I_{1/2} = 42$, as required by the local form of the anomalies.

It will be interesting to consider the non--relativistic analogue of Rarita--Schwinger
fields in Ho\v{r}ava--Lifshitz gravity, examine the structure of its quantum anomalies,
and also compare them to (super)gravity. More generally, one may consider higher spin 
fermion fields and compute their anomalies. Let $\Delta_{\pm} (M)$ be the $\pm$ chirality bundles 
on the four--dimensional space--time $M$ and let $S^r \Delta_{\pm} (M)$ denote their $r$--fold 
symmetric products. Then, following \cite{mike, exerm} (but see also \cite{eguchi3}), we introduce 
the general spin elliptic complexes $D_{m/2 , ~ n/2} ~ : \Delta_{m/2 , ~ n/2} (M) 
\rightarrow \Delta_{n/2 , ~ m/2} (M)$, where $\Delta_{m/2 , ~ n/2} (M) = S^m \Delta_+ (M) 
\otimes S^n \Delta_- (M)$ and find that their index is related to the spin $1/2$ index by 
\ba
I_{m/2 , ~ n/2} (M) & = & {1 \over 30} [n(n+2)(3n^2 +6n -14) - m(m+2)(3m^2 +6m -14)] \cdot 
\nonumber\\
& & ~~~~~~ (m+1)(n+1) I_{1/2} (M) ~. 
\ea  
It is natural to expect that this general relation extends to non--relativistic theories too, 
since $I_{1/2} \equiv I_{1/2 , ~ 0}$ is the same for both Dirac and Lifshitz fermions.

\section{Conclusions and discussion}
\setcounter{equation}{0}

We have established the universal form of the axial anomaly for massless
fermions coupled to background gauge and gravitational fields in four
space--time dimensions. The result is identical for Lifshitz fermions
with anisotropic scaling $z=3$ as it is for Dirac fermions for which $z=1$.
In either case, the gauge field contribution to the anomaly
is proportional to ${\rm Tr}(F \wedge F)$ and the gravitational
contribution is proportional to ${\rm Tr} (R \wedge R)$. This result
is not surprising in retrospect, since the anomaly obstructs
the axial current conservation law and, as such, it should be expressed by
a topological density that is locally written as total divergence. The overall
normalization of the anomaly also turned out to be the same for both
Lifshitz and Dirac fermions, as it was verified explicitly by detailed
computation based on Fujikawa's path integral method. This was also
explained on general mathematical grounds based on the integrated form of the
gravitational anomaly while showing that the $\eta$--invariant of the three--dimensional
Dirac and Lifshitz operators, which are closely related to the value of the Chern--Simons
action, are in fact equal to each other. Thus, the index of the Dirac--Lifshitz operator
coincides with the index of the relativistic Dirac operator. The index is non--zero on
space--times of the form $\mathbb{R} \times \Sigma_3$ provided that the three--dimensional
slices $\Sigma_3$ can become negatively curved to allow for level crossing.
These results are of general value and they may be used in various applications in
future work.

Let us further remark that the computation of the anomaly can be extended to more general situations,
where the fermionic Lifshitz operator is modified by a relevant term of the form
$M^2\gamma^iD_i$. This additional term coincides with  the spatial part of the conventional Dirac operator and
influences the transition of the fermion system from the $z=3$ UV fixed point to the $z=1$
Dirac fermion theory (coupled to the same background).
The anomaly is not affected by the flow, since it is attributed to  IR rather than UV effects, which explains
 the origin of our results.  Furthermore, we note in this context that we have chosen the fermions to be spinors under the spatial rotation group,
since in the IR we want to flow to ordinary Dirac fermions.

Among other issues that have been raised in the main text, the most pressing one is
the derivation of the gravitational anomaly of Lifshitz fermions by methods of supersymmetric
quantum mechanics that are analogous to the relativistic case \cite{gaume}. This will provide
a more efficient and systematic way to compute the anomaly and may also lead to new
developments in the area of non--relativistic supersymmetric field theories. A closely
related problem, which is connected to our computation of the $\eta$--invariant of the
three--dimensional Lifshitz operator, is the derivation of the effective action induced
by massless fermions in three dimensions. It is well known that the Chern--Simons action is
induced by fermions coupled to an external gauge field \cite{redlich} and this generalizes
naturally to the case of a gravitational background field in three dimensions \cite{vuo},
in agreement with general index--theoretic expectations. It was also pointed out in \cite{moore}
that the results generalize to higher odd dimensional spaces by showing, in particular, that the
imaginary part of the effective action is identified with $\eta_{\rm D}$ (up to a factor of
$\pi / 2$). Similar arguments apply to the spin $3/2$ fields that appear in supergravity.
Since we have shown that $\eta_{\mathbb{D}} = \eta_{\rm D}$, it is natural to rephrase our
result by saying that the parity violating part of the effective action induced by $z=3$ Lifshitz
fermions in the background of external fields in three dimensions is also given by the
Chern--Simons action (with the same overall normalization) after integration of the fermions.
It will be interesting to provide a diagrammatic computation of the three--dimensional effective
action in this case (as for the axial anomaly in four dimensions) in order to obtain an independent
field theoretic verification of this statement. Similar inquiries can be made about spin $3/2$
fields in non--relativistic theories of gravity, closing the general circle of investigations
that have been put forward in this paper.

We have subsequently coupled Lifshitz fermions to $(3+1)$--dimensional Ho\v{r}ava--Lifshitz
gravity and studied the role of axial anomaly in the background of instanton solutions.
Particular attention was paid to instantons with $SU(2)$ isometry group, since
there is complete classification and there are explicit forms available for
their metrics. These instanton backgrounds provide the simplest (yet non--trivial)
examples of more general solutions and they also have analogues in Einstein
gravity to which direct comparison can be made. The main difference with ordinary
instantons is their chiral nature which is inherited from the parity non--invariance
of three--dimensional topologically massive gravity. The integrated form of the
anomaly on such backgrounds, including all boundary effects, provides the
index of the fermion operator, which can be used to study chiral symmetry breaking
induced by instantons in gravitational theories. In particular, it
was found that the index can become non--zero on certain instanton backgrounds of
a unimodular phase of Ho\v{r}ava--Lifshitz gravity provided that the coupling of
the Cotton tensor term exceeds a critical value. The result was derived by
spectral flow methods and it relies heavily on the existence of harmonic spinors on
sufficiently deformed three--spheres, which are the leaves of space--time foliation.
Otherwise, the geometry of the background does not allow for bound states of the
four--dimensional fermion operator, since the index is zero, exactly as in the case of
instanton solutions of ordinary Einstein gravity with boundaries. Then, in the unimodular
phase of Ho\v{r}ava--Lifshitz gravity, it is possible to violate chiral symmetry by
non--perturbative instanton effects, provided that the volume of space exhibits a lower
bound that it is determined by the ratio of the Ricci to Cotton tensor couplings
raised to the third power. This, in turn, can lead to baryon and lepton
number violation processes in the deep ultra--violet regime of the theory. It is a
rather striking phenomenon that differentiates Ho\v{r}ava--Lifshitz gravity from
ordinary Einstein gravity once more. Certainly, further investigation is required in
order to establish this result in general, beyond the class of instantons with $SU(2)$
symmetry, and investigate its phenomenological and cosmological consequences.

A variant of this theme can be put forward by considering spherical slices of the form
$S^3 / \Gamma$, instead of $S^3$ itself, where $\Gamma$ is a finite subgroup of $SO(4)$
acting freely on $S^3$; these include lens spaces and certain other generalizations by
crystallographic groups and they can arise as prime factors in the decomposition of
more general compact three--manifolds. First of all, it will be interesting to consider
the normalized Ricci--Cotton flow on such spaces and establish an analogue of Hamilton's
space--form theorem (the latter states that if $g(0)$ is a metric of positive Ricci curvature
on a three--manifold $\Sigma_3$, then the volume--normalized Ricci flow exists for all time
and converges to the round metric on $S^3 / \Gamma$ for appropriately chosen
$\Gamma \subset SO(4)$ \cite{hamilton}, but see also \cite{anderson} for a more pedestrian
account of this and more recent results on the subject). It will also be interesting to
consider eternal solutions of the flow equations that will provide the instantons of
Ho\v{r}ava--Lifshitz gravity in this more general set up. Such generalized configurations,
if they exist, will qualify as non--relativistic analogues of the general class of
asymptotically locally Euclidean instantons of Einstein gravity that were proposed by Hitchin
\cite{kron1} and constructed by Kronheimer \cite{kron2}. These metrics, however, do not admit
any isometries, in general, and the same thing should be expected to happen here. Application
of the index theorem in these backgrounds will then require an equivariant generalization of
the $\eta$--invariant based on the spectrum of the Dirac operator on lens spaces \cite{bar}
(and generalizations thereof), as for the Eguchi--Hanson metric in Einstein gravity where
$\Sigma_3 \simeq S^3 / \mathbb{Z}_2$ (see, for instance, \cite{goette} for an up to date
exposition of this subject with several references to the mathematics literature).
The task seems formidable.

The use of spherical spaces $S^3 /\Gamma$ in the canonical $(3+1)$--dimensional formulation of
Einstein gravity is quite standard by now and it is often associated to the concept of {\em geons}.
Thus, what we are proposing here is to consider the non--relativistic analogue of geons in
the context of Ho\v{r}ava--Lifshitz gravity. Such configurations are rather exotic in that they
admit asymptotically trivial diffeomorphisms which are not deformable to the identity and they
can act non--trivially on the quantum state space (for example, they can give rise to states
with half--integral angular momentum); we refer the interested reader to the
papers \cite{fried1, fried2} for further details. The homotopy groups of the diffeomorphism
group of these spaces are directly related to the homotopy group of their isometries,
noting, in particular, that $\pi_0 [{\rm Diff} (S^3/\Gamma)] \simeq \pi_0 [N(\Gamma)/\Gamma]$,
where $N(\Gamma)$ is the normalizer of $\Gamma$ (the isometries of $S^3/\Gamma$ are those
elements $h \in SO(4)$ for which the image of an orbit of $\Gamma$ is another orbit of $\Gamma$,
i.e., when $h$ is in the normalizer $N(\Gamma)$ of $\Gamma$ defined by $h \Gamma h^{-1} = \Gamma$).
The disconnected components of the diffeomorphism group have been calculated and tabulated
in \cite{fried2} for all different type of compact spherical spaces. The large
diffeomorphisms of $\Sigma_3$ can lift to the foliation preserving diffeomorphisms of
$\mathbb{R} \times \Sigma_3$. It will be interesting to study their implications, if any, to
the gravitational anomalies in $3+1$ space--time dimensions and understand if Weyl fermions
can be consistently coupled to geometry in the quantum theory. We have not yet delved into the
details of all these generalizations, but we hope to return elsewhere and report some results
on the fate of chiral symmetry breaking and its variants on geonic backgrounds.

Another interesting generalization that is left open to future work is the
local form of the axial anomaly in $2d + 2$ space--time dimensions for
Lifshitz fermions with anisotropic scaling $z = 2n + 1$. In this case,
the Dirac--Lifshitz operator is naturally defined, modulo factor ordering ambiguities,
as $i \gamma^{\mu} \mathbb{D}_{\mu} = i\gamma^0 D_0 + i \gamma^i (\nabla^2)^n D_i$ for all
integer values of $n$ (ordinary Dirac fermions correspond to $n = 0$). It is practically
useful to consider only the values $n=0, ~ 1, ~ 2, ~ \cdots , d$ focusing, in particular,
to the "maximal" case $n=d$. Then, in this case,
it is natural to expect that the axial anomaly will be the same as for Dirac
fermions coupled to background gauge and metric fields (see, for instance, \cite{gaume}
for the relevant results in relativistic higher dimensional field theories).
One should be able to explicitly verify this statement by path integral methods or even
more easily by developing alternative computational methods based on supersymmetric
quantum mechanics, which unfortunately are lacking at the moment. Coupling the Fermi theory to
Ho\v{r}ava--Lifshitz gravity will then allow to compute the index of the fermion
operator on the corresponding instanton backgrounds (yet to be found) using the spectral
theory of the Dirac operator on odd dimensional spheres. We refer the interested reader to
the second reference in \cite{bar}, which proves the existence of metrics admitting harmonic
spinors in dimensions $2d+1 = 3 ~ {\rm mod} ~ 4$, thus generalizing Hitchin's results for
three--dimensional Berger spheres \cite{hitchin}; curiously, these more general results
single out $2d+2 = 4k$ space--time dimensions which also have prominent role in relativistic
theories and for which the index can be non--vanishing.
Higher dimensional analogues of Ho\v{r}ava--Lifshitz gravity together with the associated instanton
solutions have not been studied so far in the literature, but an interesting case should be
made in $2d+2$ dimensions with anisotropic scaling $z=2d+1$ (c.f. $n=d$). We hope to return
to these generalizations elsewhere.

Persistent boundary effects in gravitational theories arise when the spatial slices $\Sigma$  
have boundaries, as in the case of gauge theories with background monopole charge. Appropriate 
boundary terms should be added to the action, following the canonical treatment described in 
\cite{regge1, regge2} for $3+1$ Einstein gravity. Carrying the analysis over to 
Ho\v{r}ava--Lifshitz gravity should be rather straightforward and would make our analysis 
of the Euclidean action bounds applicable to more general situations. Then, the computation 
of the index would require the open space generalization of the index theorem developed 
in \cite{callias}. 

Another open question is the fate of Weyl symmetry upon quantization. It primarily
concerns the anisotropic Weyl invariant phase of Ho\v{r}ava--Lifshitz gravity, which is
thought to govern the deep ultra--violate regime of the theory, but this should
break as one flows to the infra--red by suitable running of the parameter $\lambda$ away
from the special value $1/3$. Although this issue remains largely unexplored in the literature,
it is of central importance for future developments in the subject. It also relates
to chiral symmetry violation by gravitational instanton effects, since this
possibility seems to arise only in the unimodular phase of the theory (this is a weaker
form of Weyl invariant phase that also arises at $\lambda = 1/3$). A simpler question
concerns the Weyl anomalies of matter fields in the background of gravity, such
as scalar, fermion and vector fields, which provide the quantum obstruction to the
tracelessness of their energy--momentum tensor. The Weyl anomaly only arises in
even dimensions, like the axial anomaly, and it has been exhaustively analyzed for
relativistic field theories (for an overview and history of the subject see, for
instance, \cite{duff} and references therein). The form of this anomaly remains largely
unexplored for non--relativistic field theories, apart from the case of conformally coupled
Lifshitz scalars that provide the simplest example of anisotropic Weyl invariant models at
the classical level \cite{theisen} (but see also \cite{davody}); even in this simple case there
are some questions that still remain unanswered and call for further work. One should also
be able to compute the Weyl anomaly for other fields, such as Lifshitz fermions, and compare
the results to the relativistic theories. The coupling of the fields to geometry should be
made conformal, as in \cite{theisen}. In general, there is no a priori reason to expect
that the coefficients of the anomalous terms will stay the same (in this respect the computation
of the axial anomaly is much cleaner because its form is protected by topology). Finally, the
breaking of Weyl symmetry in the gravitational sector should be addressed by appropriate
methods.

At this end, it also seems appropriate to view our results as being part of a more general
relation between non--relativistic fermions in condensed matter physics systems and Dirac
fermions in relativistic field theories, as they exhibit similar behavior. Other important
examples of this kind (though seemingly unrelated at first sight to the problems we have
considered here) include fermion number fractionization in polyacetylene (see, for instance,
\cite{BCS} and references therein) and more recently in graphene \cite{graphene}, which are
all accounted by the presence of fermion zero modes in non--trivial backgrounds. We should
delve deeper into these problems and understand the fundamental reasons for this
interplay. Definitely, Lifshitz type systems provide an interesting class of models to
further this endeavor.

\newpage

\section*{Acknowledgements}
This work was partially supported by the Cluster of Excellence "Origin and the
Structure of the Universe" in Munich, Germany. I.B. is very grateful for the hospitality
extended to him at the Arnold Sommerfeld Center for Theoretical Physics in Munich
during the early stages of the project. He also thanks the Galileo Galilei
Institute for Theoretical Physics in Florence for providing stimulating environment
in the course of this work. D.L. thanks the Theory Division at CERN for hospitality,
where part of the present work was done. Finally, we thank Marios Petropoulos for 
collaboration in the very early stages of the project and Christos Sourdis for his help
in drawing the figures.

\newpage

\appendix
\section{Dirac matrices and their trace identities}
\setcounter{equation}{0}

Recall that the Dirac gamma--matrices in the flat four--dimensional Euclidean
space $\mathbb{R}^4$ with metric $\delta_{\mu \nu}$ satisfy the anti--commutation
relations
\be
[\gamma^{\mu} , ~ \gamma^{\nu}]_+ = 2 \delta^{\mu \nu} ~,
\ee
which follow from their Lorentzian counterpart by analytic continuation
$t \rightarrow it$ and $\gamma^0 \rightarrow i \gamma^0$. Then, the chiral representation
of the Dirac algebra is provided by the $4 \times 4$ Hermitian matrices
\be
\gamma^0 = \left(\begin{array}{ccc}
0          &  & \mathbb{1} \\
           &  &            \\
\mathbb{1} &  & 0
\end{array} \right) , ~~~~~~
\gamma^i = \left(\begin{array}{ccc}
0          &  & -i \sigma_i \\
           &  &            \\
i \sigma_i &  & 0
\end{array} \right) ,
\ee
using the Pauli matrices
\be
\sigma_1 = \left(\begin{array}{ccc}
0 &  & 1 \\
  &  &   \\
1 &  & 0
\end{array} \right) , ~~~~~
\sigma_2 = \left(\begin{array}{ccc}
0 &  & -i \\
  &  &   \\
i &  & 0
\end{array} \right) , ~~~~~
\sigma_3 = \left(\begin{array}{ccc}
1 &  & 0 \\
  &  &   \\
0 &  & -1
\end{array} \right)
\label{paulim}
\ee
and $\gamma_5 = - \gamma^0 \gamma^1 \gamma^2 \gamma^3$ is represented by the
Hermitian matrix
\be
\gamma_5 = \left(\begin{array}{ccc}
\mathbb{1} &  & 0 \\
           &  &            \\
0          &  & -\mathbb{1}
\end{array} \right)
\ee
that anti--commutes with all $\gamma^{\mu}$ and squares to $\mathbb{1}$.

The trace of any product of odd number of gamma--matrices vanishes identically.
Furthermore, we have
\be
{\rm Tr} \gamma_5 = 0 = {\rm Tr} (\gamma_5 \gamma^{\mu} \gamma^{\nu}) ~,
\ee
whereas
\be
{\rm Tr} (\gamma_5 \gamma^{\mu} \gamma^{\nu} \gamma^{\kappa} \gamma^{\lambda}) =
-4 \epsilon^{\mu \nu \kappa \lambda}
\ee
written in terms of the fully anti--symmetric Levi--Civita symbol with
$\epsilon^{0123} = 1$. We also have ${\rm Tr} \left(\gamma^{\mu} \gamma^{\nu}
\right) = 4 \delta^{\mu \nu}$ and the useful relation
\be
\gamma^{\mu} \gamma^{\nu} \gamma^{\kappa} = \delta^{\mu \nu} \gamma^{\kappa} -
\delta^{\mu \kappa} \gamma^{\nu} + \delta^{\nu \kappa} \gamma^{\mu} +
{\epsilon^{\mu \nu \kappa}}_{\lambda} \gamma^{\lambda} \gamma_5
\ee
from which we obtain the following trace identity among gamma--matrices,
\be
{\rm Tr} (\gamma^{\mu} \gamma^{\nu} \gamma^{\kappa} \gamma^{\lambda})
= 4 (\delta^{\mu \nu} \delta^{\kappa \lambda} - \delta^{\mu \kappa}
\delta^{\nu \lambda} + \delta^{\mu \lambda} \delta^{\nu \kappa})
\ee
and, consequently, the identity
\ba
& & {\rm Tr} (\gamma_5 \gamma^{\mu} \gamma^{\nu} \gamma^{\kappa} \gamma^{\lambda}
\gamma^{\rho} \gamma^{\sigma}) = -4 \Big[\delta^{\mu \nu}
\epsilon^{\kappa \lambda \rho \sigma} - \delta^{\mu \kappa}
\epsilon^{\nu \lambda \rho \sigma} + \delta^{\nu \kappa}
\epsilon^{\mu \lambda \rho \sigma} + \nonumber\\
& & ~~~~~~~~~~~~~~~~~~~~~~~~ \delta^{\lambda \rho} \epsilon^{\mu \nu \kappa \sigma}
- \delta^{\lambda \sigma} \epsilon^{\mu \nu \kappa \rho} +
\delta^{\rho \sigma} \epsilon^{\mu \nu \kappa \lambda} \Big] ~.
\label{sixtra}
\ea

Finally, the trace of $\gamma_5$ with the product of eight gamma--matrices reduces
to a sum of traces of $\gamma_5$ with the product of six gamma--matrices giving, in
particular, the trace identity
\ba
& & {\rm Tr} \left(\gamma_5 ~ {1 \over 2}
[\gamma^{a^{\prime}} , ~ \gamma^{b^{\prime}}] {1 \over 2} [\gamma_a , ~ \gamma_b]
{1 \over 2} [\gamma^{c^{\prime}} , ~ \gamma^{d^{\prime}}] {1 \over 2}
[\gamma_c , ~ \gamma_d] \right) =
{3 \over 2} {\epsilon_a}^{a^{\prime} b^{\prime} c^{\prime}}
(\delta_c^{d^{\prime}} \delta_{bd} - \delta_d^{d^{\prime}} \delta_{bc}) +
\nonumber\\
& & {3 \over 2} {\epsilon_c}^{a^{\prime} c^{\prime} d^{\prime}}
(\delta_a^{b^{\prime}} \delta_{bd} - \delta_b^{b^{\prime}} \delta_{ad}) +
{3 \over 2} {\epsilon_a}^{a^{\prime} b^{\prime} d^{\prime}}
(\delta_d^{c^{\prime}} \delta_{bc} - \delta_c^{c^{\prime}} \delta_{bd}) +
{3 \over 2} {\epsilon_c}^{b^{\prime} c^{\prime} d^{\prime}}
(\delta_b^{a^{\prime}} \delta_{ad} - \delta_a^{a^{\prime}} \delta_{bd}) +
\nonumber\\
& & {3 \over 2} {\epsilon_b}^{a^{\prime} b^{\prime} c^{\prime}}
(\delta_d^{d^{\prime}} \delta_{ac} - \delta_c^{d^{\prime}} \delta_{ad}) +
{3 \over 2} {\epsilon_d}^{a^{\prime} c^{\prime} d^{\prime}}
(\delta_b^{b^{\prime}} \delta_{ac} - \delta_a^{b^{\prime}} \delta_{bc}) +
{3 \over 2} {\epsilon_b}^{a^{\prime} b^{\prime} d^{\prime}}
(\delta_c^{c^{\prime}} \delta_{ad} - \delta_d^{c^{\prime}} \delta_{ac}) +
\nonumber\\
& & {3 \over 2} {\epsilon_d}^{b^{\prime} c^{\prime} d^{\prime}}
(\delta_a^{a^{\prime}} \delta_{bc} - \delta_b^{a^{\prime}} \delta_{ac}) +
{3 \over 2} {\epsilon_{bd}}^{a^{\prime} b^{\prime}}
(\delta_a^{c^{\prime}} \delta_c^{d^{\prime}} - \delta_c^{c^{\prime}}
\delta_a^{d^{\prime}}) + {3 \over 2} {\epsilon_{bd}}^{c^{\prime} d^{\prime}}
(\delta_a^{a^{\prime}} \delta_c^{b^{\prime}} - \delta_c^{a^{\prime}}
\delta_a^{b^{\prime}}) + \nonumber\\
& & {3 \over 2} {\epsilon_{ac}}^{a^{\prime} b^{\prime}}
(\delta_b^{c^{\prime}} \delta_d^{d^{\prime}} - \delta_d^{c^{\prime}}
\delta_b^{d^{\prime}}) + {3 \over 2} {\epsilon_{ac}}^{c^{\prime} d^{\prime}}
(\delta_b^{a^{\prime}} \delta_d^{b^{\prime}} - \delta_d^{a^{\prime}}
\delta_b^{b^{\prime}}) + {3 \over 2} {\epsilon_{bc}}^{a^{\prime} b^{\prime}}
(\delta_d^{c^{\prime}} \delta_a^{d^{\prime}} - \delta_a^{c^{\prime}}
\delta_d^{d^{\prime}}) + \nonumber\\
& & {3 \over 2} {\epsilon_{ad}}^{c^{\prime} d^{\prime}}
(\delta_c^{a^{\prime}} \delta_b^{b^{\prime}} - \delta_b^{a^{\prime}}
\delta_c^{b^{\prime}}) + {3 \over 2} {\epsilon_{bc}}^{c^{\prime} d^{\prime}}
(\delta_d^{a^{\prime}} \delta_a^{b^{\prime}} - \delta_a^{a^{\prime}}
\delta_d^{b^{\prime}}) + {3 \over 2} {\epsilon_{ad}}^{a^{\prime} b^{\prime}}
(\delta_c^{c^{\prime}} \delta_b^{d^{\prime}} - \delta_b^{c^{\prime}}
\delta_c^{d^{\prime}}) + \nonumber\\
& & 4 {\epsilon_{ab}}^{a^{\prime} b^{\prime}}
(\delta_c^{c^{\prime}} \delta_d^{d^{\prime}} - \delta_d^{c^{\prime}}
\delta_c^{d^{\prime}}) + 4 {\epsilon_{cd}}^{c^{\prime} d^{\prime}}
(\delta_a^{a^{\prime}} \delta_b^{b^{\prime}} - \delta_b^{a^{\prime}}
\delta_a^{b^{\prime}}) + {\epsilon_{ac}}^{a^{\prime} c^{\prime}}
\delta_b^{b^{\prime}} \delta_d^{d^{\prime}} + {\epsilon_{bd}}^{b^{\prime}
d^{\prime}} \delta_a^{a^{\prime}} \delta_c^{c^{\prime}} + \nonumber\\
& & {\epsilon_{ad}}^{a^{\prime} d^{\prime}} \delta_b^{b^{\prime}}
\delta_c^{c^{\prime}} + {\epsilon_{bc}}^{b^{\prime} c^{\prime}}
\delta_a^{a^{\prime}} \delta_d^{d^{\prime}} + {\epsilon_{ac}}^{b^{\prime}
d^{\prime}} \delta_b^{a^{\prime}} \delta_d^{c^{\prime}} +
{\epsilon_{bd}}^{a^{\prime} c^{\prime}} \delta_a^{b^{\prime}}
\delta_c^{d^{\prime}} + {\epsilon_{ad}}^{b^{\prime} c^{\prime}}
\delta_b^{a^{\prime}} \delta_c^{d^{\prime}} + \nonumber\\
& & {\epsilon_{bc}}^{a^{\prime} d^{\prime}} \delta_a^{b^{\prime}} \delta_d^{c^{\prime}}
- {\epsilon_{ac}}^{b^{\prime} c^{\prime}} \delta_b^{a^{\prime}} \delta_d^{d^{\prime}}
- {\epsilon_{ac}}^{a^{\prime} d^{\prime}} \delta_b^{b^{\prime}} \delta_d^{c^{\prime}}
- {\epsilon_{bd}}^{b^{\prime} c^{\prime}} \delta_a^{a^{\prime}} \delta_c^{d^{\prime}}
- {\epsilon_{bd}}^{a^{\prime} d^{\prime}} \delta_a^{b^{\prime}} \delta_c^{c^{\prime}}
- \nonumber\\
& & {\epsilon_{ad}}^{b^{\prime} d^{\prime}} \delta_b^{a^{\prime}} \delta_c^{c^{\prime}}
- {\epsilon_{bc}}^{b^{\prime} d^{\prime}} \delta_a^{a^{\prime}} \delta_d^{c^{\prime}}
- {\epsilon_{ad}}^{a^{\prime} c^{\prime}} \delta_b^{b^{\prime}} \delta_c^{d^{\prime}}
- {\epsilon_{bc}}^{a^{\prime} c^{\prime}} \delta_a^{b^{\prime}} \delta_d^{d^{\prime}}
\label{eighttra}
\ea
This expression is manifestly invariant with respect to the simultaneous exchange of
indices $a \leftrightarrow c$, $b \leftrightarrow d$ and $a^{\prime} \leftrightarrow
c^{\prime}$, $b^{\prime} \leftrightarrow d^{\prime}$. It is written directly in a
form that will be used in the computations presented in section 2.3, where the
gamma--matrices are labeled by flat (tangent) Euclidean space--time indices, as for
$\mathbb{R}^4$. The expression given by \eqn{eighttra} has been put into final form
using the relation
\be
{\epsilon^{abc}}_d \epsilon^{a^{\prime} b^{\prime} c^{\prime} d} =
\delta^{a a^{\prime}} (\delta^{b b^{\prime}} \delta^{c c^{\prime}} -
\delta^{b c^{\prime}} \delta^{c b^{\prime}}) - \delta^{a b^{\prime}}
(\delta^{b a^{\prime}} \delta^{c c^{\prime}} - \delta^{b c^{\prime}}
\delta^{c a^{\prime}}) + \delta^{a c^{\prime}} (\delta^{b a^{\prime}}
\delta^{c b^{\prime}} - \delta^{b b^{\prime}} \delta^{c a^{\prime}}) ~.
\ee

These are the trace identities that will be needed for the computation of the
axial anomalies of fermions coupled to background gauge and metric fields.

\section{Some geometrical apparatus}
\setcounter{equation}{0}

We review some disperse geometrical notions and formulae that are relevant for the
coupling of fermions to background metric fields and the computation of the
axial anomaly in the presence of gravity. It also sets the notation used in the text.

\subsection{Spin connection and Lichnerowicz formula}

The coupling of fermions to a background metric field $G_{\mu \nu}$ is achieved
using Cartan's formalism (see, for instance, \cite{eguchi3}). To set up the notation,
we introduce vierbeins ${e^a}_{\mu}$ as $G_{\mu \nu} = \delta_{ab} {e^a}_{\mu}
{e^b}_{\nu}$, where $\delta_{ab}$ provides the flat tangent space--time metric in
the Euclidean domain. We also introduce the inverse vierbeins ${E_a}^{\mu}$ as
$G^{\mu \nu} = \delta^{ab} {E_a}^{\mu} {E_b}^{\nu}$, which satisfy the relations
${E_a}^{\mu} = \delta_{ab} G^{\mu \nu} {e^b}_{\nu}$ and ${E_a}^{\mu} {e^b}_{\mu}
= \delta_a^b$. The components of the spin connection ${\omega^{ab}}_{\mu}$ are
\be
{\omega^a}_{b \mu} =
{e^a}_{\nu} \left(\partial_{\mu} {E_b}^{\nu} + \Gamma_{\mu \lambda}^{\nu}
{E_b}^{\lambda} \right) = - {E_b}^{\nu} \left(\partial_{\mu} {e^a}_{\nu} -
\Gamma_{\mu \nu}^\lambda {e^a}_{\lambda} \right)
\ee
so that $\hat{\nabla}_{\mu} {e^a}_{\nu} \equiv \nabla_{\mu} {e^a}_{\nu} +
{\omega^a}_{b \mu} {e^b}_{\nu} = \partial_{\mu} {e^a}_{\nu} -
\Gamma_{\mu \nu}^\lambda {e^a}_{\lambda} + {\omega^a}_{b \mu} {e^b}_{\nu} = 0$.
Cartan's equations (in the case of no torsion) provide the
components of Riemann curvature tensor as
\be
\partial_{\mu} {\omega^{ab}}_{\nu} - \partial_{\nu} {\omega^{ab}}_{\mu} +
{\omega^a}_{c \mu} {\omega^{cb}}_{\nu} - {\omega^a}_{c \nu} {\omega^{cb}}_{\mu} =
{R^{ab}}_{\mu \nu} \equiv {e^a}_{\kappa} {e^{b}}_{\lambda}
{R^{\kappa \lambda}}_{\mu \nu} ~.
\ee

The Dirac operator $i \gamma^{\mu} D_{\mu}$ coupled to the metric is defined
using the gamma-matrices
\be
\gamma^{\mu} = \gamma^a {E_a}^{\mu} ~, ~~~~~~ [\gamma^a , ~ \gamma^b]_+ =
2 \delta_{ab} ~,
\ee
which are expressed in terms of their tangent space--time counter--parts $\gamma^a$
and the covariant derivative operator acting on spinors
\be
D_{\mu} = \partial_{\mu} + {1 \over 8} [\gamma_a , ~ \gamma_b]
{\omega_{\mu}}^{ab} ~,
\ee
which is defined through the spin connection. Then, we have
\be
[D_{\mu} , ~ D_{\nu}] = {1 \over 8} [\gamma_a , ~ \gamma_b] {R^{ab}}_{\mu \nu}
\label{lichari}
\ee
in accordance to the fact that the metric analogue of the gauge field is
provided by $(i/8) [\gamma_a , ~ \gamma_b] {\omega_{\mu}}^{ab}$ in the spinorial
representation of the tangent space--time rotation group and its field strength is
the Riemann curvature. More generally we define
\be
D_{\mu} = \nabla_{\mu} + {1 \over 8} [\gamma_a , ~ \gamma_b]
{\omega_{\mu}}^{ab}
\label{lichari2}
\ee
with the Christoffel affinity included, which also satisfies equation \eqn{lichari}
and has the property that $[D_{\mu} , ~ \gamma^{\nu}] = \gamma^a \hat{\nabla}_{\mu}
{E_a}^{\nu} = 0$.

Using \eqn{lichari2}, we also have the following relation for the square of the Dirac
operator,
\be
(i \gamma^{\mu} D_{\mu})^2 = - D_{\mu} D^{\mu} - {1 \over 32} [\gamma_a , ~ \gamma_b]
~ [\gamma_c , ~ \gamma_d] ~ R^{abcd} ~.
\ee
Contracting $\gamma_a \gamma_b \gamma_c \gamma_d$ with the identity $R^{abcd} +
R^{acdb} + R^{adbc} = 0$, it follows that
\be
[\gamma_a , ~ \gamma_b] ~ [\gamma_c , ~ \gamma_d] ~ R^{abcd} =
4 \gamma_a \gamma_b \gamma_c \gamma_d ~ R^{abcd} = -8 \gamma_a \gamma_b ~ R^{ab} =
-8 R ~,
\ee
which is solely expressed in terms of the scalar Ricci curvature. Then, one arrives at
the celebrated Lichnerowicz formula for the square of the Dirac operator
written in terms of the Laplacian operator acting on spinors, \cite{lichn},
\be
(i \gamma^{\mu} D_{\mu})^2 = - D_{\mu} D^{\mu} + {1 \over 4} R ~.
\label{finalrel}
\ee
$D^2$ is often called Bochner Laplacian to distinguish it from the Laplace--Beltrami
operator $\nabla^2$ acting on scalars.

\subsection{Geodesic interval and Synge--DeWitt tensors}

Another item that is needed for the computations (in Appendix C and in the main text)
is the {\em geodesic interval} \cite{witt1}, which is defined as one half the
square of the distance along the geodesic between any two points $x$ and $x^{\prime}$,
\be
\sigma (x, x^{\prime}) = {1 \over 2} \left(\int_x^{x^{\prime}} ds \right)^2 .
\ee
This notion also appeared in the literature under the name "world function" (in Synge's
textbook \cite{synge}) and plays important role in the theory of Green's functions on
Riemannian manifolds with metric $G_{\mu \nu}$, including manifolds of indefinite metric
\cite{hada}.

The geodesic interval is a symmetric function
$\sigma (x, x^{\prime})$ of $x$ and $x^{\prime}$ that transforms as biscalar, i.e.,
as a scalar separately at $x$ and $x^{\prime}$. It satisfies the differential
equation
\be
\sigma (x, x^{\prime}) = {1 \over 2} (\nabla_{\mu} \sigma) (\nabla^{\mu} \sigma) =
{1 \over 2} (\nabla_{\mu}^{\prime} \sigma) (\nabla^{\prime \mu} \sigma)
\label{geoda}
\ee
with boundary conditions
\be
\sigma (x, x) = 0 ~, ~~~~~~ \lim_{x \rightarrow x^{\prime}} \nabla_{\mu}
\sigma (x, x^{\prime}) = 0 =
\lim_{x \rightarrow x^{\prime}} \nabla_{\mu}^{\prime} \sigma (x, x^{\prime})
\label{boundari}
\ee
and it obeys the important relation
\be
\lim_{x \rightarrow x^{\prime}} \nabla_{\mu} \nabla_{\nu} \sigma (x, x^{\prime}) =
- \lim_{x \rightarrow x^{\prime}} \nabla_{\mu} \nabla_{\nu}^{\prime} \sigma
(x, x^{\prime}) = G_{\mu \nu} ~.
\label{boundara}
\ee
As such, it provides a natural generalization of $\sigma (x, x^{\prime}) =
(x-x^{\prime})^2 / 2$ from flat to curved space. In a general Riemannian manifold
$\sigma (x, x^{\prime})$ is not single--valued, except when $x$ and $x^{\prime}$
are sufficiently close to each other.

The geodesic interval $\sigma$ is a $C^{\infty}$ function that can be defined through
the first order non--linear equation \eqn{geoda}. This has the interpretation of the
Hamilton--Jacobi equation of a mechanical system whose evolution is described by the
geodesic equation on a given manifold. It allows to express the invariant
delta--function on a Riemannian manifold (used in the computation of Green's functions)
as
\be
{1 \over \sqrt{{\rm det} G}} \delta (x - x^{\prime}) = {1 \over \sqrt{{\rm det} G}}
\int {d^4 k \over (2 \pi)^4} e^{i k_{\mu} \nabla^{\mu} \sigma (x, x^{\prime})} ~,
\ee
which, in turn, shows that ${\rm exp}(i k_{\mu} \nabla^{\mu} \sigma (x, x^{\prime}))$
provide the analogue of plane waves in curved space. In flat Euclidean space, we have
$\nabla^{\mu} \sigma (x, x^{\prime}) = x^{\mu} - x^{\prime \mu}$ and one recovers
the ordinary plane waves, which depend linearly on the Cartesian coordinates in the
exponent. In general, $\sigma$ can not be determined in closed form, but in the
physical applications it is sufficient to know its expansion as $x$ and $x^{\prime}$
come close to each other.

The coincidence limit $x \rightarrow x^{\prime}$ of succussive covariant derivatives
of the geodesic interval, viz. $\nabla_{\mu} \nabla_{\nu} \cdots \nabla_{\kappa} \sigma$,
are called Synge--DeWitt
tensors and they are of the utmost importance in computing asymptotic expansions of
Green functions of partial differential operators. They are also very important for the
computation of the gravitational contribution to the axial anomaly. Adopting the
short--hand notation
\be
[\nabla_{\mu} \nabla_{\nu} \cdots \nabla_{\kappa} \sigma] = \lim_{x \rightarrow
x^{\prime}} \nabla_{\mu} \nabla_{\nu} \cdots \nabla_{\kappa} \sigma (x, x^{\prime}) ~,
\ee
which is commonly used in the literature, we immediately have following results for the
lowest rank Synge--DeWitt tensors, restating \eqn{boundari} and \eqn{boundara},
\be
[\sigma] = 0 ~, ~~~~~~ [\nabla_{\mu} \sigma] = 0 ~, ~~~~~~ [\nabla_{\mu} \nabla_{\nu}
\sigma] = G_{\mu \nu} ~.
\label{synd1}
\ee
For the purposes of the present work we also need the third and fourth rank tensors,
which turn out to be (in the absence of torsion)
\be
[\nabla_{\mu} \nabla_{\nu} \nabla_{\kappa} \sigma] = 0
\label{synd2}
\ee
and
\be
[\nabla_{\mu} \nabla_{\nu} \nabla_{\kappa} \nabla_{\lambda} \sigma] = {1 \over 3}
\left(R_{\mu \kappa \lambda \nu} + R_{\mu \lambda \kappa \nu} \right)
\label{synd3}
\ee
written in terms of the Riemann curvature tensor. All higher rank Synge--DeWitt
tensors involve products and derivatives of the curvature tensor but their expressions
are quite cumbersome to present here (the sixth rank tensor involves about forty different
terms and the situation becomes increasingly more complex for tensors of even higher
rank); fortunately, we do not need to go beyond rank four.

The computations are based on equation \eqn{geoda} by taking successive derivatives and
applying the Ricci and Bianchi identities as many times as necessary, depending on the
rank of the tensor. In this fashion one obtains recursive relations for the
Synge--DeWitt tensors that can be iterated starting from the basic ones \eqn{synd1}.

For example, differentiating \eqn{geoda} three times yields
\ba
\nabla_{\mu} \nabla_{\nu} \nabla_{\kappa} \sigma & = & (\nabla_{\mu} \nabla_{\nu}
\nabla^{\lambda} \sigma) (\nabla_{\kappa} \nabla_{\lambda} \sigma) +
(\nabla_{\nu} \nabla^{\lambda} \sigma) (\nabla_{\mu} \nabla_{\kappa} \nabla_{\lambda}
\sigma) + \nonumber\\
& & (\nabla_{\mu} \nabla^{\lambda} \sigma) (\nabla_{\nu} \nabla_{\kappa}
\nabla_{\lambda} \sigma) + (\nabla^{\lambda} \sigma) (\nabla_{\mu} \nabla_{\nu}
\nabla_{\kappa} \nabla_{\lambda} \sigma)
\ea
that results into the following equation (term--by--term) in the coincidence limit,
\be
[\nabla_{\mu} \nabla_{\nu} \nabla_{\kappa} \sigma] = [\nabla_{\mu} \nabla_{\nu}
\nabla_{\kappa} \sigma] + [\nabla_{\mu} \nabla_{\kappa} \nabla_{\nu} \sigma] +
[\nabla_{\nu} \nabla_{\kappa} \nabla_{\mu} \sigma] ~.
\ee
Simple rearrangement yields the equivalent expression
\ba
-2 [\nabla_{\mu} \nabla_{\nu} \nabla_{\kappa} \sigma] & = & \left([\nabla_{\mu}
\nabla_{\kappa} \nabla_{\nu} \sigma] - [\nabla_{\mu} \nabla_{\nu} \nabla_{\kappa}
\sigma]\right) + \left([\nabla_{\nu} \nabla_{\kappa} \nabla_{\mu} \sigma] -
[\nabla_{\nu} \nabla_{\mu} \nabla_{\kappa} \sigma] \right) + \nonumber\\
& & \left([\nabla_{\nu} \nabla_{\mu} \nabla_{\kappa} \sigma] - [\nabla_{\mu}
\nabla_{\nu} \nabla_{\kappa} \sigma] \right) .
\ea
The first pair of terms on the right--hand side gives zero because $\sigma$ is
a scalar function of $x$ and $[\nabla_{\kappa} , ~ \nabla_{\nu}] \sigma = 0$.
The second pair of terms also gives zero because $[\nabla_{\kappa} , ~ \nabla_{\mu}]
\sigma = 0$. Finally, the third pair of terms equal $[[\nabla_{\nu} , ~ \nabla_{\mu}]
(\nabla_{\kappa} \sigma)] = [{R^{\lambda}}_{\kappa \mu \nu} (\nabla_{\lambda}
\sigma)]$ and vanishes because $[\nabla_{\lambda} \sigma] = 0$. This proves the
relation \eqn{synd2}.
The proof of relation \eqn{synd3} proceeds along similar lines, but it is more lengthy
to present it here in detail (note, however, that our sign convention for the Riemann 
curvature tensor is opposite to the one used in \cite{witt1}). 

We also refer the interested reader to the textbook 
\cite{fulling} (see, in particular, chapter 8) for a comprehensive account of the 
recursive calculations of Synge--DeWitt tensors and their use in the asymptotic 
expansion of Green's functions on curved space--time. 

\newpage

\section{Axial anomalies for relativistic fermions}
\setcounter{equation}{0}

In this appendix we summarize the computation of the axial anomaly for
a massless Dirac fermion coupled to a background gauge and metric field,
using the elegant method of Fujikawa \cite{fuji, suzuki} in Euclidean space. Thus,
we will consider the regularized quantity
\ba
A(t, x) & = & \lim_{M \rightarrow \infty} \Big[\sum_n \varphi_n^{\dagger}
(t, x) \gamma_5 ~ e^{- (\lambda_n / M)^2} \varphi_n (t, x) \Big] \nonumber\\
& = & \lim_{M \rightarrow \infty} \Big[\sum_n \varphi_n^{\dagger}
(t, x) \gamma_5 ~ e^{- (i\gamma^{\mu} D_{\mu} / M)^2} \varphi_n (t, x) \Big]
\ea
defined in terms of the eigen--values and orthonormal eigen--functions of the
interacting Dirac operator, $(i\gamma^{\mu} D_{\mu}) \varphi_n = \lambda_n
\varphi_n$, and evaluate it. The axial anomaly simply reads $\nabla_{\mu}
J_5^{\mu} = 2 A$. An alternative (closely related) method of computing the anomaly
is provided by the point--split method of the fermion bilinear in the axial
current (see, for instance, \cite{nielsen} and references therein).

Our purpose is to set up the notation and present the main formulae and tricks
needed for the computation of the gauge and gravitational contributions to the
axial anomaly. It also simplifies the presentation in the main text of the
paper without making detours. Then, the intermediate steps and results of the
relativistic theory can be easily compared to the corresponding contributions
received by the divergence of the axial current in the non--relativistic theory
of Lifshitz fermions with anisotropic scaling $z=3$ in four space--time dimensions.

\subsection{Gauge field contribution to the anomaly}

Minimal coupling to an Abelian or non--Abelian gauge field $A_{\mu}$ in flat
space--time amounts to considering the covariant derivative $D_{\mu} =
\partial_{\mu} - iA_{\mu}$ so that $[D_{\mu} , ~ D_{\nu}] = - i F_{\mu \nu}$ is
expressed in terms of the corresponding field strength. Then, we have the
following relation,
\be
(i\gamma^{\mu} D_{\mu})^2 = - D_{\mu} D^{\mu} + {i \over 4} [\gamma^{\mu} ,
~ \gamma^{\nu}] ~ F_{\mu \nu} ~.
\label{commut}
\ee
To evaluate the anomaly it is convenient to use the eigen--values and functions
of the free Dirac operator $i\gamma^{\mu} \partial_{\mu}$, which are plane waves.
This will enable to extract the gauge field dependence of $A(t, x)$ using a basis
that has no gauge field dependence by itself.
Then, in terms of this basis, the primitive form of the local anomaly takes the
form
\be
A(t, x) = \lim_{M \rightarrow \infty} {\rm Tr} \int {d^4 k \over (2\pi)^4} ~
\gamma_5 ~ e^{-i k_{\mu} x^{\mu}} e^{- (i\gamma^{\mu} D_{\mu} / M)^2}
e^{i k_{\mu} x^{\mu}}
\label{rikoko}
\ee
taking the trace on anything available on the right--hand side.
Combining this  with relation \eqn{commut} yields immediately
\ba
& & \lim_{M \rightarrow \infty} {\rm Tr} \int {d^4 k \over (2\pi)^4} ~
\gamma_5 ~ {\rm exp} \Big[{1 \over M^2} \left((D_{\mu} + i k_{\mu})(D^{\mu} +
i k^{\mu}) - {i \over 4} [\gamma^{\mu} , ~ \gamma^{\nu}] F_{\mu \nu}\right)
\Big] = \nonumber\\
& & \lim_{M \rightarrow \infty} M^4 ~ {\rm Tr} \int {d^4 k \over (2\pi)^4} ~
e^{-k_{\mu} k^{\mu}}  \gamma_5 ~ {\rm exp} \left({1 \over M^2} D_{\mu} D^{\mu}
+ {2i \over M} k_{\mu} D^{\mu} - {i \over 4M^2} [\gamma^{\mu} , ~ \gamma^{\nu}]
F_{\mu \nu}\right) . \nonumber
\ea
In the last step $k_{\mu}$ has been rescaled to $Mk_{\mu}$.

It is now clear, by
expanding the exponential in power series of $1/M$, that only terms up to order
$1/M^4$ will survive in the limit $M \rightarrow \infty$. Furthermore, by the
trace identities of the Dirac gamma--matrices with $\gamma_5$, a non--vanishing
contribution requires having at least four such gamma--matrices and, therefore,
the only term that survives at the end is
\be
A(t, x) = {1 \over 2!} \int {d^4 k \over (2\pi)^4} ~ e^{-k_{\mu} k^{\mu}}
{\rm Tr} \left(\gamma_5 ~ {i \over 4} [\gamma^{\mu} , ~ \gamma^{\nu}] ~
{i \over 4} [\gamma^{\kappa} , ~ \gamma^{\lambda}] F_{\mu \nu}
F_{\kappa \lambda} \right) .
\ee
Recall at this point the identity ${\rm Tr}(\gamma_5 \gamma^{\mu} \gamma^{\nu}
\gamma^{\kappa} \gamma^{\lambda}) = -4 \epsilon^{\mu \nu \kappa \lambda}$ for
Euclidean gamma--matrices,
which takes care of their trace. Also, the integral over the Euclidean space momenta
$k$ is easily performed and yields immediately
\be
\int {d^4 k \over (2\pi)^4} ~ e^{-k_{\mu} k^{\mu}} = {2 \pi^2 \over (2 \pi)^4}
\int_0^{\infty} dk ~ k^3 e^{-k^2} = {1 \over 16 \pi^2}
\label{natio}
\ee
using spherical coordinates and ${\rm vol}(S^3) = 2\pi^2$ for the unit sphere.

Putting all together, the final result for the gauge field contribution to the
axial anomaly, which equals $2A$, is
\be
\partial_{\mu} J_5^{\mu} = - {1 \over 16 \pi^2} \epsilon^{\mu \nu \kappa
\lambda} {\rm Tr} (F_{\mu \nu} F_{\kappa \lambda}) \equiv -{1 \over 8 \pi^2}
{\rm Tr} (F \wedge F) ~.
\ee

\subsection{Metric field contribution to the anomaly}

Next, we summarize the computation of the gravitational
contribution to the axial anomaly for Dirac fermions, following \cite{fuji, suzuki}.
The primitive form of the anomaly is
\ba
A(t, x) & = & \lim_{M \rightarrow \infty} \sum_n \varphi_n^{\dagger}
(t, x) \gamma_5 ~ e^{- (i\gamma^{\mu} D_{\mu} / M)^2} \varphi_n (t, x)
\nonumber\\
& = & \lim_{M \rightarrow \infty} \lim_{x \rightarrow x^{\prime}}
{\rm Tr} \int {d^4 k \over (2\pi)^4} ~
\gamma_5 ~ e^{- (i\gamma^{\mu} D_{\mu} / M)^2}
e^{i k_{\mu} \nabla^{\mu} \sigma (x, x^{\prime})}
\ea
by passing to the plane wave analogue basis that naturally involves the
geodesic interval\footnote{On may choose to work in the linearized approximation
setting $G_{\mu \nu} \simeq \delta_{\mu \nu} + h_{\mu \nu}$ and assuming that
$h_{\mu \nu}$ and its derivatives are sufficiently small. Then, ${\rm exp}(i k_{\mu}
\nabla^{\mu} \sigma (x, x^{\prime})) \simeq {\rm exp}(i k_{\mu} (x^{\mu} -x^{\prime \mu}))$
and equation \eqn{rikoko} follows again by letting $x \rightarrow x^{\prime}$ after
passing ${\rm exp}(-ik_{\mu} x^{\prime \mu})$ to the left of the operator (which
only depends on $x$ and not $x^{\prime}$), as in
the gauge field case. However, the resulting gravitational contribution to the
anomaly will be only valid at the linear level, as in the original work \cite{salam},
and its non--linear generalization will be left open. Making use of the geodesic
interval remedies the situation, as it provides the means to compute the anomaly at
the full non--linear level.}.
Here, the derivatives of the Dirac operator are taken with respect to $x$
(and not $x^{\prime}$), whereas the limit $x \rightarrow x^{\prime}$ should be
taken after computing the action of the operators on ${\rm exp}(i k_{\mu}
\nabla^{\mu} \sigma (x, x^{\prime}))$. Then, taking into account Lichnerowicz's
formula \eqn{finalrel}, we obtain the following intermediate result for $A(t, x)$,
\ba
& & \lim_{M \rightarrow \infty} \lim_{x \rightarrow x^{\prime}}
{\rm Tr} \int {d^4 k \over (2\pi)^4} ~ e^{i k_{\mu} \nabla^{\mu} \sigma (x, x^{\prime})} ~
\gamma_5 ~ {\rm exp} \Big[{ 1 \over M^2} \left((D_{\mu}
+ i \Delta_{\mu}) (D^{\mu} + i \Delta^{\mu}) - {1 \over 4} R \right) \Big]
= \nonumber\\
& & \lim_{M \rightarrow \infty} M^4 ~
{\rm Tr} \int {d^4 k \over (2\pi)^4} ~ e^{-k_{\mu} k^{\mu}} \gamma_5 ~
{\rm exp} \left({1 \over M^2} D_{\mu} D^{\mu}
+ {2i \over M} k_{\mu} D^{\mu} - {1 \over 4M^2} R \right) .
\ea
The first line follows by acting with the operator $(i\gamma^{\mu} D_{\mu})^2$,
in which case $D_{\mu}$ is simply
replaced by $D_{\mu} + i \Delta_{\mu}$ with $\Delta_{\mu} (x, x^{\prime}) =
k_{\nu} \nabla_{\mu} \nabla^{\nu} \sigma (x, x^{\prime})$ (the curved space analogue
of $D_{\mu} + i k_{\mu}$), whereas the second line follows by taking the limit
$x \rightarrow x^{\prime}$, in which case $\Delta_{\mu} (x, x^{\prime})$ is replaced
by $k_{\mu}$ thanks to relation \eqn{boundara} for the rank two Synge--DeWitt tensor,
and by rescaling $k_{\mu}$ to $Mk_{\mu}$. Although the rank three Synge--DeWitt tensor
\eqn{synd2} vanishes, so that $[\nabla^{\mu} \Delta_{\mu}] = 0$, all higher rank tensors
do not vanish and can in principle give non--zero contribution in the coincidence limit
$x \rightarrow x^{\prime}$. Careful treatment of the problem show that none of these
additional terms have sufficient number of gamma matrices to survive under the trace,
and, therefore, we make no error by taking the coincidence limit right from the
beginning, as shown above.

The rest of the computation proceeds as in the gauge field case by expanding the
exponential in power series of $1/M$ and taking the limit $M \rightarrow \infty$.
The only difference is that the curvature term in the exponential contains no
gamma--matrices and so the non--vanishing contribution to the trace comes from a
different combination of terms of order $1/M^4$. Dropping all lower order terms,
which have zero trace, and all higher order terms, which vanish identically as
$M \rightarrow \infty$, we obtain the following contribution to the axial anomaly,
\ba
A(t, x) & = & {\rm Tr} \int {d^4 k \over (2\pi)^4} ~ e^{-k_{\mu} k^{\mu}} \gamma_5
\left({1 \over 2} (D^2)^2 + {2 \over 3} k_{\mu} k_{\nu}
k_{\kappa} k_{\lambda} D^{\mu} D^{\nu} D^{\kappa} D^{\lambda} \right. \nonumber\\
& & \left. - {2 \over 3} k_{\mu} k_{\nu} \left(D^2 D^{\mu} D^{\nu}
+ D^{\mu} D^2 D^{\nu} + D^{\mu} D^{\nu} D^2 \right) \right) ,
\label{onebefo}
\ea
setting here for convenience $D^2 = D_{\mu} D^{\mu}$.

The integrals over the Euclidean space momenta $k$ are easily performed by
introducing a unit vector with components $\hat{k}_{\mu}$, so that $k_{\mu}
= k \hat{k}_{\mu}$ with $k_{\mu} k^{\mu} = k^2$. Alternatively, we may choose to
work with tangent space--time indices $\hat{k}_{\mu} = {e^a}_{\mu} \hat{k}_a$,
since $k_{\mu} k^{\mu} = k_a k^a$, and note the identities
\be
\int {d^4 k \over (2\pi)^4} ~ e^{-k^2} ~ k^2 ~ \hat{k}_a \hat{k}_b =
{1 \over 4} \delta_{a b} {2 \pi^2 \over (2 \pi)^4}
\int_0^{\infty} dk ~ k^5 e^{-k^2} = {1 \over 32 \pi^2} \delta_{a b}
\ee
and
\be
\int {d^4 k \over (2\pi)^4} ~ e^{-k^2} ~ k^4 ~ \hat{k}_a \hat{k}_b
\hat{k}_c \hat{k}_d = {1 \over 64 \pi^2} \left(\delta_{a b}
\delta_{c d} + \delta_{a c} \delta_{b d} +
\delta_{a d} \delta_{b c} \right) ,
\ee
which are manifestly symmetric in all permutations of the indices.
These integrals are evaluated using spherical coordinates in the space of
four--momenta, choosing $\hat{k}_1 = {\rm sin} \theta {\rm sin} \phi {\rm sin} \psi$,
$\hat{k}_2 = {\rm sin} \theta {\rm sin} \phi {\rm cos} \psi$, $\hat{k}_3 =
{\rm sin} \theta {\rm cos} \phi$ and $\hat{k}_4 = {\rm cos} \theta$ in terms
of the corresponding angular variables that range as $0 \leq \theta \leq \pi$,
$0 \leq \phi \leq \pi$ and $0 \leq \psi \leq 2\pi$. The integration measure
over the Euler angles in momentum space is ${\rm sin}^2 \theta {\rm sin} \phi
d\theta d\phi d\psi$ so that the volume of the unit sphere is ${\rm vol}(S^3) =
2\pi^2$. Taking these results together with the integral \eqn{natio},
one finds that equation \eqn{onebefo} yields in steps
\ba
A(t, x) & = & {1 \over 96 \pi^2} {\rm Tr} \left(\gamma_5 (D_{\mu} D_{\nu} D^{\mu}
D^{\nu} - D_{\mu} D^2 D^{\mu}) \right) \nonumber\\
& = & {1 \over 192 \pi^2} {\rm Tr} \left(\gamma_5
[D_{\mu} , ~ D_{\nu}] [D^{\mu} , ~ D^{\nu}] \right) \nonumber\\
& = & {1 \over 64 \cdot 192 \pi^2} {\rm Tr} \left(\gamma_5 ~
[\gamma_a , ~ \gamma_b] ~ [\gamma_c  , ~ \gamma_d] \right)
{R^{ab}}_{\mu \nu} R^{cd \mu \nu} \nonumber\\
& = & -{1 \over 768 \pi^2} \epsilon_{abcd} {R^{ab}}_{\mu \nu} R^{cd \mu \nu} ~,
\ea
using ${\rm Tr}(\gamma_5 \gamma_a \gamma_b \gamma_c \gamma_d) = - 4 \epsilon_{abcd}$
for the tangent space--time indices. We may pass to the fully covariant totally
anti--symmetric symbol defined as $\epsilon_{\mu \nu \kappa \lambda} =
{e^a}_{\mu} {e^b}_{\nu} {e^c}_{\kappa} {e^d}_{\lambda} \epsilon_{abcd}$.

Thus, the final result for the metric field contribution to the axial anomaly,
which equals $2A$, is
\be
\nabla_{\mu} J_5^{\mu} = - {1 \over 384 \pi^2} \epsilon^{\mu \nu \kappa
\lambda} {R^{\rho \sigma}}_{\mu \nu} R_{\rho \sigma \kappa \lambda}
\equiv -{1 \over 192 \pi^2} {\rm Tr} (R \wedge R)
\ee
written in terms of the corresponding Ricci curvature two--form.

Here and in the main text $\epsilon^{\mu \nu \kappa \lambda}$ is defined as 
$\epsilon^{0123} = 1 / \sqrt{{\rm det} G}$ and $\epsilon_{0123} = \sqrt{{\rm det} G}$, 
unless stated otherwise.

\section{Dirac operator on Berger spheres}
\setcounter{equation}{0}

In this appendix we consider the Dirac operator defined on three--dimensional spheres $S^3$
that are endowed with $SU(2)$ homogeneous metrics. Following \cite{hitchin} (but see also
\cite{bar}), we solve the eigen--value problem of $i \gamma^i D_i$ ($i = 1, 2, 3$) acting on
two--component spinors,
\be
i \gamma^i D_i \Psi (x) = \zeta ~ \Psi (x) ~ ; ~~~~~~
\Psi (x) = \left(\begin{array}{c}
\psi_1 \\
 \\
\psi_2 \\
\end{array} \right) ,
\ee
in the special case of Berger spheres that exhibit an additional $U(1)$ isometry.
This provides a tractable example where all computations can be carried out to the end
and the results are relevant for evaluating the integrated form of the axial anomaly on
gravitational instanton backgrounds with $SU(2) \times U(1)$ isometry group. Along the
way, we point out the limitations to compute the spectrum in closed form for totally
anisotropic geometries on $S^3$.

It turns out that there can be zero modes (often called harmonic spinors \cite{lichn})
whose number depends on the metric on $S^3$ and can grow without bound. This should
be contrasted to results for the space of harmonic spinors on a two--dimensional compact
Riemann surface of genus $g$. In that case, the number of zero modes can not grow very
large as it is bounded by the topological invariant $g+1$;
also, on a Riemann surface there are several inequivalent spin structures, whereas in
the case we are considering here there is only one spin structure. The presence of
harmonic spinors allows for level crossing under the spectral flow and it is reflected in
the $\eta$--invariant of the Dirac operator. All these problems will be examined in
detail in the following.

\subsection{Spectrum of Dirac operator}

Let us consider the Bianchi IX class of three--dimensional metrics, as in section 3.3,
\be
ds^2 = \gamma_1 (\sigma^1)^2 + \gamma_2 (\sigma^2)^2 + \gamma_3 (\sigma^3)^2 ~,
\label{bianchi9}
\ee
using the left--invariant one--forms $\sigma^I$ of the group $SU(2)$ that satisfy the
relations
\be
d\sigma^I + {1 \over 2} {\epsilon^I}_{JK} \sigma^J \wedge \sigma^K = 0
\ee
and coefficients $\gamma_1$, $\gamma_2$, $\gamma_3$ that are, in general, unequal.
There is a dual basis of vector fields $f_I$ associated to $\sigma^I$ with $<f_I , \sigma^J>
= \delta_I^J$ that satisfy the commutation relations
\be
[f_I , ~ f_J] = - {\epsilon_{IJ}}^K f_K
\label{algcommu}
\ee
and represent the Killing vector fields of this particular class of geometries. Both
$\sigma^I$ and $f_I$ can be written in terms of three angular coordinates $x^i =
(\theta, \varphi, \psi)$, but the explicit expressions will not be needed here.

We also introduce the orthonormal coframe $e^I = \sqrt{\gamma_I} ~ \sigma^I$
associated to the dreibeins ${e^I}_i$ with $e^I = {e^I}_i dx^i$ and the orthonormal
frame $E_I = f_I / \sqrt{\gamma_I}$ associated to the inverse dreibeins ${E_I}^i$
with $E_I = {E_I}^i \partial_i$, which allow us to define the Dirac operator on $S^3$
as
\be
i \gamma^i D_i = i \gamma^I {E_I}^i \left(\partial_i + {1 \over 8}
[\gamma_J , ~ \gamma_K] {\omega^{JK}}_i \right) .
\label{thredira}
\ee
The formulae are identical with those found in Appendix B1. The only difference is that
the space--time index $\mu$ is now replaced by the space index $i$ and the Dirac matrices
$\gamma^a$ are now replaced $\gamma^I$; the latter satisfy the anti--commutation
relations $[\gamma^I , ~ \gamma^J] = 2 \delta^{IJ}$ (since $I$ and $J$ are tangent space
indices) and they should be identified with the Pauli matrices \eqn{paulim} as
$\gamma^1 = \sigma_1$, $\gamma^2 = - \sigma_2$ and $\gamma^3 = \sigma_3$ (the Pauli
matrices should not be confused with the one--forms $\sigma^I$ of $SU(2)$ used above).
Then, $\epsilon_{IJK} [\gamma^I , ~ \gamma^J]/8$ provide the spinorial
representation of the Lie algebra generators $f_I$ in \eqn{algcommu}.

Explicit calculation shows that the connection one--forms of Bianchi IX metrics are
\ba
{\omega^1}_2 & = & {1 \over 2 \sqrt{\gamma_1 \gamma_2 \gamma_3}}
(-\gamma_1 - \gamma_2 + \gamma_3) e^3 ~, \\
{\omega^1}_3 & = & {1 \over 2 \sqrt{\gamma_1 \gamma_2 \gamma_3}}
(\gamma_1 - \gamma_2 + \gamma_3) e^2 ~, \\
{\omega^2}_3 & = & {1 \over 2 \sqrt{\gamma_1 \gamma_2 \gamma_3}}
(\gamma_1 - \gamma_2 - \gamma_3) e^2 ~.
\ea
Using the fact that the square of the Pauli matrices is the identity matrix, it
follows that the spin connection term of the Dirac equation \eqn{thredira} equals to
\be
{i \over 8} \gamma^I {E_I}^i [\gamma_J , ~ \gamma_K] {\omega^{JK}}_i =
{1 \over 4 \sqrt{\gamma_1 \gamma_2 \gamma_3}} (\gamma_1 + \gamma_2 +
\gamma_3) \mathbb{1} ~.
\ee
Then, the Dirac operator for all homogeneous metrics on $S^3$ is the following
first order differential operator\footnote{Comparison with \cite{hitchin} (see, in particular,
p. 28) can be readily made by setting $f_1 = - e_2 /2$, $f_2 = - e_3 /2$, $f_3 = - e_1 /2$
and $\gamma_1 = \lambda_2 / 4$, $\gamma_2 = \lambda_3 / 4$, $\gamma_3 =
\lambda_1 / 4$ in Hitchin's basis of the Lie algebra generators.}
written in terms of the three vector fields $f_I$,
\be
i \gamma^i D_i = \left(\begin{array}{ccc}
i f_3/\sqrt{\gamma_3} &  & i f_1/\sqrt{\gamma_1} - f_2/\sqrt{\gamma_2} \\
  &  &   \\
i f_1/\sqrt{\gamma_1} + f_2/\sqrt{\gamma_2} &  & - i f_3/\sqrt{\gamma_3}
\end{array} \right) +
{1 \over 4 \sqrt{\gamma_1 \gamma_2 \gamma_3}} (\gamma_1 + \gamma_2 + \gamma_3) ~.
\ee

The eigen--values of this operator can only be found explicitly when there is an
additional $U(1)$ isometry associated to axially symmetric geometries.
Letting any two coefficients of the metric become equal, say $\gamma_1 = \gamma_2
\equiv \gamma$ (all other choices are mutually related by a $\mathbb{Z}_3$ permutation
symmetry of the three principal axes of $S^3$), we obtain
\be
ds^2 = \gamma \Big[(\sigma^1)^2 + (\sigma^2)^2 + \delta^2 (\sigma^3)^2 \Big] ~,
\ee
where $\delta^2 = \gamma_3/\gamma$ measures the degree of anisotropy of the associated
Berger sphere. Since the Dirac operator scales uniformly as $\sqrt{\gamma}$, we may
drop the conformal factor $\gamma$ and only consider the Dirac operator on $S^3$ with
line element $(\sigma^1)^2 + (\sigma^2)^2 + \delta^2 (\sigma^3)^2$,
\be
i \gamma^i D_i = \left(\begin{array}{ccc}
i f_3/\delta &  & i f_1 - f_2 \\
  &  &   \\
i f_1 + f_2 &  & - i f_3/\delta
\end{array} \right) +
{1 \over 4 \delta} (\delta^2 + 2) ~.
\label{isodiraco}
\ee

The following simple but very important observation allows to determine the
spectrum of the Dirac operator \eqn{isodiraco} on Berger spheres. Writing
$i \gamma^i D_i = {\cal Q} + (\delta^2 + 2)/4\delta$ and setting $f_{\pm} = if_1 \mp f_2$,
we note that the operator ${\cal Q}^2 + {\cal Q}/\delta$ takes the diagonal form
\be
{\cal Q}^2 + {1 \over \delta} {\cal Q} = \left(\begin{array}{ccc}
f_+ f_- - (f_3^2 - if_3)/ \delta^2 &  & 0 \\
     &  &   \\
0    &  & f_- f_+ - (f_3^2 + if_3)/ \delta^2
\end{array} \right)
\label{diaago}
\ee
after making essential use of the Lie algebra commutation relations \eqn{algcommu}.
The eigen--value problem of the resulting second order operators
\be
{\cal F}_{\pm} = f_{\pm} f_{\mp} - {1 \over \delta^2} (f_3^2 \mp if_3)
\ee
can be easily solved by noting that they commute with the Laplace operator
$\Delta$ acting on functions on the Berger sphere (actually, $2 \Delta =
{\cal F}_+ + {\cal F}_-$). As such, they all have a common system of eigen--states
provided by the appropriate representation of $SU(2)$ associated to $f_i$.
If we were using unitary irreducible representations of $SU(2)$ with definite spin
$j = 0, 1/2, 1, 3/2, \cdots $, the action of the generators on the base states
$\{ |j, m> ; ~ -j \leq m \leq j \}$ would be
$f_{\pm} f_{\mp} |j, m> = (j \pm m) (j \mp m +1) |j, m>$ and
$f_3 |j, m> = im |j, m>$, giving
\be
{\cal F}_{\pm} |j, m> = \Big[(j \pm m)(j \mp m +1) + {1 \over \delta^2} m (m \mp 1)
\Big] ~ |j, m> ~.
\ee
However, one has to be slightly more careful in applying group theoretical methods
here as the appropriate representation is not irreducible.

The problem at hand is closely related to the determination of the energy levels and
eigen--states of a three--dimensional rigid rotor with Hamiltonian
\be
H = {1 \over 2I_1} L_1^2 + {1 \over 2I_2} L_2^2 + {1 \over 2I_3} L_3^2
\ee
in the special case\footnote{The more general case $I_1 \neq I_2 \neq I_3$ corresponds
to a totally anisotropic Bianchi IX metric, which can not be solved exactly in closed
form.} that the moments of inertia are $I_1 = I_2 \neq I_3$. We simply have to set
$f_I = iL_I$ to turn \eqn{algcommu} into the commutation relations of the quantum
angular momenta operators (with Planck's constant normalized to $1$). But there is
an important difference between the two problems related to the representations of
the $SU(2)$ symmetry group. The angular momenta
$L_I$ are differential operators of two angular variables $(\theta , \varphi)$ that
parametrize $S^2$ in a spherical coordinate system in $\mathbb{R}^3$. As a result,
the eigen--states of the rigid rotor are spherical harmonics provided by $|j, m>$
with integer quantum number $j = 0, 1, 2, \cdots$. Setting $I_1 = I_2 \equiv I$, we
find the energy eigen--values $[j(j+1) - m^2]/2I + m^2/2I_3$ which appear with
multiplicity $1$ if $m=0$ and multiplicity $2$ if $m \neq 0$. On the other hand,
the Killing vector fields $f_I$ are differential operators of three angular variables
$(\theta, \varphi , \psi)$ that parametrize $S^3$, and, as such, they are the
generators of $SU(2)$ in the {\em regular} representation\footnote{The regular
representation of a group $G$ is defined as $U(g) f(x) = f(g x)$ on the space of all
square integrable functions on the group manifold endowed with the Haar measure. It satisfies
$U(g_1) U(g_2) = U(g_1 g_2)$ and it is unitary. For $SU(2)$, the local coordinates
on the group manifold are $x^i = (\theta , \varphi , \psi)$ and $gx$ is the left action
of the group generated by the corresponding left--invariant vector fields $f_I$.
Similar considerations apply to the regular representation $U(g) f(x) = f(x g^{-1})$ acting
on the other side, which also satisfies $U(g_1) U(g_2) = U(g_1 g_2)$ and it is unitary.}.
Fourier analysis on $SU(2)$ decomposes the regular representation into direct sum
of unitary irreducible representations with all $j = 0, 1/2, 1, 3/2, \cdots$ including
half--integer ones. Furthermore, each irreducible components appears with multiplicity
equal to its dimension $2j+1$ (see, for instance, \cite{carme} and in particular
p.57--60). This, in turn, determines the multiplicity of eigen--values
of the Laplace operator $\Delta$, and, hence, of the operators ${\cal F}_{\pm}$ which
commute with it on Berger spheres. The resulting spectrum and multiplicities differ
from those of rigid rotor despite of the similarities.

Returning back to expression \eqn{diaago}, we find that the eigen--values $z$ of the
operator ${\cal Q}$ must satisfy the quadratic equation
\be
z^2 + {z \over \delta} = (j \pm m)(j \mp m +1) + {1 \over \delta^2} m (m \mp 1)
\ee
for all $j = 0, 1/2, 1, 3/2, \cdots$. Since the two sign options are mutually related by
sending $m$ to $-m$, we may choose the lower sign (without loss of generality) and obtain
the eigen--values of ${\cal Q}$
\be
z = {1 \over 2 \delta} \left(- 1 \pm \sqrt{4 \delta^2 (j-m) (j+m+1) + (2m+1)^2} \right)
\label{mirmaka}
\ee
that appear with multiplicity $2j+1$. Both sign options are admissible in \eqn{mirmaka}
as long as $j \neq m$, in which case both components of the eigen--spinor $\Psi (x)$ are
non--zero. The choice $j=m$ is special and requires separate investigation. In this case,
the eigen--value problem for ${\cal Q}$ is solved directly as
\be
{\cal Q} \left(\begin{array}{c}
|j, -j> \\
 \\
0 \\
\end{array} \right) = {j \over \delta}
\left(\begin{array}{c}
|j, -j> \\
 \\
0 \\
\end{array} \right)
\ee
Comparison with \eqn{mirmaka} shows that only the upper sign yields the correct
eigen--value $z = j/ \delta$. Likewise, one finds
\be
{\cal Q} \left(\begin{array}{c}
0 \\
 \\
|j, j> \\
\end{array} \right) = {j \over \delta}
\left(\begin{array}{c}
0 \\
 \\
|j, j> \\
\end{array} \right)
\ee
showing that the eigen--value $z = j/ \delta$ arises with multiplicity $2(2j+1)$.

Setting $p = j + m + 1$ and $q = j - m$, we find that the spectrum of the Dirac operator
\eqn{isodiraco} on Berger spheres is given by
\be
\zeta_{\pm} = {\delta \over 4} \pm {1 \over 2 \delta} \sqrt{4 \delta^2 pq + (p-q)^2} ~;
~~~~~~ (p, q) \in \mathbb{N}
\label{sasha1}
\ee
as long as $q \neq 0$. The multiplicity of these eigen--values is $p+q = 2j+1$ for each
pair $(p, q)$. The additional eigen--values arising for $q = 0$ are
\be
\zeta_0 = {\delta \over 4} + {p \over 2 \delta} ~; ~~~~~~ p \in \mathbb{N}
\label{sasha2}
\ee
and their multiplicity is $2p = 2(2j+1)$. Thus, the complete spectrum of the Dirac
operator\footnote{Compared to \cite{hitchin}, there is a discrepancy by an overall factor
of $2$ in the allowed values of $\zeta$ that results from the different normalization of
the $SU(2)$ one--forms $\sigma^I$, but this is irrelevant.}
is provided by $\zeta_{\pm}$ and $\zeta_0$ forming a two--dimensional state lattice labeled
by the quantum numbers $(p, q)$. The eigen--values $\zeta_+$ and $\zeta_0$ are positive
definite, whereas $\zeta_-$ can also become zero or negative by tuning the parameters.
Finally, we note that the eigen--values $\zeta_{\pm}$ simplify for $p=q$ as they depend
linearly on $\delta$, i.e., $\zeta_{\pm} = \delta /4 \pm p$.

\noindent
{\bf Special cases:}
Three special cases arise from the general discussion that are worth noting. They reduce
the general problem to simpler ones that are better known in the literature and can be
used for comparison and consistency checks. These special cases are also important for
understanding the behavior of the eigen--values under spectral flow as $\delta$ varies
from $0$ to $\infty$.

First, we consider $\delta =1$ that corresponds to the homogeneous and isotropic constant
curvature metric on $S^3$. It can be easily seen by little combinatorics that the spectrum
of the Dirac operator becomes
\be
\delta = 1 : ~~~~~~ \zeta = \pm {1 \over 4} (2n+1) ~, ~~~~ n \in \mathbb{N} ~~~ {\rm with} ~
{\rm multiplicity} ~~ n(n+1) ~.
\label{cameen1}
\ee
In this case the eigen--values are equally distributed to positive and negative values
and there is no spectral asymmetry in the problem.

Next, we consider $\delta = 0$ that corresponds to a
fully squashed three--sphere along its third principal axis. The resulting configuration
is non--singular as it exhibits no curvature singularity. It can be related to a
two--dimensional sphere endowed with constant curvature metric. Then, the eigen--values
$\zeta_0$ become infinite and the same is true for $\zeta_{\pm}$, which tend to
$\pm \infty$, respectively, as long as $p \neq q$. Finite spectrum arises only when
$p=q \equiv n$ and it is given by
\be
\delta = 0 : ~~~~~~ \zeta = \pm n ~, ~~~~ n \in \mathbb{N} ~~~ {\rm with} ~
{\rm multiplicity} ~~ 2n ~.
\label{cameen2}
\ee
This is the spectrum of the Dirac operator for the round $S^2$ which is equally
distributed to positive and negative values and shows no spectral asymmetry.
However, there is spectral asymmetry in the three--dimensional problem, because
$\zeta_0$ provide the excess of positive eigen--values that become infinite as
$S^3$ collapses to $S^2$ (it explains later why the $\eta$--invariant of the
three--dimensional Dirac operator does not vanish for $\delta = 0$).

Finally, we consider the limiting case $\delta \rightarrow \infty$ that yields
a singular metric on $S^3$ as the sphere becomes infinitely stretched along its third
principal axis. Then, all eigen--values become infinite exhibiting a universal
behavior,
\be
\delta \rightarrow \infty : ~~~~~~ \zeta \simeq {\delta \over 4} ~~~~ {\rm with} ~
{\rm infinite} ~ {\rm multiplicity} ~.
\label{asymptf}
\ee

\subsection{Harmonic spinors and level crossing}

Harmonic spinors are normalizable zero modes of the Dirac operator. They appear when
$\zeta_- = 0$, i.e., when there are positive integer solutions $(p, q)$ to the equation
\be
\delta^2 = 2 \sqrt{4 pq \delta^2 + (p-q)^2} ~.
\label{spevalh}
\ee
Thus, a necessary condition for the existence of harmonic spinors is the following bound
for the anisotropy parameter $\delta$ of Berger spheres,
\be
\delta \geq 4 ~,
\label{lowbouna}
\ee
in which case the metric becomes sufficiently stretched to allow for negative curvature.
This is in accord to Lichnerowich's formula \eqn{finalrel} for the square of the Dirac operator,
which now takes the form $(i \gamma^i D_i)^2 = - D_i D^i + R/4$. It tells us that if the Ricci
scalar curvature $R$ is positive definite the Dirac operator will not admit zero modes. Zero
curvature is also excluded in this case, as it corresponds to the special value $\delta = 2$
that falls well below the bound \eqn{lowbouna} for the existence of harmonic spinors.

Note that the harmonic spinors and their multiplicities are invariant under rescaling of
the metric on $S^3$. They have clear geometrical interpretation but no topological origin.
Thus, for each pair of positive integers $(p, q)$ that solve equation \eqn{spevalh}
there is an associated zero mode with multiplicity $p+q$. The simplest ones arise for $p=q$
in which case $\delta = 4p$. For $p=q=1$, in particular, we have $\delta = 4$ and the space
of harmonic spinors has dimension $2$. More generally, for $\delta = 4p$ the dimension of the
space of harmonic spinors is at least $2p$. An example that exhibits two different solutions arises
for $\delta = 260$ in which case $p=q=65$ and $p=528$, $q=8$ are both admissible choices
and the multiplicity of zero modes turns out to be $2 \cdot 65 + 2 \cdot (528 + 8)
= 1202$ in total (noting that $p=8$, $q = 528$ has also to be counted).
Conversely, the values of $\delta$ that allow for harmonic spinors are
provided by $8pq + 2 \sqrt{16p^2 q^2 + (p-q)^2}$ for each pair of integers $(p, q)$, without
restriction, and they are generally irrational numbers in the interval $[4, \infty)$.

The possibility to have zero modes implies level crossing under the spectral
flow, which is very important for the integrated form of the axial anomaly. Let us examine how
the spectrum of the Dirac operator changes as $\delta$ varies. Note that the eigen--values
$\zeta_+$ and $\zeta_0$ given by \eqn{sasha1} and \eqn{sasha2}, respectively, are always positive
irrespective of $\delta$. On the other hand, $\zeta_-$ are all negative for $\delta < 4$, but for
$\delta \geq 4$ some become positive or zero. The first level crossing occurs when the anisotropy
parameter passes the critical value $\delta = 4$. More level crossings occur when $\delta$
becomes bigger and bigger. As soon as an eigen--value crosses zero, it will stay positive
for all higher values of $\delta$ and never cross back to negative values, according to the
general formulae. Eventually, as $\delta \rightarrow \infty$, all eigen--values turn positive
leaving no negative ones behind.

Fig.6 provides the spectral flow of the eigen--values $\zeta_-$ as functions of $\delta$ for
the lowest values of the positive integers $p$ and $q$. Obviously, all eigen--values depend
linearly upon $\delta$ whenever $p=q$; these lines pass from the values $-1, -2, \dots$ when
$\delta = 0$ (they are the negative eigen--values of the Dirac operator on $S^2$), whereas all
other spectral lines are pushed to infinite as $S^3$ collapses to $S^2$. Here, we plot the
results for $(p, q) = (1, 1)$, $(1, 2)$, $(1,3)$, $(2, 2)$, $(1, 4)$ and $(2, 3)$ for which
level crossing occurs at $\delta = 4$, $5.67$, $6.95$, $8$, $8.03$ and $9.80$, respectively,
where harmonic spinors make their appearance. Similar plots arise for all higher values
of $(p, q)$. Obviously, the lines $(p, q)$ and $(q, p)$ are identical. Note that all curves
reach asymptotically the slope line $\zeta = \delta /4$ thanks
to the universal behavior \eqn{asymptf} as $\delta \rightarrow \infty$. We have omitted
from the figure the spectral flow of the eigen--values $\zeta_+$ and $\zeta_0$ as they never
cross zero, and, hence, they are uninteresting for our purposes.

\begin{figure}[h]
\centering
\epsfig{file=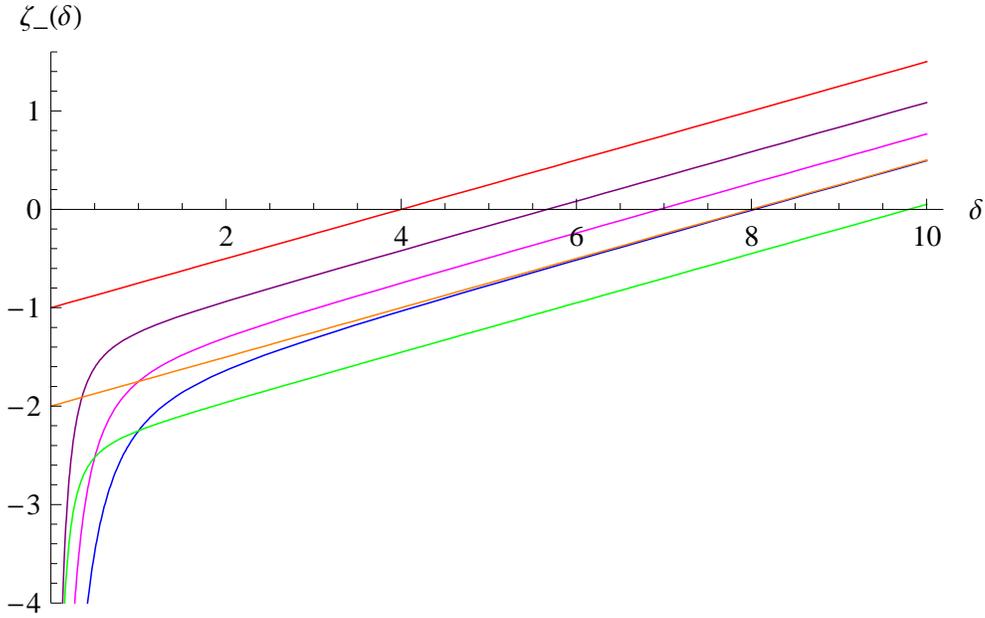,width=13cm}
\caption{Level crossing of the modes $\zeta_-(\delta)$ from negative to positive values}
\end{figure}

\subsection{The $\eta$--invariant}

The results for the spectrum can been used to compute the $\eta$--invariant of the
Dirac operator on Berger spheres. This invariance measures the asymmetry between the number of
positive and negative modes in the spectrum, and, as such, it vanishes when $\delta =1$.
This is seen by computing the $\eta$--invariant in closed form. The result is very
important for evaluating the index of the Dirac operator on gravitational instanton
backgrounds by the Atiyah--Patodi--Singer index theorem \cite{APS} (see also \cite{gilka, melrose}).

First, we examine the case $\delta < 4$ which is the simplest as it does not
involve level crossing. There are infinitely many positive modes $\zeta_+$ and $\zeta_0$ and
infinitely many negative modes $\zeta_-$. Considering the difference of the spectral
Riemann zeta--function between the positive and negative eigen--values, excluding zero
modes,
\be
\eta_{{\rm D}} (s) = \sum_{{\rm eigenvalues}} ({\rm sign} ~ \zeta) ~ |\zeta|^{-s} ~,
\ee
allows to define the $\eta$--invariant of the Dirac operator by appropriate analytic
continuation at $s=0$ as
\be
\eta_{{\rm D}} = \eta_{{\rm D}} (0) ~.
\ee
The $\eta$--invariant is ill--defined without the zeta--function regularization as it is the
difference of two infinite numbers.
Note that under rescaling of the metric the spectrum of the Dirac operator scales uniformly
but $\eta_{{\rm D}}$ remains invariant since it is evaluated at $s=0$.

Then, taking into account the multiplicities of the various eigen--values, we have explicitly
that
\be
\eta_{{\rm D}} (s) =
\sum_{p, q >0} (p+q) \Big[\left({\delta^2 \over 2} + X \right)^{-s}
- \left(- {\delta^2 \over 2} + X \right)^{-s}\Big]
+ \sum_{p>0} 2p \left({\delta^2 \over 2} + p \right)^{-s}
\ee
setting
\be
X = \sqrt{4 \delta^2 pq + (p-q)^2}
\ee
for notational convenience.
The first term refers to the contribution of $\zeta_+$, the second to $\zeta_-$ and the third
to $\zeta_0$. The first two terms are a bit tricky to evaluate, whereas the third one is more
straightforward. We will present the essential tricks and details since they are also needed
in the main text to evaluate more complicated sums.

The computations are most easily done by expanding all terms in power series of
$\delta$. For the last one we have
\be
\left({\delta^2 \over 2} + p \right)^{-s} = {1 \over p^s} \left(1 - {s \delta^2 \over 2p} +
{s(s+1) \delta^4 \over 8 p^2} + \cdots \right)
\ee
that yields the following expansion in terms of Riemann zeta--functions,
\be
I_0 (s) = \sum_{p>0} 2p \left({\delta^2 \over 2} + p \right)^{-s} = 2 \zeta (s-1) - \delta^2 s
\zeta (s) + {1 \over 4} \delta^4 s(s+1) \zeta (s+1) + \cdots ~.
\ee
The higher order terms are irrelevant because they vanish at $s=0$ and they are omitted
(they contain the factor $s\zeta (s+n)$ for all integers $n \geq 2$).
Since $\zeta (-1) = -1/12$, $\zeta (0) = -1/2$ and $s\zeta(s+1)$ equals\footnote{Recall that
$\zeta(s+1)$ has a simple pole at $s=0$ with residue $1$ since the Riemann zeta--function
satisfies the functional relation
$\zeta (s+1) = 2^{s+1} \pi^{s} {\rm sin} ({\pi (s+1) / 2}) \Gamma (-s) \zeta (-s)$.
The pole structure at $s=0$ follows from the identity $(-s)\Gamma (-s) = \Gamma (1-s)$
and $\Gamma (1) = 0$.}
$1$ at $s=0$, we obtain immediately
\be
I_0 (0) = -{1 \over 6} + {\delta^4 \over 4} ~.
\ee

Next, we consider the power series expansion
\be
\left(\pm {\delta^2 \over 2} + X \right)^{-s} = {1 \over X^s} \left(1 \mp {s \delta^2
\over 2X} + {s(s+1) \delta^4 \over 8X^2} \mp {s(s+1)(s+2) \delta^6 \over 48X^3} + \cdots \right)
\ee
that yields
\ba
& & I_+(s) - I_-(s) = \sum_{p, q >0} (p+q) \Big[\left({\delta^2 \over 2} + X \right)^{-s}
- \left(- {\delta^2 \over 2} + X \right)^{-s}\Big] \nonumber\\
&  & ~~~~~~~~ = - \delta^2 s f\left({s+1 \over 2}\right) -
{1 \over 24} \delta^6 s(s+1)(s+2) f\left({s+3 \over 2}\right) + \cdots ~,
\label{iamah}
\ea
where
\be
f(s) = \sum_{p, q >0} {p+q \over X^{2s}} = \sum_{p, q >0} {p+q \over [4 \delta^2 pq +
(p-q)^2]^s} ~.
\ee
The function $f(s)$ converges absolutely for ${\rm Re}s > 3/2$ on the complex $s$--plane,
and, therefore, the omitted terms in \eqn{iamah} are irrelevant as they vanish at $s=0$.
Furthermore, the meromorphic continuation of $f(s)$ to the entire complex plane has only
simple poles whose residues at $s = 1/2$ and $s=3/2$ turn out to be
\be
{\rm res} f(s)|_{s=1/2} = {\delta^2 -1 \over 6} ~, ~~~~~~
{\rm res} f(s)|_{s=3/2} = {1 \over 2 \delta^2} ~.
\label{resqar}
\ee
They provide the values of $(s/2)f((s+1)/2)$ and $(s/2)f((s+3)/2)$ at $s=0$. All other
terms appearing in the expression \eqn{iamah} vanish at $s=0$, and, thus, we obtain
\be
I_+(0) - I_-(0) = {\delta^2 \over 3} - {5 \delta^4 \over 12} ~.
\ee
The mathematical properties of $f(s)$ can be studied using the Euler--Maclaurin summation
formula\footnote{If a function $f(x)$ has continuous derivatives on the interval $[0, ~ N]$,
then the sum of $f(n)$ with $n \in \mathbb{N}$ is converted into integral as follows,
\be
\sum_{n=0}^N f(n) = \int_0^N dx ~ f(x) + {1 \over 2} \left(f(N) + f(0)\right) +
\sum_{k=1}^p {B_{2k} \over (2k)!} \left(f^{(2k-1)}(N) - f^{(2k-1)}(0) \right) + {\cal R} ~.
\nonumber
\ee
Here, $B_{2k}$ are the Bernoulli numbers ($B_2 = 1/6$, $B_4 = -1/30$, $B_6 = 1/42$, etc) and
$f^{(2k-1)}(x)$ denotes the $(2k-1)$--th derivative of the function $f(x)$. The remainder
(error) term ${\cal R}$ is normally small for suitable value of $p$ and it is bounded as
\be
| {\cal R} | \leq {2 \zeta(2p) \over (2\pi)^{2p}} \int_0^N dx ~ |f^{(2p)} (x) | ~.
\ee}
that converts infinite sums into integrals -- in this case of two variables (see, for
instance, \cite{tenen}).

There is no point in presenting the technical details of the residue formulae since they
are somewhat involved. We refer the interested reader to the comprehensive report
\cite{habel} for the relevant computations as well as certain generalizations to higher
dimensional Berger spheres.
We only note here that when $\delta =1$ the computation of the residues can be done by
elementary methods. In this case $f(s)$ simplifies and it is expressed in terms of Riemann's
zeta--function as follows,
\be
f(s) = \sum_{p, q > 0} {1 \over (p+q)^{2s-1}} = \zeta (2s-2) - \zeta (2s-1) ~.
\ee
Then, it can be easily seen that $f(1/2) = \zeta (-1) - \zeta (0) = 5/12$ is finite, and,
hence, there is no pole, i.e., the residue at $s=1/2$ vanishes. Evaluating $f(3/2) = \zeta (1)
- \zeta (2)$ shows that there a simple pole at $s=3/2$ given by the pole of $\zeta (2s+1)$ at
$s=0$ whose residue is $1/2$ (it is convenient to compare $f(s+3/2)$ with $\zeta (2s+1)$ at
$s=0$). These results agree with the general formulae \eqn{resqar}
for $\delta =1$.

Finally, putting all these together, we add up the individual terms $I_0(0) + I_+(0) - I_-(0)$
and arrive at the following result for the $\eta$--invariant of the Dirac operator on Berger
spheres with anisotropy parameter $\delta < 4$, as in \cite{hitchin},
\be
\eta_{{\rm D}} = - {1 \over 6} (\delta^2 - 1)^2 ~.
\label{myetai}
\ee
It can be readily seen that $\eta_{{\rm D}}$ vanishes for the isotropic metric $\delta =1$,
as expected. Also, in the fully squashed limit $\delta = 0$, $\eta_{{\rm D}} = - 1/6$ counting
the asymmetry of positive and negative modes that become infinite\footnote{For $\delta = 0$ this
asymmetry is accounted by the excess positive modes $\zeta_0$. Their zeta--function
regularization yields the simple expression $\eta_{\rm D} (s) = 2 \zeta (s-1)$. Then,
setting $s=0$, we find by elementary means that $\eta_{\rm D} = 2 \zeta (-1) = -1/6$ as
required on general grounds.} (the finite eigen--values are equally distributed to positive
and negative values).

At this point, it is instructive to compare the $\eta$--invariant with the gravitational
Chern--Simons action \cite{simons}
\be
W_{\rm CS} = {1 \over 2} \int_{\Sigma_3} {\rm Tr} \left(\omega \wedge d \omega + {2 \over 3}
\omega \wedge \omega \wedge \omega \right)
\ee
evaluated for the class of homogeneous metric on $S^3$. Using the Bianchi IX form
of the metric \eqn{bianchi9} and the expressions for the connection one--forms ${\omega^I}_J$
derived earlier, we find (up to a sign convention that depends on the orientation of $S^3$)
\be
W_{\rm CS} = 16 \pi^2 \Big[1 + {1 \over 2 \gamma_1 \gamma_2 \gamma_3}
(\gamma_1 + \gamma_2 - \gamma_3)(\gamma_1 - \gamma_2 + \gamma_3)
(\gamma_1 - \gamma_2 - \gamma_3) \Big] ~.
\ee
The result is scale invariant, as required on general grounds. Then, choosing $\gamma_1 =
\gamma_2 = 1$ and $\gamma_3 = \delta^2$, we obtain immediately
\be
{1 \over 48 \pi^2} W_{\rm CS} ({\rm Berger}) = {1 \over 6} (\delta^4 - 2 \delta^2 + 2) =
- \eta_{{\rm D}} + {1 \over 6}
\label{relaspr}
\ee
that relates $W_{\rm CS}$ and $\eta_{{\rm D}}$ for Berger spheres. This is a special case of a
more general relation among the Chern--Simons action and the $\eta$--invariant found in the
context of the Atiyah--Patodi--Singer theory \cite{APS} (for a compact three--manifold without
boundaries it suffices to consider a spin four--manifold bounded by it, extend the product
metric near the boundary and integrate the Pontrjagin form relative to this metric).

Next, we examine the case $\delta > 4$ that involves level crossing. Every time an eigenvalue
$\zeta_-$ crosses from negative to positive values, $\eta_{{\rm D}}$ jumps by $2$ to
$\eta_{{\rm D}} + 2$, since the spectral asymmetry, as calculated for $\delta < 4$, changes
by $2$. Let us denote by
\be
{\cal C}(\delta_{\rm c}) = \{(p, q) \in \mathbb{N}^2 ; ~~ \delta_{\rm c}^2 = 2 \sqrt{4pq
\delta_{\rm c}^2 + (p-q)^2} \}
\ee
the set of positive integers $(p, q)$ that account for harmonic spinors at each one of the
special values $\delta_{\rm c} < \delta$. This set has at least two elements associated to the
two harmonic spinors at $\delta_{\rm c} = 4$. Since the zero modes at $\delta_{\rm c}$ have
multiplicity $p+q$, the quantity
\be
S(\delta) = \sum_{\delta_{\rm c} < \delta} \Big[\sum_{(p, q) \in {\cal C}(\delta_{\rm c})}
(p+q) \Big]
\label{oumarta2}
\ee
provides the total number of modes that crossed from negative to positive values as $\delta$
was varying from $\delta < 4$ to any given value $\delta > 4$.
Thus, for $\delta > 4$, the $\eta$--invariant of the Dirac operator for Berger spheres is
shifted by twice the number of these modes and equals
\be
\eta_{{\rm D}} = - {1 \over 6} (\delta^2 - 1)^2 + 2 S(\delta)
\label{geneta}
\ee
provided that $\delta$ does not assume the special values \eqn{spevalh}.

On the other hand, if $\delta$ is tuned to any one of the values \eqn{spevalh}, including
$\delta = 4$, the corresponding zero modes have to be removed from the spectrum while
computing the sum $\eta_{{\rm D}} (s)$. We also have to add $1$ for each one of these modes
(i.e., shift $\eta_{{\rm D}}$ by the dimension of the space of harmonic spinors arising at
those values of $\delta$), and, consequently, the net effect is zero. Thus, there is no change
of $\eta_{{\rm D}}$ in this case. Changes can only occur when there is level crossing.

Summarizing, the formula \eqn{geneta} for the $\eta$--invariant is the most general one
and it is valid for all values of the anisotropy parameter $\delta$ of Berger spheres.
The result is finite for all finite values of $\delta$ and it only becomes infinite in
the extreme limit $\delta \rightarrow \infty$ where the geometry becomes singular. In the
latter case there are no negative modes left behind and the spectral asymmetry is infinite,
as expected.


\newpage


\begin{thebibliography}{99}
\bibitem{bell}
J.S. Bell and R. Jackiw, ``A PCAC puzzle: $\pi^0 \rightarrow \gamma \gamma$ in
the $\sigma$--model", Nuovo Cimento \underline{60A} (1969) 47.
\bibitem{adler}
S.L. Adler, ``Axial--vector vertex in spinor electrodynamics", Phys. Rev.
\underline{177} (1969) 2426.
\bibitem{jackiw}
R. Jackiw, ``Axial anomaly", Int. J. Mod. Phys. \underline{A25} (2010) 659.
\bibitem{bardeen}
W.A. Bardeen and A.R. White eds., {\em Symposium on Anomalies, Geometry and Topology},
World Scientific, Singapore, 1986.
\bibitem{fuji}
K. Fujikawa, ``Path--integral measure for gauge--invariant fermion theories",
Phys. Rev. Lett. \underline{42} (1979) 1195; ``Path integral for gauge theories
with fermions", Phys. Rev. \underline{D21} (1980) 2848; Erratum--ibid.
\underline{22} (1980) 1499.
\bibitem{suzuki}
K. Fujikawa and H, Suzuki, {\em Path Integrals and Quantum Anomalies}, Oxford
University Press, Oxford, 2004.
\bibitem{salam}
R. Delbourgo and A. Salam, ``The gravitational correction to PCAC", Phys.
Lett. \underline{B40} (1972) 381.
\bibitem{eguchi1}
T. Eguchi and P.G.O. Freund, ``Quantum gravity and world topology", Phys.
Rev. Lett. \underline{37} (1976) 1251.
\bibitem{nielsen}
N.K. Nielsen, H. R\"omer and B. Schroer, ``Classical anomalies and local version
of the Atiyah--Singer theorem", Phys. Lett. \underline{B70} (1977) 445;
``Anomalous currents in curved space", Nucl. Phys. \underline{B136} (1978) 475.
\bibitem{gaume}
L. Alvarez--Gaum\'e and E. Witten, ``Gravitational anomalies", Nucl. Phys.
\underline{B234} (1983) 269.
\bibitem{hooft}
G. 't Hooft, ``Symmetry breaking through Bell--Jackiw anomalies", Phys. Rev. Lett.
\underline{37} (1976) 8; ``Computation of the quantum effects due to a
four--dimensional pseudoparticle", Phys. Rev. \underline{D14} (1976) 3432;
Erratum--ibid. \underline{D18} (1978) 2199.
\bibitem{AS}
M.F. Ayiyah and I.M. Singer, ``Index of elliptic operators", Ann. Math.
\underline{87} (1968) 485; ibid. \underline{87} (1968) 531; ibid.
\underline{87} (1968) 546; ibid. \underline{93} (1971) 119; ibid.
\underline{93} (1971) 139.
\bibitem{gilka}
P.B. Gilkey, {\em Invariance Theory, the Heat Equation, and the Atiyah--Singer
Index Theorem}, Studies in Advanced Mathematics, second edition, CRC Press,
Boca Raton, 1995.
\bibitem{ezra}
N. Berline, E. Getzler and M. Vergne, {\em Heat Kernels and Dirac Operators},
Springer--Verlag, Berlin, 2004.
\bibitem{K3a}
N. Hitchin, ``Compact four--dimensional Einstein manifolds", J. Diff. Geom.
\underline{9} (1974) 435.
\bibitem{K3b}
S.T. Yau, ``On Calabi's conjecture and some new results on algebraic geometry",
Proc. Nat. Acad. Sci. USA \underline{74} (1977) 1798.
\bibitem{lichn}
A. Lichnerowicz, ``Spineurs harmoniques", C.R. Acad. Sci. \underline{A257}
(1963) 7.
\bibitem{gibbons}
G.W. Gibbons and C.N. Pope, ``The positive action conjecture and asymptotically
Euclidean metrics in quantum gravity", Commun. Math. Phys. \underline{66}
(1979) 267.
\bibitem{APS}
M.F. Atiyah, V.K. Patodi and I.M. Singer, ``Spectral asymmetry and Riemannian
geometry", Proc. Camb. Philos. Soc. \underline{77} (1975) 43; ibid.
\underline{78} (1975) 405; ibid. \underline{79} (1976) 71.
\bibitem{melrose}
R. Melrose, {\em The Atiyah--Padoti--Singer Index Theorem}, A.K. Peters Ltd,
Wellesley, 1993.
\bibitem{hawking}
S.W. Hawking, ``Gravitational instantons", Phys. Lett. \underline{A60} (1977) 81.
\bibitem{extra1}
T. Eguchi and A.J. Hanson, ``Asymptotically flat self--dual solutions to Euclidean
gravity", Phys. Lett. \underline{B74} (1978) 249.
\bibitem{romer}
H. R\"omer and B. Schroer, ``Fractional winding numbers and surface effects",
Phys. Lett. \underline{B71} (1977) 182.
\bibitem{eguchi2}
T. Eguchi, P.B. Gilkey and A.J. Hanson, ``Topological invariants and absence of
an axial anomaly for a Euclidean Taub--NUT (Newman--Unti--Tamburino) metric",
Phys. Rev. \underline{D17} (1978) 423.
\bibitem{pope}
C.N. Pope, ``Axial--vector anomalies and the index theorem in charged
Schwarzschild and Taub--NUT spaces", Nucl. Phys. \underline{B141} (1978) 432.
\bibitem{extra2}
G.W. Gibbons, C.N. Pope and H. R\"omer, ``Index theorem boundary terms for
gravitational instantons", Nucl. Phys. \underline{B157} (1979) 377.
\bibitem{extra3}
S.W. Hawking and C.N. Pope, ``Symmetry breaking by instantons in supergravity",
Nucl. Phys. \underline{B146} (1978) 381.
\bibitem{extra4}
A.J. Hanson and H. R\"omer, ``Gravitational instanton contribution to spin $3/2$
axial anomaly", Phys. Lett. \underline{B80} (1978) 58.
\bibitem{eguchi3}
T. Eguchi, P.B. Gilkey and A.J. Hanson, ``Gravitation, gauge theories and differential
geometry", Phys. Rept. \underline{66} (1980) 213.
\bibitem{horava1}
P. Ho\v{r}ava, ``Membranes at quantum criticality", JHEP \underline{0903} (2009) 020
[arXiv:0812.4287[hep--th]].
\bibitem{horava2}
P. Ho\v{r}ava, ``Quantum gravity at a Lifshitz point", Phys. Rev. \underline{D79}
(2009) 084008 [arXiv:0901.3775[hep--th]].
\bibitem{blas}
D. Blas, O. Pujolas and S. Sibiryakov, ``Models of non--relativistic quantum gravity:
The good, the bad and the healthy", preprint [arXiv:1007.3503 [hep--th]].
\bibitem{thomas}
T. Sotiriou, ``Ho\v{r}ava--Lifshitz gravity: a status report", J. Phys. Conf. Ser.
\underline{283} (2011) 012034 [arXiv:1010.3218[hep--th]].
\bibitem{hennea}
M. Henneaux, A. Kleinschmidt and G. Lucena Gomez, ``A dynamical inconsistency of
Ho\v{r}ava gravity", Phys. Rev. \underline{D81} (2010) 064002 [arXiv:0912.0399[hep--th]].
\bibitem{horava3}
P. Ho\v{r}ava and C.M. Melby--Thompson, ``General covariance in quantum gravity at
a Lifshitz point", Phys. Rev. \underline{D82} (2010) 064027 [arXiv:1007.2410[hep--th]];
P. Ho\v{r}ava, ``General covariance in gravity at a Lifshitz point", preprint
[arXiv:1101.1081[hep--th]].
\bibitem{wadia}
A. Dhar, G. Mandal and S.R. Wadia, ``Asymptotically free four--fermi theory in 4
dimensions at the $z=3$ Lifshitz--like fixed point", Phys. Rev. \underline{D80}
(2009) 105018 [arXiv:0905.2928[hep--th]].
\bibitem{horava4}
P. Ho\v{r}ava, ``Quantum criticality and Yang--Mills gauge theory", Phys. Lett.
\underline{B694} (2010) 172 [arXiv:0811.2217[hep--th]].
\bibitem{BBLP}
I. Bakas, F. Bourliot, D. L\"ust and M. Petropoulos, ``Geometric flows in
Ho\v{r}ava--Lifshitz gravity", JHEP \underline{1004} (2010) 131
[arXiv:1002.0062[hep--th]]; I. Bakas, ``Gradient flows and instantons at a
Lifshitz point", J. Phys. Conf. Ser. \underline{283} (2011) 012004
[arXiv:1009.6173[hep--th]].
\bibitem{hitchin}
N.J. Hitchin, ``Harmonic spinors", Adv. Math. \underline{14} (1974) 1.
\bibitem{bar}
C. B\"ar, ``The Dirac operator on homogeneous spaces and its spectrum on
3--dimensional lens spaces", Arch. Math. \underline{59} (1992) 65;
``Metrics with harmonic spinors", Geom. Funct. Anal. \underline{6}
(1996) 899.
\bibitem{witt1}
B.S. DeWitt, ``Dynamical Theory of Groups and Fields" in {\em Relativity, Groups and
Topology}, eds. C. DeWitt and B. DeWitt, Gordon and Breach, New York,
1965; ``Quantum theory of gravity. III. Applications of the covariant
theory", Phys. Rev. \underline{162} (1967) 1239.
\bibitem{synge}
J.L. Synge, {\em Relativity: The General Theory}, North Holland, Amsterdam, 1960.
\bibitem{hada}
J. Hadamard, {\em Lectures on Cauchy's Problem in Linear Partial Differential
Equations}, Yale University Press, New Haven, 1923.
\bibitem{adm}
C.W. Misner, K.S. Thorne and J.A. Wheeler, {\em Gravitation}, Freeman, San Francisco,
1973.
\bibitem{hagen}
C.R. Hagen, ``Scale and conformal transformations in Galilean--covariant field
theory", Phys. Rev. \underline{D5} (1972) 377.
\bibitem{julian}
J. Schwinger, ``On gauge invariance and vacuum polarization", Phys. Rev.
\underline{82} (1951) 664.
\bibitem{zuber}
C. Itzykson and J.--B. Zuber, {\em Quantum Field Theory}, McGraw--Hill, New York, 1980.
\bibitem{xue}
W. Xue, ``Non--relativistic supersymmetry", preprint [arXiv:1008.5102[hep--th]].
\bibitem{witt2}
B.S. DeWitt, ``Quantum theory of gravity. I. The canonical theory", Phys. Rev.
\underline{160} (1967) 1113.
\bibitem{DJT}
S. Deser, R. Jackiw and S. Templeton, ``Three-Dimensional Massive Gauge Theories",
Phys. Rev. Lett.  \underline{48} (1982) 975; ``Topologically massive gauge theories",
Annals Phys. \underline{140} (1982) 372; Erratum--ibid. \underline{185} (1988) 406.
\bibitem{simons}
S.S. Chern and J. Simons, ``Characteristic forms and geometric invariants", Ann. Math.
\underline{99} (1974) 48.
\bibitem{witten}
J.H. Horne and E. Witten, ``Conformal gravity in three dimensions as a gauge theory",
Phys. Rev. Lett. \underline{62} (1989) 501.
\bibitem{styau}
H.D. Cao, B. Chow, S.--C. Chu and S.--T. Yau eds., {\em Collected Papers on Ricci
Flow}, Series in Geometry and Physics, International Press, Somerville, 2003.
\bibitem{topping}
P. Topping, {\em Lectures on the Ricci Flow}, London Mathematical Society Lecture
Note Series, Cambridge University Press, Cambridge, 2006.
\bibitem{lni}
B. Chow, P. Lu and L. Ni, {\em Hamilton's Ricci Flow}, Graduate Texts in Mathematics,
American Mathematical Society, Science Press, Providence, 2006.
\bibitem{cotton}
A.U.O. Kisisel, O. Sarioglu and B. Tekin, ``Cotton flow", Class. Quant. Grav.
\underline{25} (2008) 165019 [arXiv:0803.1603[hep--th]].
\bibitem{park}
M.--I. Park, ``Ho\v{r}ava gravity and gravitons at a conformal point", preprint
[arXiv:0910.5117[hep--th]].
\bibitem{japi}
R. Jackiw, ``A nonrelativistic chiral soliton in one dimension", J. Nonlin. Math. Phys.
\underline{4} (1997) 261 [arXiv:hep--th/9611185].
\bibitem{seminara}
L. Griguolo and D. Seminara, ``Chiral solitons from dimensional reduction of Chern--Simons
gauged non--linear Schr\"odinger equation: classical and quantum aspects", Nucl. Phys.
\underline{B516} (1998) 467 [arXiv:hep--th/9709075].
\bibitem{mixmast1}
C.W. Misner, ``Mixmaster universe", Phys. Rev. Lett. \underline{22} (1969) 1071;
``Quantum cosmology 1", Phys. Rev. \underline{186} (1969) 1319.
\bibitem{mixmast2}
I. Bakas, F. Bourliot, D. L\"ust and M. Petropoulos, ``Mixmaster universe in
Ho\v{r}ava--Lifshitz gravity", Class. Quant. Grav. \underline{27} (2010) 045013
[arXiv:0911.2665 [hep-th]].
\bibitem{jackson}
J. Isenberg and M. Jackson, ``Ricci flow of locally homogeneous geometries on
closed manifolds", J. Diff. Geom. \underline{35} (1992) 723.
\bibitem{rama1}
R. Jackiw, ``Quantum meaning of classical field theory", Rev. Mod. Phys.
\underline{49} (1977) 681.
\bibitem{rama2}
R. Rajaraman, {\em Solitons and Instantons}, North Holland, Amsterdam, 1987.
\bibitem{callias}
C. Callias, ``Axial anomalies and index theorems on open spaces", Commun. Math.
Phys. \underline{62} (1978) 213; R. Bott and R. Seeley, ``Some remarks on the paper
of Callias", Commun. Math. Phys. \underline{62} (1978) 235.
\bibitem{rubakov}
V. Rubakov, {\em Classical Theory of Gauge Fields}, Princeton University Press,
Princeton, 2002.
\bibitem{calca}
G. Calcagni, ``Cosmology of the Lifshitz universe", JHEP \underline{0909} (2009)
112 [arXiv:0904.0829[hep--th]].
\bibitem{kiritsis}
E. Kiritsis and G. Kofinas, ``Ho\v{r}ava--Lifshitz cosmology", Nucl. Phys.
\underline{B821} (2009) 467 [arXiv:0904.1334[hep--th]].
\bibitem{brande}
R. Brandenberger, ``Matter bounce in Ho\v{r}ava--Lifshitz cosmology", Phys. Rev.
\underline{D80} (2009) 043516 [arXiv:0904.2835[hep--th]].
\bibitem{ergin}
R. Percacci and E. Sezgin, ``One loop beta functions in topologically massive
gravity", Class. Quant. Grav. \underline{27} (2010) 155009
[arXiv:1002.2640[hep--th]].
\bibitem{nigel}
N.J. Hitchin, ``Einstein metrics and the eta--invariant", Bolletino U.M.I. (7)
\underline{11--B} Suppl. Fasc. 2 (1997) 95.
\bibitem{giha}
G.W. Gibbons and S.W. Hawking, ``Action integrals and partition functions in
quantum gravity", Phys. Rev. \underline{D15} (1977) 2752.
\bibitem{york}
J. York, ``Role of conformal three--geometry in the dynamics of gravitation", 
Phys. Rev. Lett. \underline{28} (1972) 1082. 
\bibitem{roman1}
I.I. Cotaescu, S. Moroianu and M. Visinescu, ``Quantum anomalies for generalized
Euclidean Taub--NUT metrics", J. Phys. \underline{A38} (2005) 7005;
S. Moroianu and M. Visinescu, ``$L^2$--index of the Dirac operator of generalized
Euclidean Taub--NUT metrics", preprint [arXiv:math--ph/0511025].
\bibitem{roman2}
A. Moroianu and S. Moroianu, ``The Dirac operator on generalized Taub--NUT
spaces", preprint [arXiv:1003.5364[math.DG]].
\bibitem{mike}
S.M. Christensen and M.J. Duff, ``Axial and conformal anomalies for arbitrary spin
in gravity and supergravity", Phys. Lett. \underline{B76} (1978) 571; ``New
gravitational index theorems and super theorems", Nucl. Phys. \underline{B154}
(1979) 301.
\bibitem{vann}
M.T. Grisaru, H. R\"omer, N.K. Nielsen and P. van Nieuwenhuisen, ``Approaches
to the gravitational spin--$3/2$ axial anomaly", Nucl. Phys. \underline{B140}
(1978) 477.
\bibitem{malcom}
M.J. Perry, ``Anomalies in supergravity", Nucl. Phys. \underline{B143} (1978) 114.
\bibitem{exerm}
H. R\"omer, ``Axial anomaly and boundary terms for spinor fields", Phys. Lett.
\underline{B83} (1979) 172.
\bibitem{redlich}
A.N. Redlich, ``Parity violation and gauge non--invariance of the effective gauge
field action in three dimensions", Phys. Rev. \underline{D29} (1984) 2366.
\bibitem{vuo}
I. Vuorio, ``Parity violation and the effective gravitational action in three
dimensions", Phys. Lett. \underline{B175} (1986) 176.
\bibitem{moore}
L. Alvarez--Gaum\'e, S. Della Pietra and G. Moore, ``Anomalies and odd dimensions",
Annals Phys. \underline{163} (1985) 288.
\bibitem{hamilton}
R.S. Hamilton, ``Three manifolds of positive Ricci curvature", J. Diff. Geom.
\underline{17} (1982) 255.
\bibitem{anderson}
J. Milnor, ``Towards the Poincar\'e conjecture and the classification of $3$--manifolds",
Notices of Amer. Math. Soc. \underline{50} (2003) 1226;
M.T. Anderson, ``Geometrization of $3$--manifolds via the Ricci flow", Notices of
Amer. Math. Soc. \underline{51} (2004) 184.
\bibitem{kron1}
N. Hitchin, ``Polygons and gravitons", Math. Proc. Camb. Phil. Soc.
\underline{85} (1979) 465.
\bibitem{kron2}
P.B. Kronheimer, ``The construction of ALE spaces as hyper--K\"ahler quotients",
J. Diff. Geom. \underline{29} (1989) 665.
\bibitem{goette}
S. Goette, ``Computations and applications of $\eta$ invariants", preprint
[arXiv:1011.4766[math.DG]].
\bibitem{fried1}
J.L. Friedman and R. Sorkin, ``Spin $1/2$ from gravity", Phys. Rev. Lett.
\underline{44} (1980) 1100.
\bibitem{fried2}
J.L. Friedman and D.M. Witt, ``Internal symmetry groups of quantum geons", Phys. Lett.
\underline{B120} (1983) 324.
\bibitem{regge1}
T. Regge and C. Teitelboim, ``Role of surface integrals in the Hamiltonian formulation 
of general relativity", Annals Phys. \underline{88} (1974) 286. 
\bibitem{regge2}
S.W. Hawking and G.T. Horowitz, ``The gravitational Hamiltonian, action, entropy and 
surface terms", Class. Quant. Grav. \underline{13} (1996) 1487 [arXiv:gr--qc/9501014]. 
\bibitem{duff}
M.J. Duff, ``Twenty years of the Weyl anomaly", Class. Quant. Grav. \underline{11}
(1994) 1387 [arXiv:hep--th/9308075].
\bibitem{theisen}
I. Adam, I.V. Melnikov and S. Theisen, ``A non--relativistic Weyl anomaly", JHEP
\underline{0909} (2009) 130 [arXiv:0907.2156[hep--th]].
\bibitem{davody}
A. Davody, ``Weyl anomaly in non--relativistic CFTs", Phys. Lett. \underline{B685}
(2010) 341 [arXiv:0909.3705[hep--th]].
\bibitem{BCS}
R. Jackiw and J.R. Schrieffer, ``Solitons with fermion number $1/2$ in condensed
matter and relativistic field theories", Nucl. Phys. \underline{B190}[FS3] (1981)
253.
\bibitem{graphene}
R. Jackiw and S.--Y. Pi, ``Chiral gauge theory for graphene", Phys. Rev. Lett.
\underline{98} (2007) 266402 [arXiv:cond--mat/0701760].
\bibitem{fulling}
S.A. Fulling, {\em Aspects of Quantum Field Theory in Curved Space--Time}, 
Cambridge University Press, Cambridge, 1989. 
\bibitem{carme}
M. Carmeli, {\em Group Theory and General Relativity}, McGraw--Hill, New York, 1977.
\bibitem{tenen}
G. Tenenbaum, {\em Introduction to Analytic and Probabilistic Number Theory},
Cambridge University Press, Cambridge, 1996.
\bibitem{habel}
M Habel and M. Peter, ``The eta invariant of Berger spheres and hypergeometric
identities", preprint, Mathematisches Institut, Albert--Ludwigs--Universit\"at, 2002.
\end{thebibliography}
\end{document}